%% file: aanda.tex
\documentclass[longauth]{aa}

\usepackage{graphicx}
\usepackage[dvipsnames]{xcolor}
\usepackage{txfonts}
\usepackage{subfigure}

\usepackage{float}
\usepackage{longtable,booktabs}
\usepackage{natbib,twoopt}
\usepackage{xspace}
\usepackage{balance}
\usepackage{upgreek}
\usepackage{multirow}
\usepackage{array}
\usepackage[normalem]{ulem}
\usepackage{soul}
\usepackage[colorlinks=true, allcolors=blue]{hyperref}
\usepackage{outlines}
\usepackage{makecell}
\usepackage{siunitx}

\usepackage{euclid}

\usepackage{tikz}
\usetikzlibrary{shapes.geometric, arrows}

\tikzstyle{startstop} = [rectangle, rounded corners, 
minimum width=3cm, 
minimum height=1cm,
text centered, 
draw=black, 
fill=red!30]

\tikzstyle{io} = [trapezium, 
trapezium stretches=true, % A later addition
trapezium left angle=70, 
trapezium right angle=110, 
minimum width=2.cm, 
minimum height=1cm, text centered, 
draw=black, fill=blue!20]

\tikzstyle{process} = [rectangle, 
minimum width=4.2cm, 
minimum height=1cm, 
text centered, 
text width=4.2cm, 
draw=black, 
fill=gray!7]

\tikzstyle{decision} = [diamond, 
minimum width=3cm, 
minimum height=1cm, 
text centered, 
draw=black, 
fill=green!30]
\tikzstyle{arrow} = [thick,->,>=stealth]
\usepackage[switch,mathlines]{lineno}

\defcitealias{Bisigello20}{B20}

\let\oldequation\equation
\let\oldendequation\endequation
\renewenvironment{equation}
  {\linenomathNonumbers\oldequation}
  {\oldendequation\endlinenomath}
  
\let\oldalign\align
\let\oldendalign\endalign
\renewenvironment{align}
  {\linenomathNonumbers\oldalign}
  {\oldendalign\endlinenomath}

\usepackage[flushleft]{threeparttable}

\def\apjl{ApJ Lett.}
\def\apjs{ApJ Suppl.}

\def\aap{A\&A}

\bibpunct{(}{)}{;}{a}{}{,}
\DeclareGraphicsExtensions{.eps,.ps,.pdf}

\makeatletter
\renewcommand*\aa@pageof{, page \thepage{} of \pageref*{LastPage}}
\makeatother

\begin{document}
%\linenumbers
\title{\Euclid preparation. XXXVII.}
\subtitle{Galaxy colour selections with \Euclid and ground photometry for cluster weak-lensing analyses}

\input{authors}

\date{Received --; accepted --}

\abstract
   {}
   {
   We derived galaxy colour selections from \textit{Euclid} and ground-based photometry, aiming to accurately define background galaxy samples in cluster weak-lensing analyses. These selections have been implemented in the \textit{Euclid} data analysis pipelines for galaxy clusters.
   }
   {
   Given any set of photometric bands, we developed a method for the calibration of optimal galaxy colour selections that maximises the selection completeness, given a threshold on purity. Such colour selections are expressed as a function of the lens redshift. 
   }
   {
   We calibrated galaxy selections using simulated ground-based $griz$ and \textit{Euclid} $Y_{\scriptscriptstyle\rm E}J_{\scriptscriptstyle\rm E}H_{\scriptscriptstyle\rm E}$ photometry. Both selections produce a purity higher than $97\%$. The $griz$ selection completeness ranges from 30\% to $84\%$ in the lens redshift range $z_{\rm l}\in[0.2,0.8]$. With the full $grizY_{\scriptscriptstyle\rm E}J_{\scriptscriptstyle\rm E}H_{\scriptscriptstyle\rm E}$ selection, the completeness improves by up to $25$ percentage points, and the $z_{\rm l}$ range extends up to $z_{\rm l}=1.5$. The calibrated colour selections are stable to changes in the sample limiting magnitudes and redshift, and the selection based on $griz$ bands provides excellent results on real external datasets. Furthermore, the calibrated selections provide stable results using alternative photometric aperture definitions obtained from different ground-based telescopes. The $griz$ selection is also purer at high redshift and more complete at low redshift compared to colour selections found in the literature. We find excellent agreement in terms of purity and completeness between the analysis of an independent, simulated \textit{Euclid} galaxy catalogue and our calibration sample, except for galaxies at high redshifts, for which we obtain up to 50 percent points higher completeness. The combination of colour and photo-$z$ selections applied to simulated \textit{Euclid} data yields up to 95\% completeness, while the purity decreases down to 92\% at high $z_{\rm l}$. We show that the calibrated colour selections provide robust results even when observations from a single band are missing from the ground-based data. Finally, we show that colour selections do not disrupt the shear calibration for stage III surveys. The first \textit{Euclid} data releases will provide further insights into the impact of background selections on the shear calibration.
   }
   {}

\keywords{Galaxies: statistics -- Galaxies: photometry -- Galaxies: distances and redshifts -- Galaxies: clusters: general -- Cosmology: observations -- large-scale structure of Universe}

\authorrunning{Euclid collaboration: G. F. Lesci et al.}

\titlerunning{Galaxy colour selections with \Euclid and ground photometry for cluster weak-lensing}

\maketitle

\section{Introduction}
In the last decade, galaxy clusters have proven to be excellent probes for cosmological analyses \citep[see, e.g.][]{Mantz15,Sereno15_cosm,Planck_counts,Costanzi19,Marulli21,Lesci22}, also allowing for the investigation of dark matter interaction models \citep{Peter13,Robertson17,Eckert22} and gas astrophysics \citep{Vazza17,CHEX-MATE,Zhu21,Sereno21}. As galaxy clusters are dominated by dark matter, the functional form of their matter density profiles can be derived from $N$-body dark-matter-only simulations \citep{NFW,BMO,DK14}. This allows one to estimate the mass of observed clusters, which is essential for both astrophysical and cosmological studies \citep{Teyssier11,Pratt19}. Currently, weak gravitational lensing is one of the most reliable methods to accurately and precisely measure cluster masses \citep{citlens2,citlens3,citlens4,ser+al17_psz2lens,citlens6,Schrabback21,Zohren22}. Consequently, weak-lensing cluster mass estimates are widely used in current photometric galaxy surveys, such as the Kilo Degree Survey \citep[KiDS;][]{KiDS1000,Bellagamba19}, the Dark Energy Survey \citep[DES;][]{DES_counts,DES}, and the Hyper Suprime-Cam survey \citep[HSC;][]{hsc_med+al18b,HSC}. \\
\indent An accurate selection of lensed background galaxies is crucial to derive a reliable cluster weak-lensing signal. Including the contribution from foreground and cluster member galaxies may significantly dilute the weak-lensing signal \citep{Broadhurst05, Medezinski07,Sifon15,McClintock19}. For example, background selections with 90\% purity dilute the cluster reduced shear measurements by 10\% \citep[see, e.g.][]{Dietrich19}, in the absence of intrinsic alignments \citep{Heymans03}. Highly pure background selections are required to properly account for this effect in weak-lensing measurements, in order to minimise the variance in the selection purity. Selection incompleteness, instead, impacts the weak-lensing noise and, in turn, the signal-to-noise ratio (S/N), which depends on the density of background sources along with the intrinsic ellipticity dispersion and measurement noise \citep[see, e.g.][]{Schrabback18,Umetsu20}. The effect of low background densities can be partially mitigated by increasing the size of the cluster-centric radial bins used in the analysis, or through the stacking of the weak-lensing signal of cluster ensembles.\\
\indent Background selections based on the galaxy photometric redshift (photo-$z$) posteriors are commonly used in the literature \citep{Gruen14,Applegate14,Melchior17,ser+al17_psz2lens,Bellagamba19}, as well as galaxy colour selections \citep{Medezinski10,Oguri12,hsc_med+al18b,Klein19}. These selections can also be combined to significantly improve the background sample completeness and, in turn, the weak-lensing S/N. In fact, colour selections have been demonstrated to help identify galaxies with poorly defined photometric redshifts that would not have been classified as background sources through photo-$z$ selection alone \citep{cov+al14,ser+al17_psz2lens,Bellagamba19}. \\
\indent The aim of this paper is to develop a method to obtain optimal colour selections, namely with a maximal completeness given a threshold on purity, given any set of photometric filters. We provide, for the first time, colour selections expressed as a continuous function of the lens limiting redshift. This allows for a finer background definition compared to colour selections found in the literature \citep{Medezinski10,Oguri12,hsc_med+al18b}, implying a significant improvement in the weak-lensing source statistics. In view of \Euclid and stage IV surveys, we derived colour selections on simulated data. We exploited the galaxy catalogue developed by \citet{Bisigello20}, hereafter referred to as \citetalias{Bisigello20}, and extended by \citet{Bisigello22}, which includes simulated Sloan Digital Sky Survey \citep[SDSS;][]{Gunn98} $griz$ magnitudes and simulated \textit{Euclid} observations in the $Y_{\scriptscriptstyle\rm E}J_{\scriptscriptstyle\rm E}H_{\scriptscriptstyle\rm E}$ bands. In addition, we tested the efficiency of these colour selections on real public external data and on simulations, combining them with photo-$z$ selections. \\
\indent This paper is part of a series presenting and discussing mass measurements of galaxy clusters using the \textit{Euclid} combined clusters and
weak-lensing pipeline \texttt{COMB-CL}. \texttt{COMB-CL} forms part of the
global \textit{Euclid} data processing pipeline and is responsible for measuring weak-lensing shear profiles and masses for photometrically detected clusters. A comprehensive description of the code structure and methods employed by \texttt{COMB-CL} will be presented in a forthcoming paper, but a brief overview of the pipeline can be found in the appendix of \citet{Sereno23}. The galaxy colour selections presented in this paper are already implemented in \texttt{COMB-CL}. \\
\indent The paper is organised as follows. In Sect.\ \ref{sec:data}, we describe the dataset used for the calibration of galaxy colour selections, and in Sect.\ \ref{sec:method} we detail a general method to derive optimal colour selections. In Sect.\ \ref{sec:results}, we show the selections obtained for $griz$ and $grizY_{\scriptscriptstyle\rm E}J_{\scriptscriptstyle\rm E}H_{\scriptscriptstyle\rm E}$ filter sets, validating them on external datasets. In Sect.\ \ref{sec:literature} we compare the $griz$ selection calibrated in this work with selections from the literature. Finally, in Sect.\ \ref{sec:summary}, we draw our conclusions.

\section{Calibration sample}\label{sec:data}
\begin{figure}[t!]
\centering
\includegraphics[width = \hsize, height = 6.7cm] {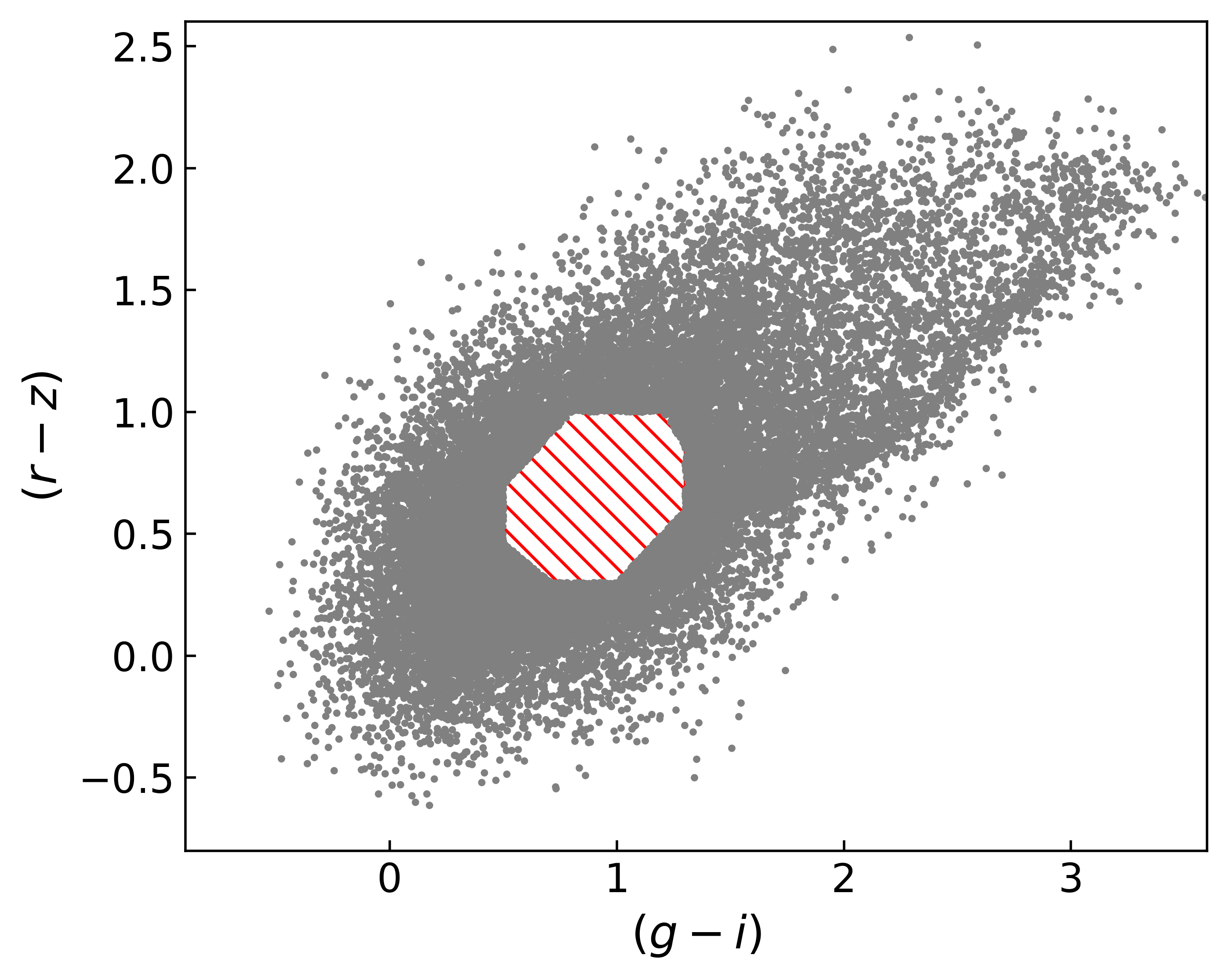}
\caption{Example of an uncalibrated selection in the $(r-z)$ - $(g-i)$ colour-colour space. The grey dots represent the selected galaxy colours. Galaxies within the octagonal hatched region are excluded by applying Eq.\ \eqref{eq:all_cond}. Specifically, $(r-z)$ and $(g-i)$ correspond to $x$ and $y$ in Eq.\ \eqref{eq:all_cond}, respectively. }
\label{fig:example_CCspace}
\end{figure}
We based our analysis on the photometric catalogue developed by \citetalias{Bisigello20} and extended by \citet{Bisigello22}. This catalogue contains simulated \textit{Euclid} $I_{\scriptscriptstyle\rm E}Y_{\scriptscriptstyle\rm E}J_{\scriptscriptstyle\rm E}H_{\scriptscriptstyle\rm E}$ aperture magnitudes\footnote{$I_{\scriptscriptstyle\rm E}$ band observations are supplied by the \textit{Euclid} Visible Imager \citep[VIS;][]{Cropper16}, while $Y_{\scriptscriptstyle\rm E}J_{\scriptscriptstyle\rm E}H_{\scriptscriptstyle\rm E}$ photometry is provided by the Near-Infrared Spectrometer and Photometer \citep[NISP;][]{Schirmer22}.}, covering the spectral range 5500--20\,000 \AA, along with the Canada-France Imaging Survey \citep[CFIS;][]{CFIS} $u$ band, for the galaxies contained in the COSMOS catalogue by \citet[][COSMOS15]{Laigle16}. Specifically, such photometry is based on \ang{;;3} fixed-aperture magnitudes. Despite the $u$ band already being present in COSMOS15, \citetalias{Bisigello20} derived it using the same approach adopted for the other filters in order to avoid colour biases. \citetalias{Bisigello20} verified that this provides results that are consistent with the observed fluxes. Simulated SDSS $griz$ magnitudes, spanning the wavelength range 4000--11\,000 \AA, are also provided in the catalogue, since observations in similar filters, such as those in \textit{Vera C. Rubin Observatory} \citep[\textit{Rubin}/LSST;][]{LSST} and DES, will be available to complement \textit{Euclid} observations \citep{Pocino21,Scaramella22}. Corrections for photometric offsets due to flux outside the fixed-aperture, systematic offsets, and Galactic extinction, as suggested in \citet{Laigle16}, have been applied. \citetalias{Bisigello20} derive simulated magnitudes through two alternative approaches. The first is a linear interpolation of the 30 medium-band and broad-band filters available in the COSMOS15 catalogue, based on the effective wavelength of the filters. The second approach is based on the best theoretical template that describes the spectral energy distribution (SED) of each galaxy, assuming the COSMOS15 redshifts as the ground truth. The SED fitting is performed based on COSMOS15 bands and the template resulting in the minimum $\chi^{2}$ is used to predict the expected fluxes. We refer to \citetalias{Bisigello20} for the details of the SED templates used, based on the model by \citet{Bruzual03}. The expected fluxes are then randomised 10 times considering a Gaussian distribution centred on the true flux and with standard deviation equal to the expected photometric uncertainities, scaled considering the depths listed in Table 1 of \citet{Bisigello22}. In this process, the $I_{\scriptscriptstyle\rm E}Y_{\scriptscriptstyle\rm E}J_{\scriptscriptstyle\rm E}H_{\scriptscriptstyle\rm E}$ magnitude errors expected for the Euclid Wide Survey are considered. Despite the fact that the $griz$ photometry is based on SDSS filter transmissions, the corresponding uncertainties are based on depths that are consistent with those of DES and the Ultraviolet Near-Infrared Optical Northern Survey (UNIONS).\footnote{UNIONS is carried out with the Subaru Telescope \citep{Iye04}, the Canada-France-Hawaii Telescope \citep[CFHT;][]{Gwyn12}, and the Panoramic Survey Telescope and Rapid Response System \citep[Pan-STARRS;][]{Chambers16}. More information at \url{https://www.skysurvey.cc/news/}.} The $ugriz$ photometry provided by LSST is expected to go from 1 to 2.5 magnitudes deeper at the end of the \Euclid mission, depending on the photometric filter. Throughout this paper, we focus on the magnitudes derived from the best theoretical SED templates, as these estimates better reproduce absorption and emission lines that are not covered by COSMOS15 bands. We neglect $u$ magnitudes since, due to the low $u$-band throughput, a 5$\sigma$ depth of 25.6 mag will only be reached after 10 years of LSST observations\footnote{\url{https://www.lsst.org/scientists/keynumbers}}. In addition, the $u$ band is not available in DES wide fields. We emphasise that the \citetalias{Bisigello20} catalogue contains all the galaxies present in the COSMOS15 sample, which is deeper than the shear samples derived from current surveys \citep[see, e.g.][]{Giblin21,Gatti21} and expected from the Euclid Wide Survey \citep{Scaramella22}. As we discuss in the following, the colour selections calibrated in this study yield robust results against alternative magnitude cuts, including those that reproduce the selections adopted in current and \Euclid cosmic shear analyses.

\section{Method}\label{sec:method}
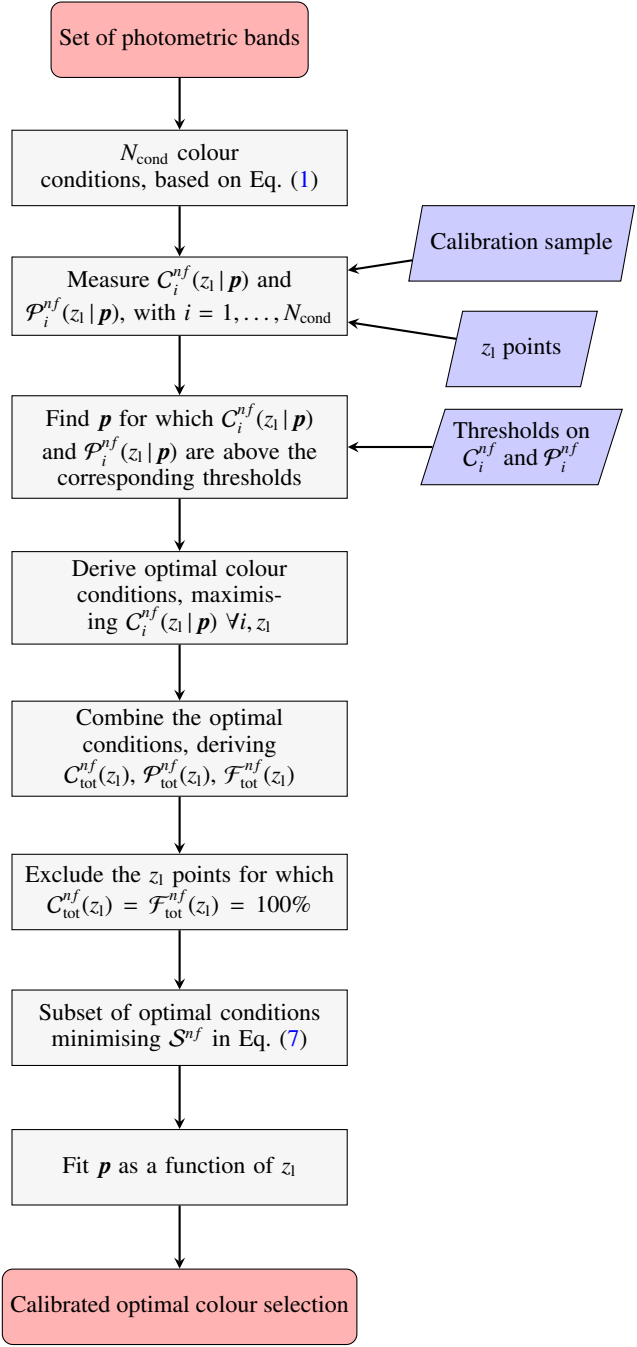
\begin{figure}[ht]
\begin{tikzpicture}[node distance=2cm, align=center]
\small

% Boxes
\node (start) [startstop] {Set of photometric bands};

\node (in1) [process, below of=start, yshift=0.3cm] {$N_{\rm cond}$ colour\\ conditions, based on Eq.\ \eqref{eq:all_cond}};

\node (pro1) [process, below of=in1, yshift=0.3cm] {Measure $\mathcal{C}_i^{nf}(z_{\rm l}\,|\,\vec{p})$ and $\mathcal{P}_i^{nf}(z_{\rm l}\,|\,\vec{p})$, with $i=1,\ldots,N_{\rm cond}$};

\node (in2) [io, right of=pro1, xshift=2.5cm, yshift=0.7cm] {Calibration sample};

\node (in3) [io, right of=pro1, xshift=2.5cm, yshift=-0.7cm] {$z_{\rm l}$ points};

\node (pro2) [process, below of=pro1, yshift=0cm] {Find $\vec{p}$ for which $\mathcal{C}_i^{nf}(z_{\rm l}\,|\,\vec{p})$ and $\mathcal{P}_i^{nf}(z_{\rm l}\,|\,\vec{p})$ are above the corresponding thresholds};

\node (in4) [io, right of=pro2, xshift=2.5cm] {Thresholds on \\ $\mathcal{C}_i^{nf}$ and $\mathcal{P}_i^{nf}$};

\node (pro3) [process, below of=pro2, yshift=0cm] {Derive optimal colour conditions, maximising $\mathcal{C}_i^{nf}(z_{\rm l}\,|\,\vec{p})$ $\forall i,z_{\rm l}$};

\node (pro4) [process, below of=pro3, xshift=0cm, yshift=0.cm] {Combine the optimal conditions, deriving $\mathcal{C}^{nf}_{\rm tot}(z_{{\rm l}})$, $\mathcal{P}^{nf}_{\rm tot}(z_{{\rm l}})$, $\mathcal{F}^{nf}_{\rm tot}(z_{{\rm l}})$};

\node (pro5) [process, below of=pro4, yshift=0.1cm] {Exclude the $z_{\rm l}$ points for which $\mathcal{C}^{nf}_{\rm tot}(z_{{\rm l}})=\mathcal{F}^{nf}_{\rm tot}(z_{{\rm l}})=100\%$};

\node (pro6) [process, below of=pro5, yshift=0.2cm] {Subset of optimal conditions minimising $\mathcal{S}^{nf}$ in Eq.\ \eqref{eq:S}};

\node (pro7) [process, below of=pro6, yshift=0.15cm] {Fit $\vec{p}$ as a function of $z_{\rm l}$};

\node (end) [startstop, below of=pro7, yshift=0.15cm] {Calibrated optimal colour selection};

% Arrows
\draw [arrow] (start) -- (in1);
\draw [arrow] (in1) -- (pro1);
\draw [arrow] (in2) -- (pro1);
\draw [arrow] (in3) -- (pro1);
\draw [arrow] (pro1) -- (pro2);
\draw [arrow] (in4) -- (pro2);
\draw [arrow] (pro2) -- (pro3);
\draw [arrow] (pro3) -- (pro4);
\draw [arrow] (pro4) -- (pro5);
\draw [arrow] (pro5) -- (pro6);
\draw [arrow] (pro6) -- (pro7);
\draw [arrow] (pro7) -- (end);

\end{tikzpicture}
\caption{Flowchart summarising the calibration process described in Sect.\ \ref{sec:method}. Round red rectangles represent the start and end points of the calibration process. Grey rectangles represent processing steps, while blue trapezoids correspond to the inputs.}
\label{fig:flowchart}
\end{figure}
In order to find a set of optimal galaxy colour-redshift relations that maximises the selection completeness given a threshold on the foreground contamination, we considered the colours given by any combination of photometric bands. This includes bands that are not adjacent in wavelength. Thus, for each colour-colour space, given a redshift lower limit, $z_{\rm l}$, corresponding to the lens redshift, we considered the following set of conditions,
\begin{align}\label{eq:all_cond}
&x > c_1\,\lor\,\nonumber\\
&x < c_2\,\lor\,\nonumber\\
&y > c_3\,\lor\,\nonumber\\
&y < c_4\,\lor\,\nonumber\\
&x > s_1 y + c_5\,\lor\,\nonumber\\
&x > s_2 y + c_6\,\lor\,\nonumber\\
&x < s_3 y + c_7\,\lor\,\nonumber\\
&x < s_4 y + c_8,
\end{align}
where $\lor$ is the logical `or' operator, $x$ and $y$ are two different colours, and $c_i$ and $s_i$ are colour selection parameters. Specifically, $c_1,\ldots,c_8\in(-\infty,+\infty)$, $s_1$ and $s_3\in(0,+\infty)$, while $s_2$ and $s_4\in(-\infty,0)$. The edges of the aforementioned parameter ranges are excluded, and Eq.\ \eqref{eq:all_cond} defines an irregular octagon that contains the foreground galaxies, as we show in Fig.\ \ref{fig:example_CCspace}. As we shall see, since we only select the colour conditions that satisfy given requirements, not all the sides of the irregular octagon may be considered. In addition, since we considered the conditions in Eq.\ \eqref{eq:all_cond} as independent, the $c_1,\ldots,c_8$ and $s_1,\ldots,s_4$ parameters are not related to each other. In particular, for each condition in Eq.\ \eqref{eq:all_cond}, we derived the completeness, 
\begin{equation}\label{eq:C}
\mathcal{C}_i^{nf}(z_{\rm l}\,|\,\vec{p}) := \frac{N_{{\rm sel},i}(z_{\rm g}>z_{\rm l}\,|\,\vec{p})}{N_{\rm tot}(z_{\rm g}>z_{\rm l})} \,,
\end{equation}
and the purity,
\begin{equation}\label{eq:P}
\mathcal{P}_i^{nf}(z_{\rm l}\,|\,\vec{p}) := \frac{N_{{\rm sel},i}(z_{\rm g}>z_{\rm l}\,|\,\vec{p})}{N_{{\rm sel},i}(z_{\rm g}\geq0\,|\,\vec{p})} \,,
\end{equation}
where $i$ is the $i$th colour condition index, $z_{\rm g}$ is the galaxy redshift, $\vec{p}$ is the set of colour condition parameters, $N_{{\rm sel},i}$ is the number of galaxies selected with the $i$th colour condition, $N_{\rm tot}$ is the total number of galaxies in the calibration sample, while the $nf$ superscript represents quantities derived from colour conditions not fitted as a function of $z_{\rm l}$. As we shall see, we do not adopt any superscripts for the quantities derived from fitted colour conditions. In Eqs. \eqref{eq:C} and \eqref{eq:P}, we have $i=1...\,N_{\rm cond}$, where $N_{\rm cond}$ is the number of all possible colour conditions, given Eq.\ \eqref{eq:all_cond}, expressed as
\begin{equation}
N_{\rm cond} = 8\,\frac{N_{\rm col}!}{(N_{\rm col}-2)!\,2!},
\end{equation}
where $N_{\rm col}$ is the number of colours, given by
\begin{equation}
N_{\rm col} = \frac{N_{\rm band}!}{(N_{\rm band}-2)!\,2!},
\end{equation}
where $N_{\rm band}$ is the number of photometric bands. \\
\indent We set requirements on completeness and purity to be satisfied by each colour condition in Eq.\ \eqref{eq:all_cond}. Specifically, for a given $z_{\rm l}$, we selected the colour conditions having at least one $\vec{p}$ set providing $\mathcal{C}_i^{nf}(z_{\rm l}\,|\,\vec{p})$ and $\mathcal{P}_i^{nf}(z_{\rm l}\,|\,\vec{p})$ larger than their corresponding thresholds. We remark that $\vec{p}$ does not explicitly depend on $z_{\rm l}$ at this stage, and that $z_{\rm l}$ values are arbitrarily sampled. Setting a threshold on $\mathcal{C}_i^{nf}(z_{\rm l}\,|\,\vec{p})$ is important for excluding colour conditions that do not significantly contribute to the total completeness, and that may appear as optimal only due to statistical fluctuations. Thus, the threshold on $\mathcal{C}_i^{nf}(z_{\rm l}\,|\,\vec{p})$ is meant to be low compared to that on $\mathcal{P}_i^{nf}(z_{\rm l}\,|\,\vec{p})$. Indeed, as we shall detail in Sect.\ \ref{sec:shear_measurements}, impurities in the background selection imply systematic uncertainties in galaxy cluster reduced shear measurements. Highly pure selections are required to properly account for this effect, in order to minimise the scatter in purity. We discuss the choice of the thresholds on $\mathcal{C}_i^{nf}(z_{\rm l}\,|\,\vec{p})$ and $\mathcal{P}_i^{nf}(z_{\rm l}\,|\,\vec{p})$ in greater detail in Sect.\ \ref{sec:results:calibration}. For each colour condition in Eq.\ \eqref{eq:all_cond}, with parameter values for which the conditions on $\mathcal{C}_i^{nf}(z_{\rm l}\,|\,\vec{p})$ and $\mathcal{P}_i^{nf}(z_{\rm l}\,|\,\vec{p})$ are satisfied, we selected the $\vec{p}$ set providing the highest completeness at a given $z_{\rm l}$. In this way, we derived the set of optimal colour conditions maximising the selection completeness, given the chosen threshold on purity. \\
\indent We note that the maximum $z_{\rm l}$ of the calibrated colour selections depends on the $\mathcal{C}_i^{nf}(z_{\rm l}\,|\,\vec{p})$ and $\mathcal{P}_i^{nf}(z_{\rm l}\,|\,\vec{p})$ limits, while the minimum $z_{\rm l}$ is derived by excluding the $z_{\rm l}$ points for which $\mathcal{C}^{nf}_{\rm tot}(z_{{\rm l}})=\mathcal{F}^{nf}_{\rm tot}(z_{{\rm l}})=100\%$. Here, $\mathcal{C}^{nf}_{\rm tot}$ and $\mathcal{F}^{nf}_{\rm tot}$ are the completeness and the foreground failure rate given by the full set of optimal colour conditions, respectively. For simplicity, we drop the dependence on $\vec{p}$ in the text. The foreground failure rate is defined as follows:
\begin{equation}\label{eq:F}
\mathcal{F}_{\rm tot}^{nf}(z_{\rm l}) := \frac{N_{{\rm sel}}(z_{\rm g}<z_{\rm l})}{N_{\rm tot}(z_{\rm g}<z_{\rm l})} = \frac{N_{{\rm sel}}(z_{\rm g}>z_{\rm l})}{N_{\rm tot}(z_{\rm g}<z_{\rm l})} \,\frac{1-\mathcal{P}_{\rm tot}^{nf}(z_{\rm l})}{\mathcal{P}_{\rm tot}^{nf}(z_{\rm l})} \,,
\end{equation}
where $N_{{\rm sel}}$ is the number of galaxies selected with all the optimal colour conditions, given a condition on $z_{\rm g}$, and $\mathcal{P}_{\rm tot}^{nf}(z_{\rm l})$ is the purity given by the full set of optimal conditions. On the right-hand side of Eq.\ \eqref{eq:F}, derived from Eqs.\ \eqref{eq:C} and \eqref{eq:P}, we can see that $\mathcal{F}_{\rm tot}^{nf}(z_{\rm l})$ diminishes with increasing $z_{\rm l}$ if high lower limits on purity are chosen. We stress that $\mathcal{F}_{\rm tot}^{nf}(z_{\rm l})\leq1$ by definition. \\
\begin{figure}[t!]
\centering
\begin{subfigure}{}
    \includegraphics[width = \hsize] {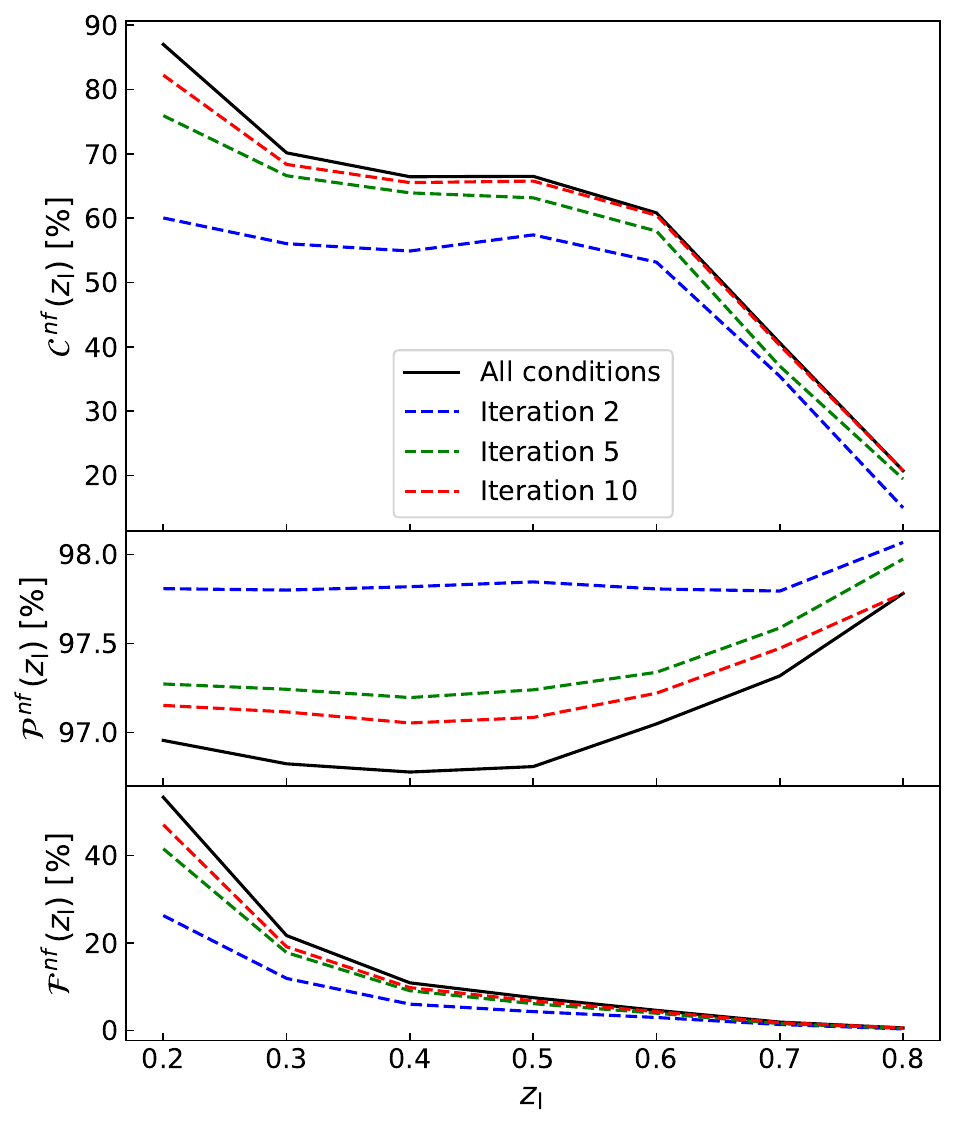}
    \caption{Selection completeness (top panel), purity (middle panel), and foreground failure rate (bottom panel), derived from subsets of optimal colour conditions not fitted as a function of $z_{\rm l}$, for the case of $griz$ photometry. The solid black lines represent the selection given by the full set of optimal colour conditions, while the dashed lines show the selection at different steps of the iterative process detailed in Sect.\ \ref{sec:method}, given by subsets of optimal conditions.}
    \label{fig:examples1}
\end{subfigure}
\begin{subfigure}{}
    \includegraphics[width = \hsize] {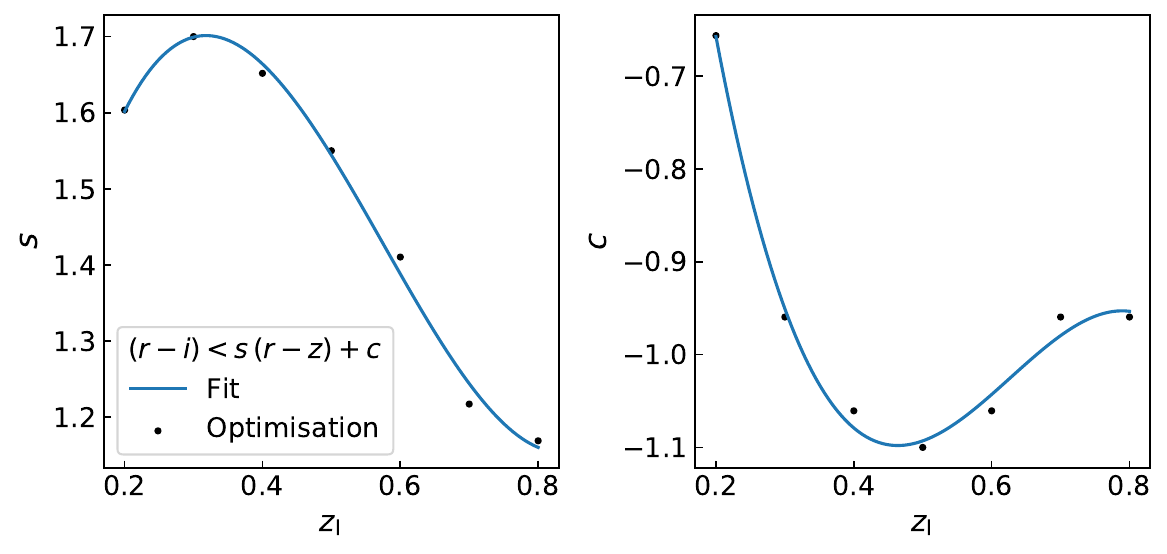}
    \caption{Values of $s$ (left panel) and $c$ (right panel) parameters, from Eq.\ \eqref{eq:all_cond}, as a function of $z_{\rm l}$ for the colour condition quoted in the left panel legend. The black dots represent the optimal values of $s$ and $c$, while the blue curves represent the polynomial fits.}
    \label{fig:examples2}
\end{subfigure}
\end{figure}
\indent In the selection process described above, some colour conditions may be redundant. Thus, we iteratively searched for an optimal subset of colour conditions to find the minimum number of conditions sufficient to approximately reproduce the required completeness. Specifically, at each step of this iterative process, we computed the following quantity:
\begin{equation}\label{eq:S}
\mathcal{S}^{nf} = \sum_{j=1}^{N} \mathcal{C}^{nf}_{\rm tot}(z_{{\rm l},j}) - \mathcal{C}^{nf}(z_{{\rm l},j}),
\end{equation}
where $N$ is the number of $z_{\rm l}$ points, $\mathcal{C}^{nf}_{\rm tot}(z_{{\rm l},j})$ is the completeness given by all optimal conditions, while $\mathcal{C}^{nf}(z_{{\rm l},j})$ is the completeness given by a subset of optimal conditions, computed at the $j$th $z_{\rm l}$ value. As the first step of this iterative process, we found the optimal colour condition minimising $\mathcal{S}^{nf}$. Then, at each iteration, we added the colour condition that, combined with the conditions selected in the previous steps, minimises $\mathcal{S}^{nf}$. We repeated this process until $\mathcal{S}^{nf}$ was lower than a given tolerance. We remark that the logical operator between colour conditions is $\lor$. \\
\indent Lastly, we applied a nonlinear least squares analysis to find the best fit to the $\vec{p}$ parameters as a function of $z_{\rm l}$ for the subset of optimal colour conditions. We chose the fitting formulae which best reproduce the $z_{\rm l}$ dependence, namely polynomials, while aiming at minimising the number of free parameters in the fit. \\
\indent In Fig.\ \ref{fig:flowchart} we show a flowchart summarising the calibration process described in this section. In Fig.\ \ref{fig:examples1} we show an example of the iterative process detailed above, while Fig.\ \ref{fig:examples2} displays an example of parameter dependence on $z_{\rm l}$. Hereafter, we refer to the completeness, purity, and foreground failure rate, derived from sets of fitted colour conditions, as $\mathcal{C}(z_{\rm l})$, $\mathcal{P}(z_{\rm l})$, and $\mathcal{F}(z_{\rm l})$, respectively. For better clarity, in Table \ref{tab:symbols} we summarise the symbols referring to the completeness functions introduced in this section.

\section{Results}\label{sec:results}
\begin{table}[t]
\caption{\label{tab:symbols}Description of the completeness functions introduced in Sect.\ \ref{sec:method}. Analogous descriptions hold for purity and foreground failure rate.}
  \centering
    \begin{tabular}{ | c | c | } 
      \hline
      \rule{0pt}{2.7ex}
      Symbol & Description \\[2.4pt]

      \hline
      \rule{0pt}{3.6ex}
      $C_i^{nf}$ & \makecell{Completeness of the $i$th colour condition,\\ given a set of sampled parameters.} \\[6.5pt]

      \hline
      \rule{0pt}{3.6ex}
      $C_{\rm tot}^{nf}$ & \makecell{Completeness derived through the combination\\ of all the optimal colour conditions.} \\[6.5pt]

      \hline
      \rule{0pt}{3.6ex}
      $C^{nf}$ & \makecell{Completeness given by the combination\\ of a subset of optimal colour conditions.} \\[6.5pt]

      \hline
      \rule{0pt}{3.6ex}
      $C$ & \makecell{Completeness obtained from a subset of optimal\\ colour conditions fitted as a function of $z_{\rm l}$.} \\[6.5pt]
      
      \hline
    \end{tabular}
  \tablefoot{The $nf$ superscript represents quantities derived from colour conditions not fitted as a function of $z_{\rm l}$. Optimal colour conditions satisfy the thresholds on purity and completeness, and provide maximal completeness.}
\end{table}

\subsection{Calibration of colour selections}\label{sec:results:calibration}
\begin{figure*}[t]
\centering
\includegraphics[width = \hsize-2.7cm, height = 6cm] {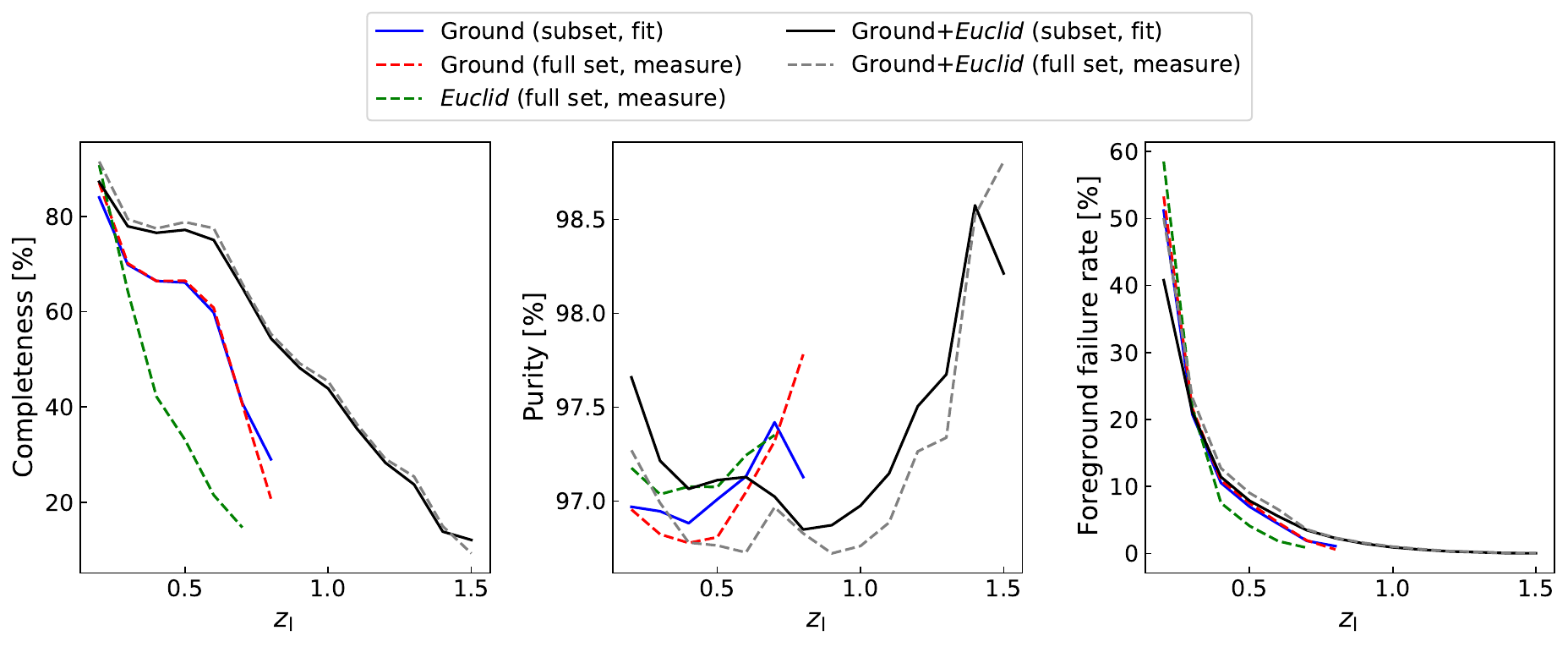}
\includegraphics[width = \hsize-2.7cm, height = 5cm] {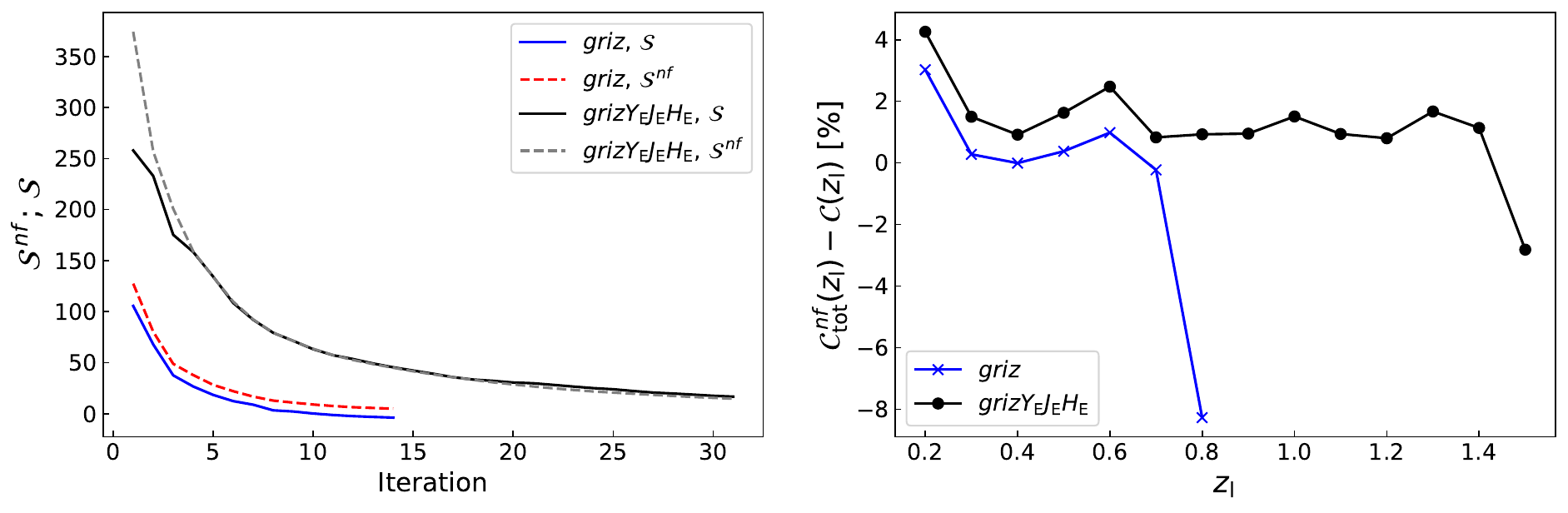}
\caption{Summary of the results on the colour selection optimisation, based on the \citetalias{Bisigello20} galaxy sample. \textit{Top panels}: Selection completeness (left panel), purity (central panel), and foreground failure rate (right panel), as a function of $z_{\rm l}$. The dashed lines represent the selections derived from the full sets of optimal conditions not fitted as a function of $z_{\rm l}$, in the case of ground-only (red), \textit{Euclid}-only (green), and the combination of ground-based and \textit{Euclid} bands (grey). The solid lines represent the selections obtained from the subsets of optimal conditions, with parameters fitted as a function of $z_{\rm l}$, in the case of ground-only (blue) and for the combination of ground-based and \textit{Euclid} bands (black). \textit{Bottom panels}: In the left panel, $\mathcal{S}$ and $\mathcal{S}^{nf}$ are shown as a function of the iteration number. For the ground-based selection, using $griz$ filters, $\mathcal{S}$ and $\mathcal{S}^{nf}$ are represented by solid blue and dashed red lines, respectively. For the selection derived from the combination of ground-based and \textit{Euclid} filters, namely $grizY_{\scriptscriptstyle\rm E}J_{\scriptscriptstyle\rm E}H_{\scriptscriptstyle\rm E}$, $\mathcal{S}$ and $\mathcal{S}^{nf}$ are represented by solid black and dashed grey lines, respectively. In the right panel, the difference between $\mathcal{C}^{nf}_{\rm tot}$ and $\mathcal{C}$ is shown, for the $griz$ (blue lines) and $grizY_{\scriptscriptstyle\rm E}J_{\scriptscriptstyle\rm E}H_{\scriptscriptstyle\rm E}$ (black lines) selections.}
\label{fig:all_comparison}
\end{figure*}
\begin{figure*}[t]
\centering
\includegraphics[width = \hsize-2.7cm, height = 5cm] {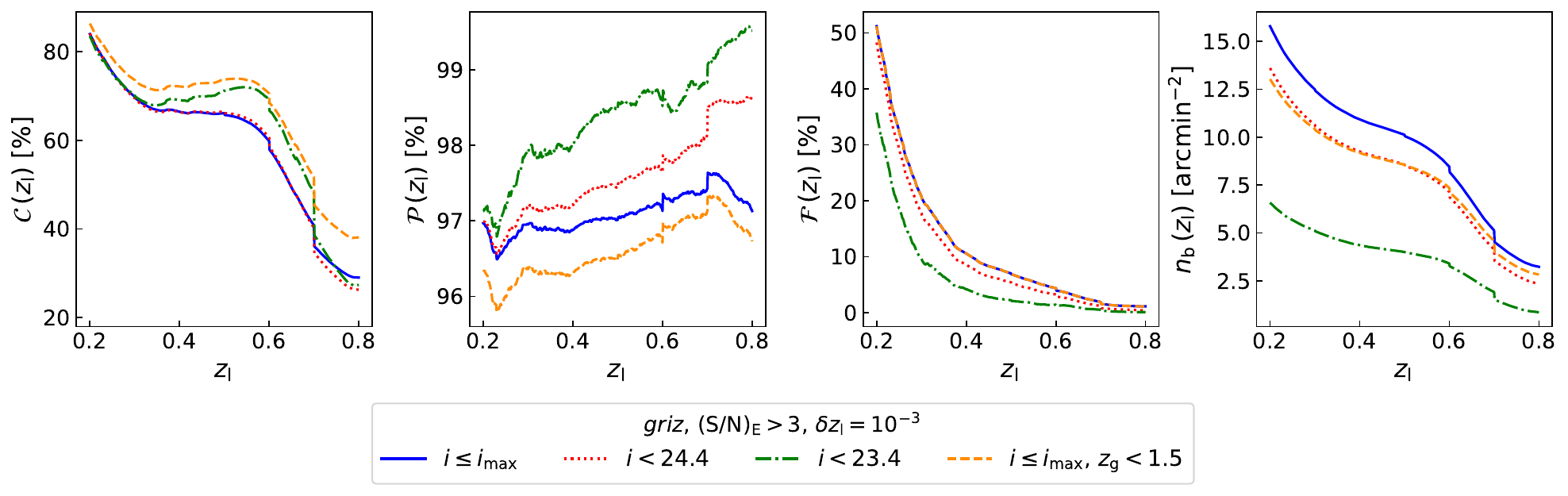}

\vspace{0.6cm}

\includegraphics[width = \hsize-2.7cm, height = 5cm] {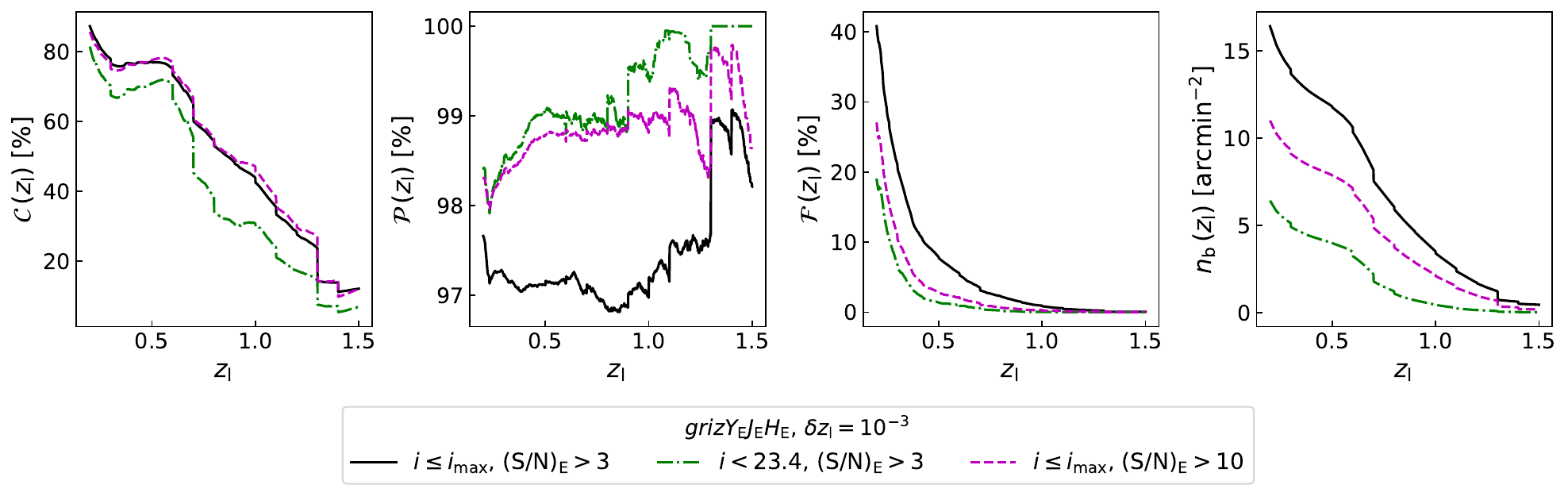}
\caption{Results from the fitted colour selections derived in Sect.\ \ref{sec:results:calibration}, assuming the alternative magnitude and redshift selections described in Sect.\ \ref{sec:limits}. From left to right: completeness, purity, foreground failure rate, and background density as a function of $z_{\rm l}$. The assumed $z_{\rm l}$ precision is $\delta z_{\rm l}=10^{-3}$. \textit{Top panels:} efficiency of the $griz$ selection, detailed in Table \ref{tab:griz}, applied to the \citetalias{Bisigello20} catalogue with $({\rm S/N})_{\rm E}>3$ and $i\leq i_{\rm max}$ (blue solid lines), to its subsample including galaxies with $i<24.4$ (red dotted lines), to the case with $i<23.4$ (green dash-dotted lines), and to the sample with $z_{\rm g}<1.5$ (orange dashed lines). \textit{Bottom panels:} efficiency of the $grizY_{\scriptscriptstyle\rm E}J_{\scriptscriptstyle\rm E}H_{\scriptscriptstyle\rm E}$ selection, detailed in Table \ref{tab:grizYJH}, applied to the \citetalias{Bisigello20} catalogue with $({\rm S/N})_{\rm E}>3$ and $i\leq i_{\rm max}$ (black solid lines), to its subsample with $i<23.4$ (green dash-dotted lines), and to the subsample with $({\rm S/N})_{\rm E}>10$ (magenta dashed lines).}
\label{fig:magLims}
\end{figure*}
By applying the methods detailed in Sect.\ \ref{sec:method} and adopting the \citetalias{Bisigello20} calibration sample described in Sect.\ \ref{sec:data}, we calibrated galaxy colour selections using ground-based and \textit{Euclid} photometry, namely SDSS $griz$ and \textit{Euclid} $Y_{\scriptscriptstyle\rm E}J_{\scriptscriptstyle\rm E}H_{\scriptscriptstyle\rm E}$ filters, respectively. These selections are implemented in \texttt{COMB-CL}, and will be available for weak-lensing analyses of galaxy clusters. We considered the following cases: ground-only, \textit{Euclid}-only, and the combination of ground-based and \textit{Euclid} photometry. For the cases including \textit{Euclid} photometry, we adopted an ${\rm S/N}$ threshold for \textit{Euclid} near-infrared observations of $({\rm S/N})_{\rm E}>3$, which corresponds to $Y_{\scriptscriptstyle\rm E}<24.85$, $J_{\scriptscriptstyle\rm E}<25.05$, and $H_{\scriptscriptstyle\rm E}<24.95$ \citep{Scaramella22}. In addition, we considered $z_{\rm l}$ points in the range $z_{\rm l}\in[0.1,2.5]$, assuming a precision of $\delta z_{\rm l}=0.1$ for the sampling. To derive the full set of optimal colour conditions, we imposed $\mathcal{C}_i^{nf}(z_{\rm l}\,|\,\vec{p})>10\%$ for the $i$th colour condition. For the ground-only and \textit{Euclid}-only cases, we imposed that the purity of each colour condition is $\mathcal{P}_i^{nf}(z_{\rm l}\,|\,\vec{p})>99\%$. We adopted a more restrictive threshold on purity for the combination of ground-based and \textit{Euclid} photometry, corresponding to $\mathcal{P}_i^{nf}(z_{\rm l}\,|\,\vec{p})>99.7\%$. This threshold is chosen as the larger number of colour combinations leads to a higher summation of impurities. We obtained $\mathcal{P}^{nf}(z_{\rm l})> 97\%$ for any $z_{\rm l}$, when combining all the optimal colour conditions, as shown in Fig.\ \ref{fig:all_comparison}. As we shall discuss in Sect.\ \ref{sec:real_data}, the purity derived from different real datasets is stable, showing sub-percent changes, on average.\\
\indent The $\mathcal{F}^{nf}(z_{\rm l})$ decrease with increasing $z_{\rm l}$, shown in Fig.\ \ref{fig:all_comparison}, is expected, as discussed in Sect.\ \ref{sec:method}. In addition, for any combination of photometric bands, we found $\mathcal{C}^{nf}_{\rm tot}(z_{\rm l})=\mathcal{F}^{nf}_{\rm tot}(z_{\rm l})=100\%$ for $z_{\rm l}=0.1$. Consequently, we set $z_{\rm l}=0.2$ as the minimum lens redshift for the calibrated colour selections. As shown in Fig.\ \ref{fig:all_comparison}, from $griz$ photometry we derived a selection within $z_{\rm l}\in[0.2,0.8]$, with 84\% completeness at $z_{\rm l}=0.2$, decreasing to 29\% at $z_{\rm l}=0.8$. In the \textit{Euclid}-only case, namely $Y_{\scriptscriptstyle\rm E}J_{\scriptscriptstyle\rm E}H_{\scriptscriptstyle\rm E}$ and $I_{\scriptscriptstyle\rm E}$ bands, results are not competitive with those derived from $griz$ photometry. On the other hand, by combining ground-based and \textit{Euclid} photometry, the completeness significantly increases in the $z_{\rm l}$ range covered by the $griz$ selection, by up to 25 percent points. Also the $z_{\rm l}$ range of the selection is significantly extended compared to the $griz$ case, corresponding to $z_{\rm l}\in[0.2,1.5]$. Specifically, in this case we exclude the \textit{Euclid} $I_{\scriptscriptstyle\rm E}$ band, as it covers a large wavelength interval, namely $\sim\,$5000--10\,000 $\AA$, corresponding to the wavelength range already covered by $griz$ photometry. Furthermore, the use of very broad photometric bands is not the most optimal choice for calibrating galaxy colour selections, which share similarities with photo-$z$ estimates. \\
\indent We excluded any possible redundant colour condition, as detailed in Sect.\ \ref{sec:method}. In Table \ref{tab:griz} we show the subset of optimal colour conditions for the ground-only case, namely $griz$ photometry, along with the corresponding parameter fits. The first condition quoted in Table \ref{tab:griz} corresponds to the one derived in the first step of the iterative process described in Sect.\ \ref{sec:method}. This is analogous for the subsequent conditions. We remark that the quoted conditions have different ranges of validity in $z_{\rm l}$. Analogous information is listed in Table \ref{tab:grizYJH} for the combination of ground-based and \textit{Euclid} photometry, corresponding to $grizY_{\scriptscriptstyle\rm E}J_{\scriptscriptstyle\rm E}H_{\scriptscriptstyle\rm E}$ filters. We neglected the optimisation and parameter fitting for the \textit{Euclid}-only case, as we have already shown that it does not provide competitive completeness values. \\
\indent In Fig.\ \ref{fig:all_comparison} we show the results for the selections obtained from the subsets of optimal conditions, with parameters fitted as a function of $z_{\rm l}$. For both $griz$ and $grizY_{\scriptscriptstyle\rm E}J_{\scriptscriptstyle\rm E}H_{\scriptscriptstyle\rm E}$ photometry, such fitted selections well reproduce those given by the full sets of optimal conditions. To quantify the goodness of the colour condition parameter fits, we defined a parameter analogous to $\mathcal{S}^{nf}$ in Eq.\ \eqref{eq:S}, namely $\mathcal{S}$. This parameter quantifies the difference between $\mathcal{C}^{nf}_{\rm tot}$, that is the completeness given by the full set of optimal conditions not fitted as a function of $z_{\rm l}$, and $\mathcal{C}$, which is the completeness given by the subset of optimal colour conditions fitted as a function of $z_{\rm l}$. As shown in Fig.\ \ref{fig:all_comparison}, $\mathcal{S}$ does not perfectly match $\mathcal{S}^{nf}$, for both $griz$ and $grizY_{\scriptscriptstyle\rm E}J_{\scriptscriptstyle\rm E}H_{\scriptscriptstyle\rm E}$ selections. This is due to the fact that the $c_1,\ldots,c_8$, $s_1,\ldots,s_4$ parameters in Eq.\ \eqref{eq:all_cond} do not always show a simple dependence on $z_{\rm l}$. Despite the fact that better parameter fits could be achieved by adopting an arbitrarily high order polynomial as the model, we set a 4th order polynomial as the highest-degree functional form for describing these parameters (see Tables \ref{tab:griz} and \ref{tab:grizYJH}). As shown in Fig.\ \ref{fig:all_comparison}, $\mathcal{C}$ is underestimated by at most 4 percent points. We verified that adding further conditions to these selections, that is, lowering the $\mathcal{S}$ threshold down to $0$, provides sub-percent level improvements in the selection completeness, on average. We remark that, in order to derive colour selections not defined in $z_{\rm l}$ bins, the final selection completeness is slightly degraded compared to $\mathcal{C}^{nf}_{\rm tot}$ for some $z_{\rm l}$ values. In realistic cluster weak-lensing analyses, however, we expect this to statistically increase the galaxy background completeness. When colour selections are defined on finite sets of $z_{\rm l}$ points, the background galaxies are excluded based on the $z_{\rm l}$ precision adopted in the colour selection calibration.

\subsection{Dependence on magnitude and redshift selections}\label{sec:limits}
\begin{figure*}[t]
\centering
\includegraphics[width = \hsize-2.7cm, height = 4.8cm] {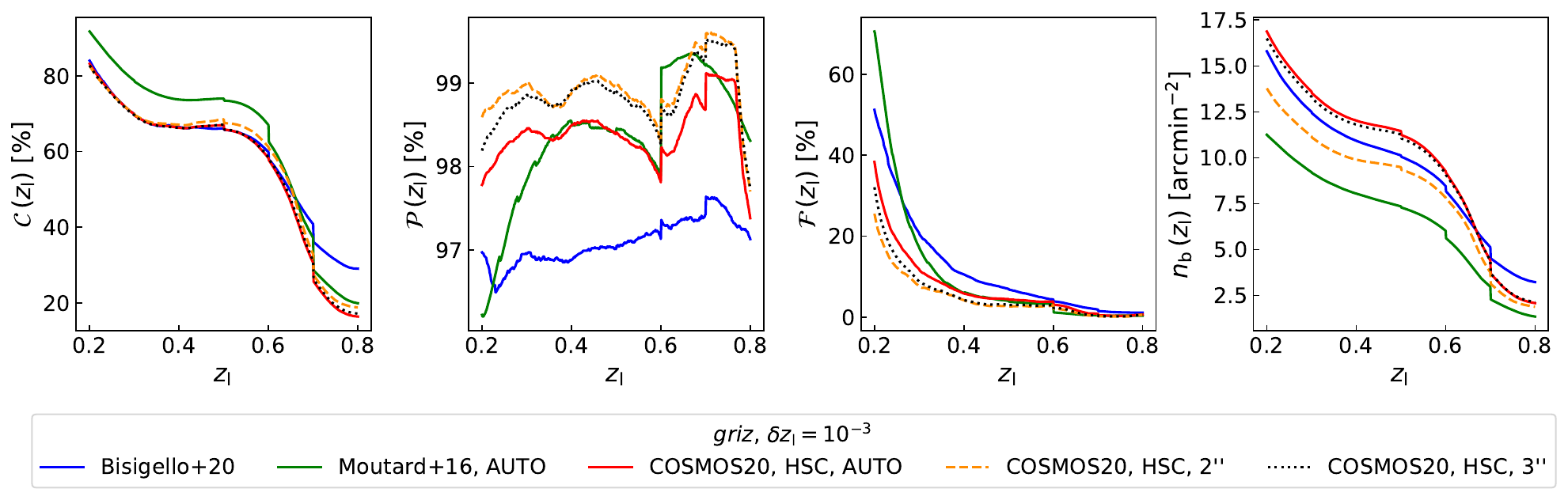}
\caption{Results of the application of the fitted colour selection based on $griz$ photometry, reported in Table \ref{tab:griz}, to the datasets introduced in Sect.\ \ref{sec:real_data}. From left to right: completeness, purity, foreground failure rate, and background density as a function of $z_{\rm l}$. The assumed $z_{\rm l}$ precision is $\delta z_{\rm l}=10^{-3}$. The $griz$ selection is applied to the \citetalias{Bisigello20} catalogue with magnitude limits corresponding to those used in the calibration process (blue solid lines), to the full depth \citet{Moutard16} catalogue (green solid lines), and to the \citet{Weaver22} catalogue with HSC Kron, \ang{;;2} and \ang{;;3} aperture magnitudes (solid red, dashed orange and dotted black lines, respectively), for which we imposed $i<25$.}
\label{fig:real_data}
\end{figure*}
To verify the robustness of the $griz$ selection with respect to alternative magnitude cuts, we applied the selection $i<24.4$, corresponding to the peak value of the $i$ magnitude distribution in the \citetalias{Bisigello20} catalogue. We also investigate the selection for the subsample with $i<23.4$, which is a threshold similar to the DES $i$ band limit \citep{DES}. In both cases, we derived higher $\mathcal{P}(z_{\rm l})$ and lower $\mathcal{F}(z_{\rm l})$, compared to what we found from the calibration sample used in Sect.\ \ref{sec:results:calibration}, namely the one with $({\rm S/N})_{\rm E}>3$ and $i\leq i_{\rm max}$, where $i_{\rm max}=24.9$ is the maximum $i$ magnitude in the sample (see Fig.\ \ref{fig:magLims}). In the case with $i<24.4$, $\mathcal{C}(z_{\rm l})$ is close to that from the calibration sample, while for $i<23.4$ we derived higher completeness, on average. In addition, as the bulk of the redshift distribution in the calibration sample, described in Sect.\ \ref{sec:data}, extends up to $z_{\rm g}\sim4$, we applied the $griz$ selection to the galaxy sample with redshift $z_{\rm g}<1.5$, $({\rm S/N})_{\rm E}>3$, and $i\leq i_{\rm max}$. In Fig.\ \ref{fig:magLims}, we can see that this redshift limit provides $\mathcal{F}(z_{\rm l})$ values that are identical to those derived from the calibration sample, which is expected since $\mathcal{F}(z_{\rm l})$ does not depend on the maximum redshift of the sample, while the completeness increases by up to $10$ percent points and the purity is at most 1 percent point lower. We note that the computation of $\mathcal{C}(z_{\rm l})$ and $\mathcal{F}(z_{\rm l})$ is made relative to the sample under consideration. In other words, they refer to galaxy populations defined by given magnitude and redshift limits. We measured the aforementioned colour selections by assuming a $z_{\rm l}$ precision of $\delta z_{\rm l}=10^{-3}$. This $\delta z_{\rm l}$ value is one order of magnitude lower (i.e.\ one order of magnitude higher precision) than the typical galaxy cluster photometric redshift uncertainty in current surveys \citep[see, e.g.][]{Rykoff16,Maturi19} and \textit{Euclid} \citep{Adam19}. Consequently, the $\delta z_{\rm l}=10^{-3}$ precision ensures the reliability of the colour condition fits for galaxy cluster background selections. We remark that we assumed $\delta z_{\rm l}=0.1$ for the $z_{\rm l}$ sampling in the calibration process. \\
\indent In Fig.\ \ref{fig:magLims} we show the efficiency of the $grizY_{\scriptscriptstyle\rm E}J_{\scriptscriptstyle\rm E}H_{\scriptscriptstyle\rm E}$ selection, computed by adopting $\delta z_{\rm l}=10^{-3}$, applied to the \citetalias{Bisigello20} calibration sample, with $({\rm S/N})_{\rm E}>3$ and $i\leq i_{\rm max}$. We found analogous selections from the subsample with $i<23.4$ and from the one with $({\rm S/N})_{\rm E}>10$. Specifically, in both cases, we derived higher $\mathcal{P}(z_{\rm l})$ and lower $\mathcal{F}(z_{\rm l})$, in agreement with what we found from the $griz$ selection. In addition, the increase in the minimum \textit{Euclid} ${\rm S/N}$ does not significantly change the completeness, while the $i<23.4$ limit decreases $\mathcal{C}(z_{\rm l})$ by at most $18$ percent points. As we obtained excellent $\mathcal{P}(z_{\rm l})$ and $\mathcal{F}(z_{\rm l})$ estimates from these tests, we conclude that both $griz$ and $grizY_{\scriptscriptstyle\rm E}J_{\scriptscriptstyle\rm E}H_{\scriptscriptstyle\rm E}$ selections are stable and reliable with respect to changes in the sample limiting magnitude and redshift. In addition, we note that brighter galaxy samples provide lower foreground contamination. This is expected, as faint galaxies have more scattered colour-redshift relations.\\
\indent In Fig.\ \ref{fig:magLims} we show the density of background galaxies, $n_{\rm b}(z_{\rm l})$, defined as the number of selected galaxies with $z_{\rm g}>z_{\rm l}$ per square arcmin. For both $griz$ and $grizY_{\scriptscriptstyle\rm E}J_{\scriptscriptstyle\rm E}H_{\scriptscriptstyle\rm E}$ selections, $n_{\rm b}(z_{\rm l})=16$ arcmin$^{-2}$ at $z_{\rm l}=0.2$ for $i\leq i_{\rm max}$ and $({\rm S/N})_{\rm E}>3$, decreasing with increasing $z_{\rm l}$. In both colour selections, the $i<23.4$ limit implies the largest decrease in $n_{\rm b}(z_{\rm l})$, providing $n_{\rm b}(z_{\rm l})<7$ arcmin$^{-2}$. In addition, for the $griz$ selection, the $i<24.4$ and $z_{\rm g}<1.5$ limits provide consistent results on $n_{\rm b}(z_{\rm l})$, showing a difference of at most 3 arcmin$^{-2}$ compared to that derived from the calibration sample. With regard to the $grizY_{\scriptscriptstyle\rm E}J_{\scriptscriptstyle\rm E}H_{\scriptscriptstyle\rm E}$ selection, the $({\rm S/N})_{\rm E}>10$ limit implies a decrease in $n_{\rm b}(z_{\rm l})$ of up to 5 arcmin$^{-2}$ at low $z_{\rm l}$, while $n_{\rm b}(z_{\rm l})$ becomes compatible with that derived from the calibration sample for $z_{\rm l}>1$. 

\subsection{$griz$ selection validation on real data}\label{sec:real_data}
\begin{figure*}[t]
\centering
\includegraphics[width = \hsize-2.7cm, height = 4.8cm] {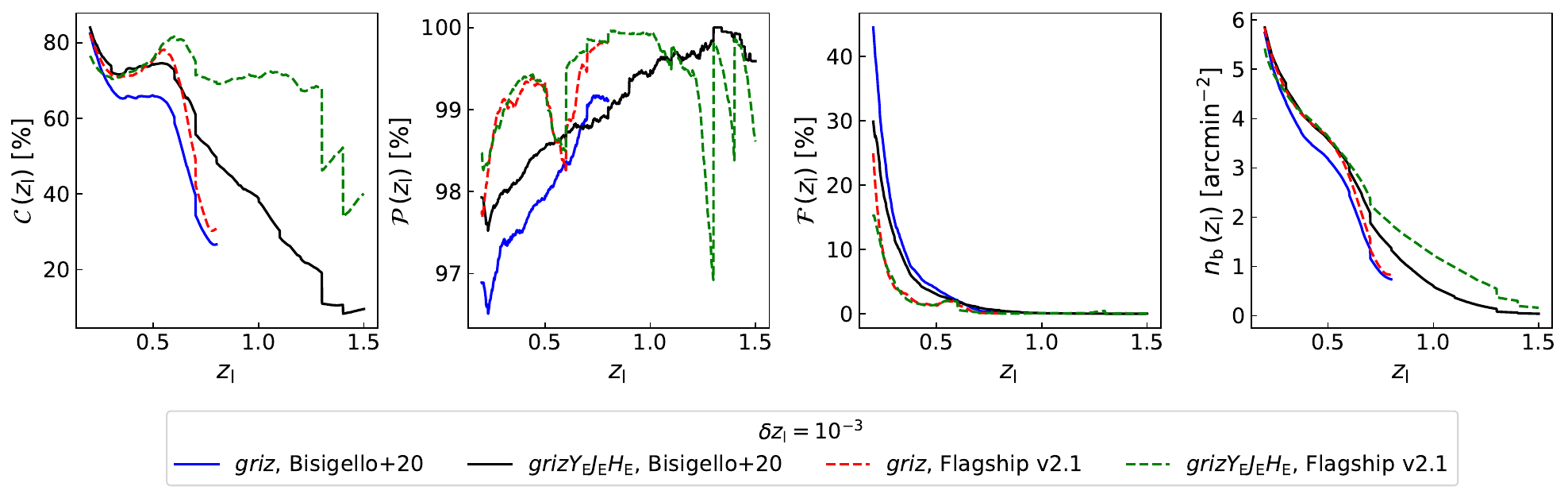}
\caption{Application of the calibrated colour selections to the Flagship simulated sample described in Sect.\ \ref{sec:flagship}. From left to right: completeness, purity, foreground failure rate, and background density as a function of $z_{\rm l}$, from the fitted colour selections based on $griz$ and $grizY_{\scriptscriptstyle\rm E}J_{\scriptscriptstyle\rm E}H_{\scriptscriptstyle\rm E}$ bands, adopting 5$\sigma$ magnitude limits. The assumed $z_{\rm l}$ precision is $\delta z_{\rm l}=10^{-3}$. The solid blue and solid black lines represent the $griz$ and $grizY_{\scriptscriptstyle\rm E}J_{\scriptscriptstyle\rm E}H_{\scriptscriptstyle\rm E}$ selections, respectively, applied to the \citetalias{Bisigello20} catalogue. The dashed red and dashed green curves represent the $griz$ and $grizY_{\scriptscriptstyle\rm E}J_{\scriptscriptstyle\rm E}H_{\scriptscriptstyle\rm E}$ selections, respectively, applied to the Flagship v2.1 catalogue (Euclid Collaboration in prep.).}
\label{fig:flagship}
\end{figure*}
To further assess the reliability of the $griz$ colour selection detailed in Sect.\ \ref{sec:results:calibration}, we applied it to external datasets obtained from real observations. In particular, we considered the VIMOS Public Extragalactic Redshift Survey \citep[VIPERS;][]{Guzzo14} Multi-Lambda Survey (VMLS) photometric catalogue by \citet{Moutard16}, including Canada-France-Hawaii Telescope Legacy Survey \citep[CFHTLS;][]{CFHTLS} $griz$ Kron aperture magnitudes \citep{Kron}. This catalogue covers $22$ deg$^2$ and provides reliable photometric redshifts for more than one million galaxies with a typical accuracy of $\sigma_z \leq 0.04$, and a fraction of catastrophic failures lower than $2\%$ down to $i\sim23$. These statistics are based on VIPERS data, complemented with the most secure redshifts selected from other spectroscopic surveys. We remind that in VIPERS a colour-colour pre-selection was employed to enhance the effective sampling of the VIMOS spectrograph. Nevertheless, the VIPERS selection does not introduce any significant colour bias above z $\sim0.6$ \citep{Guzzo14}. In addition, as we shall see in the following, the selection completeness and purity obtained from the VMLS dataset do not exhibit remarkable deviations from those obtained from other galaxy samples. In Fig.\ \ref{fig:real_data} we can see that, by applying the $griz$ selection to the VMLS sample, we derived higher $\mathcal{P}(z_{\rm l})$ and lower $\mathcal{F}(z_{\rm l})$ compared to what we found from the \citetalias{Bisigello20} catalogue, on average. This agrees with what we found in Sect.\ \ref{sec:limits}, as the \citet{Moutard16} catalogue is shallower than the \citetalias{Bisigello20} sample. For the same reason, $n_{\rm b}(z_{\rm l})$ is $3$ arcmin$^{-2}$ lower, on average. In addition, the completeness is up to $8$ percent points higher for $z_{\rm l}<0.6$, becoming lower for higher $z_{\rm l}$ values. \\
\indent We also applied the $griz$ selection to the COSMOS \texttt{CLASSIC} catalogue by \citet[][COSMOS20]{Weaver22}, which reaches the same photometric redshift precision as COSMOS15, namely $\sigma_z/(1+z)=0.007$, at almost one magnitude deeper. We considered $griz$ Kron, \ang{;;2} and \ang{;;3} aperture magnitudes from HSC. In addition, we selected galaxies with a photometric redshift derived from at least 30 bands, and with $i<25$, in order to consider a sample with highly reliable redshift estimates. By adopting more complex selection criteria, which may involve galaxies with photometric redshifts derived from a shared set of photometric bands, we do not expect remarkable differences in the results. Similar results for the cases with Kron, \ang{;;2} and \ang{;;3} aperture magnitudes are shown in Fig.\ \ref{fig:real_data}. Compared to what we derived from the \citetalias{Bisigello20} sample, the completeness is similar, with the largest differences at $z_{\rm l}>0.6$. In addition, $\mathcal{F}(z_{\rm l})$ is lower and $\mathcal{P}(z_{\rm l})$ is higher for any $z_{\rm l}$. For Kron and \ang{;;3} aperture magnitudes, $n_{\rm b}(z_{\rm l})$ is slightly higher compared to that obtained from the \citetalias{Bisigello20} sample, on average. Lower $n_{\rm b}(z_{\rm l})$ values show up for the \ang{;;2} aperture magnitudes, which is expected as we applied the same magnitude limit for each photometric aperture definition. Indeed, for these tests we did not include aperture correction terms. Lastly, comparing the purity derived from the COSMOS20 and VMLS samples, we note that for $z_{\rm l}>0.3$ the differences are below 1 percent point, on average. Thus, we conclude that the $griz$ selection provides robust and reliable results on real data.

\subsection{Validation on Flagship v2.1}\label{sec:flagship}
We tested the colour selections calibrated in Sect.\ \ref{sec:results:calibration} on the \textit{Euclid} Flagship galaxy catalogue v2.1.10 (Euclid Collaboration in prep.), which is currently the best simulated \textit{Euclid} galaxy catalogue available. This catalogue is based on an $N$-body simulation with around 4 trillion particles with mass $m_{\rm p}\sim 10^9$ $h^{-1}$M$_\odot$. 
%The simulation was run using \texttt{PKDGRAV3} \citep{Potter17}, in a box with side $L = 3780$ $h^{-1}$Mpc. 
A flat $\Lambda$ cold dark matter ($\Lambda$CDM) cosmological model was assumed, with matter density parameter $\Omega_{\rm m} = 0.319$, baryon density parameter $\Omega_{\rm b} = 0.049$, dark energy density parameter $\Omega_\Lambda = 0.681$, scalar spectral index $n_{\rm s} = 0.96$, Hubble parameter $h = H_0/(100 \, \text{ km s$^{-1}$ Mpc$^{-1}$}) = 0.67$, and standard deviation of linear density fluctuations on 8 $h^{-1}$Mpc scales $\sigma_8 = 0.83$. The haloes were identified using \texttt{Rockstar} \citep{Behroozi13}, and then populated with a halo occupation distribution model which was calibrated to reproduce observables such as clustering statistics as a function of galaxy luminosity. The galaxy SED templates used are the COSMOS templates from \citet{Ilbert09}, based on the models by \citet{Bruzual03} and \citet{Polletta07}. In addition, galaxy photo-$z$ probability distribution functions, namely $p(z_{\rm g})$, are included in Flagship, derived through a Nearest Neighbours Photometric Redshifts (NNPZ) pipeline \citep{Desprez20}. \\
\indent From the Flagship catalogue, we extracted a lightcone within ${\rm RA}\in[\ang{158},\ang{160}]$ and ${\rm Dec}\in[\ang{12},\ang{15}]$, considering the galaxies in the whole redshift range covered by the simulation, namely $z_{\rm g}\in[0,3]$. Specifically, $z_{\rm g}$ is the galaxy true redshift, and we verified that the contribution of peculiar velocities does not significantly change the results. We focused on \ang{;;2} aperture LSST $ugrizy$ and \textit{Euclid} $I_{\scriptscriptstyle\rm E}Y_{\scriptscriptstyle\rm E}J_{\scriptscriptstyle\rm E}H_{\scriptscriptstyle\rm E}$ photometry, as the simulated fluxes estimated for other ground-based surveys do not account for observational noise. Specifically, the photometric noise takes into account the depth expected in the southern hemisphere at the time of the third data release (DR3) for the Euclid Wide Survey. The LSST and \textit{Euclid} $10\sigma$ magnitude limits, which are proxies for extended sources, correspond to $u<24.4$, $g<25.6$, $r<25.7$, $i<25.0$, $z<24.3$, $y<23.1$, $I_{\scriptscriptstyle\rm E}<25$, $Y_{\scriptscriptstyle\rm E}<23.5$, $J_{\scriptscriptstyle\rm E}<23.5$, and $H_{\scriptscriptstyle\rm E}<23.5$. The fluxes we considered are not reddened due to Milky Way extinction, consistent with the analyses performed in the previous sections. \\
\begin{figure*}[t]
\centering
\includegraphics[width = \hsize-2.7cm, height = 4.8cm] {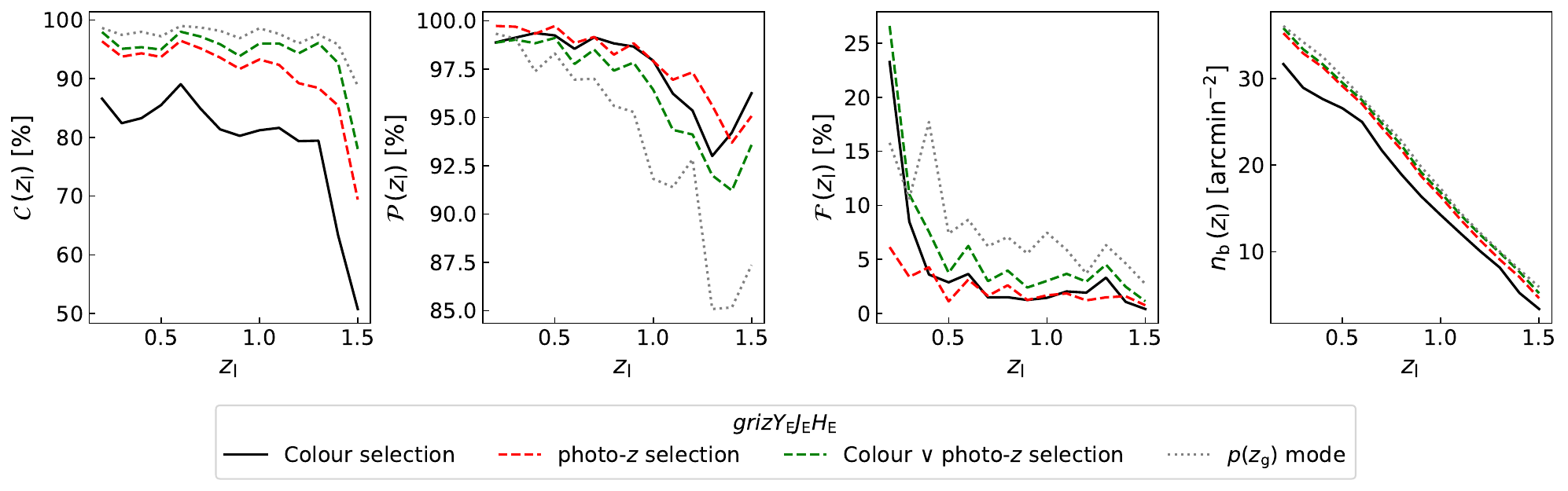}
\caption{Comparison of colour and photo-$z$ selections. From left to right: completeness, purity, foreground failure rate, and background density as a function of $z_{\rm l}$, obtained from Flagship v2.1. The solid black lines represent the $grizY_{\scriptscriptstyle\rm E}J_{\scriptscriptstyle\rm E}H_{\scriptscriptstyle\rm E}$ selection. The dashed green lines show the combination, through the $\lor$ logical operator, of $grizY_{\scriptscriptstyle\rm E}J_{\scriptscriptstyle\rm E}H_{\scriptscriptstyle\rm E}$ and photo-$z$ selection (Eq.\ \ref{eq:photoz_sel}). The dashed red lines represent the photo-$z$ selection, while the dotted black lines represent the selection based on the $p(z_{\rm g})$ mode.}
\label{fig:combined_photoz}
\end{figure*}
\indent In Fig.\ \ref{fig:flagship}, we show the application of $griz$ and $grizY_{\scriptscriptstyle\rm E}J_{\scriptscriptstyle\rm E}H_{\scriptscriptstyle\rm E}$ selections to Flagship. For this test, we assumed 5$\sigma$ magnitude cuts for LSST $ugrizy$ and \textit{Euclid} $I_{\scriptscriptstyle\rm E}Y_{\scriptscriptstyle\rm E}J_{\scriptscriptstyle\rm E}H_{\scriptscriptstyle\rm E}$ bands. In addition, we show results from the \citetalias{Bisigello20} sample in Fig.\ \ref{fig:flagship}, for which we assumed 5$\sigma$ magnitude cuts rescaled from the 10$\sigma$ limits listed in \citet[][Table 1]{Bisigello22}. We found that $n_{\rm b}(z_{\rm l})$ derived from Flagship agrees with that obtained from the \citetalias{Bisigello20} sample. The largest differences, of about 1 arcmin$^{-2}$, arise when the $grizY_{\scriptscriptstyle\rm E}J_{\scriptscriptstyle\rm E}H_{\scriptscriptstyle\rm E}$ selection is applied. We note that $n_{\rm b}(z_{\rm l})\sim0$ for $z_{\rm l}\sim1.5$, implying that lenses at these values of $z_{\rm l}$ may not exhibit significant weak-lensing signals. Nevertheless, we verified that $n_{\rm b}(z_{\rm l})$ is enhanced at any $z_{\rm l}$ when the selection defined for \textit{Euclid} weak-lensing analyses \citep{Laureijs11,Scaramella22} is assumed. This selection consists in a 10$\sigma$ cut in the $I_{\scriptscriptstyle\rm E}$ band, corresponding to $I_{\scriptscriptstyle\rm E}<25$ for a \ang{;;2} aperture, yielding a galaxy density of around 39 arcmin$^{-2}$ when applied to the Flagship dataset. In fact, in this case $n_{\rm b}(z_{\rm l})$ ranges from 30 arcmin$^{-2}$ at low $z_{\rm l}$ to 3 arcmin$^{-2}$ at $z_{\rm l}=1.5$. \\
\indent As shown in Fig.\ \ref{fig:flagship}, on average we obtained higher $\mathcal{P}(z_{\rm l})$ and lower $\mathcal{F}(z_{\rm l})$ for $z_{\rm l}<1$ from Flagship, compared to what we derived from the \citetalias{Bisigello20} sample. For the $griz$ selection case, $\mathcal{C}(z_{\rm l})$ agrees with that derived from the \citetalias{Bisigello20} sample, with the largest differences, of up to 16 percent points, at $z_{\rm l}\sim0.5$. Larger differences in $\mathcal{C}(z_{\rm l})$ are obtained from the $grizY_{\scriptscriptstyle\rm E}J_{\scriptscriptstyle\rm E}H_{\scriptscriptstyle\rm E}$ selection. From Flagship we obtained $\mathcal{C}(z_{\rm l})$ up to $10$ percent points larger for $z_{\rm l}<0.6$, and up to $50$ percent points larger for higher $z_{\rm l}$. We verified that this discrepancy in the completeness, in the case of the $grizY_{\scriptscriptstyle\rm E}J_{\scriptscriptstyle\rm E}H_{\scriptscriptstyle\rm E}$ selection, is not significantly attenuated through the assumption of 3$\sigma$ and 10$\sigma$ magnitude limits on both \citetalias{Bisigello20} and Flagship catalogues. Analogous results were obtained by assuming limits corresponding to the magnitude distribution peaks derived from the \citetalias{Bisigello20} catalogue, namely $g<24.9$, $r<24.6$, $i<24.3$, $z<24.1$, $Y_{\scriptscriptstyle\rm E}<23.8$, $J_{\scriptscriptstyle\rm E}<23.6$, and $H_{\scriptscriptstyle\rm E}<23.5$. Moreover, we verified that the $grizY_{\scriptscriptstyle\rm E}J_{\scriptscriptstyle\rm E}H_{\scriptscriptstyle\rm E}$ selection completeness does not remarkably vary by assuming the \Euclid weak-lensing selection defined above, namely $I_{\scriptscriptstyle\rm E}<25$. Similar results are obtained by considering the photometric errors expected for the DR2 of the Euclid Wide Survey, assuming the corresponding 3$\sigma$, 5$\sigma$, and 10$\sigma$ magnitude cuts. For each of the alternative magnitude cuts discussed in this section, we found that the $grizY_{\scriptscriptstyle\rm E}J_{\scriptscriptstyle\rm E}H_{\scriptscriptstyle\rm E}$ selection yields a purity up to 3 percent points higher at $z_{\rm l}>1.2$ when it is applied to the \citetalias{Bisigello20} catalogue, compared to what is derived from Flagship. At $z_{\rm l}<1.2$, instead, the purity obtained from \citetalias{Bisigello20} is 1 percent point lower, on average. Furthermore, the alternative magnitude cuts do not remarkably impact the selection purity at any $z_{\rm l}$.\\
\indent We additionally adopted SDSS fluxes, which do not include photometric noise, in place of LSST fluxes in Flagship. In this case, the completeness is up to 35 percent points larger than that derived from the \citetalias{Bisigello20} sample, and the purity approaches 100\% for $z_{\rm l}>1$, which is similar to what we derived from the \citetalias{Bisigello20} sample (see Fig.\ \ref{fig:flagship}). Thus, the selection based on SDSS photometry is less complete and purer compared to that obtained from LSST magnitudes. \\
\indent Differences in the completeness derived from the  Flagship and \citetalias{Bisigello20} samples may originate from distinct assumptions on the physical properties of the galaxies, such as dust extinction, stellar age, nebular emission lines, or on the assumed intrinsic spectral energy distributions. This could be indicated by a different fraction of star forming galaxies in the two samples. Following \citetalias{Bisigello20}, galaxies are classified as star forming if the following condition is satisfied,
\begin{equation}
\log_{10}({\rm sSFR} / {\rm yr^{-1}}) > -10.5,
\end{equation}
where sSFR is the specific star formation rate, derived from the best SED template in the catalogue by \citetalias{Bisigello20}. We verified that, for $z_{\rm g}>1$, the fraction of star forming galaxies in Flagship is consistent within 1 percent point with that derived from the catalogue by \citetalias{Bisigello20}. Thus, we conclude that the completeness differences between the Flagship and \citetalias{Bisigello20} samples are not due to different star forming galaxy populations. We also verified that the $\log_{10}({\rm sSFR} / {\rm yr^{-1}})$ distributions derived from the two datasets are compatible, having peaks at $\sim-8.13$ and $\sim-8.35$ in \citetalias{Bisigello20} and Flagship, respectively. The agreement of these peak values is well within 1$\sigma$ of the $\log_{10}({\rm sSFR} / {\rm yr^{-1}})$ distributions. We will be able to further investigate such completeness differences through the analysis of the first data release of the Euclid Deep Survey.

\subsection{Comparison with photo-$z$ selections}\label{sec:photoz_sel}
To compare the colour selections derived in this work to selections based on the galaxy $p(z_{\rm g})$, commonly referred to as photo-$z$ selections, we analysed the Flagship sample described in Sect.\ \ref{sec:flagship}. We considered only the galaxies with a $p(z_{\rm g})$ estimate obtained with the NNPZ pipeline \citep{Desprez20}. The NNPZ photo-$z$s are designed to work well for galaxies that are expected to be used in core \textit{Euclid} weak-lensing science, namely with 5$\sigma$ limits on the $I_{\scriptscriptstyle\rm E}$ band. Thus we imposed $I_{\scriptscriptstyle\rm E}<25.75$, along with 5$\sigma$ limits on the $Y_{\scriptscriptstyle\rm E}J_{\scriptscriptstyle\rm E}H_{\scriptscriptstyle\rm E}$ bands, namely 
$Y_{\scriptscriptstyle\rm E}<24.25$, $J_{\scriptscriptstyle\rm E}<24.25$, $H_{\scriptscriptstyle\rm E}<24.25$. Specifically, we adopted the following photo-$z$ selection,
\begin{equation}\label{eq:photoz_sel}
z_{\rm g}^{\rm min} > z_{\rm l} \,,
\end{equation}
where $z_{\rm g}^{\rm min}$ is the minimum of the interval containing 95\% of the probability around the first mode of $p(z_{\rm g})$, namely $\overline{z}_{\rm g}$. We chose $z_{\rm g}^{\rm min}$ in order to derive $\mathcal{P}(z_{\rm l})$ values which are compatible with those obtained from colour selections. We verified that adding a condition on the width of $p(z_{\rm g})$ in Eq.\ \eqref{eq:photoz_sel} does not impact the results. Specifically, for the latter test, we considered the additional condition $\mathcal{A}>\mathcal{A}_{\rm min}$, where $\mathcal{A}$ is the integrated probability around $\overline{z}_{\rm g}$, computed within the redshift points, which are the closest to $\overline{z}_{\rm g}$, having an associated probability of $0.2 p(\overline{z}_{\rm g})$. We verified that  imposing $\mathcal{A}_{\rm min}=0$ or $\mathcal{A}_{\rm min}=0.8$ leads to compatible purity values with sub-percent differences on average. However, $\mathcal{A}_{\rm min}=0.8$ lowers the photo-$z$ selection completeness by around 20 percent points at all $z_{\rm l}$. Consequently, we assumed $\mathcal{A}_{\rm min}=0$. \\
\indent To perform a fair comparison of colour and photo-$z$ selections, we considered only the $grizY_{\scriptscriptstyle\rm E}J_{\scriptscriptstyle\rm E}H_{\scriptscriptstyle\rm E}$ colour selection in this section. This is because photo-$z$s in Flagship were derived from the combination of ground-based and \textit{Euclid} photometry. In Fig.\ \ref{fig:combined_photoz}, we show that the $grizY_{\scriptscriptstyle\rm E}J_{\scriptscriptstyle\rm E}H_{\scriptscriptstyle\rm E}$ selection provides, on average, a completeness 15 percent points lower than that of the photo-$z$ selection, with similar contamination. By combining $grizY_{\scriptscriptstyle\rm E}J_{\scriptscriptstyle\rm E}H_{\scriptscriptstyle\rm E}$ and photo-$z$ selections, through the logical operator $\lor$, the completeness increases by up to 10 percent points with respect to the case of photo-$z$ selection alone, amounting to $\mathcal{C}(z_{\rm l})\sim95\%$ for $z_{\rm l}<1.4$. These preliminary tests confirm the importance of the combination of colour and photo-$z$ selections, as it leads to significantly more complete background galaxy samples. We also remark that increasing the selection completeness is key to reduce biases in the shear calibration parameters due to background selections, as we shall detail in Sect.\ \ref{sec:shear_measurements}. The analysis of \Euclid data will allow for a detailed investigation of the optimal photo-$z$ selections for galaxy cluster weak-lensing analyses, outlining the synergies with colour selections. For example, colour selections applied to \Euclid data could provide more robust background samples for massive or nearby galaxy clusters, as derived by \citet{hsc_med+al18b}. Leveraging colour selections also serves as a valuable cross-validation method for addressing the effect of unknown systematic uncertainties in photo-$z$ estimates. Lastly, Fig.\ \ref{fig:combined_photoz} shows the selection based only on the first mode of $p(z_{\rm g})$. Specifically, in this case we selected the galaxies with $\overline{z}_{\rm g}>z_{\rm l}$. Despite $\mathcal{C}(z_{\rm l})>90\%$ at all $z_{\rm l}$, the purity is up to $10$ percent points lower than that obtained from the $grizY_{\scriptscriptstyle\rm E}J_{\scriptscriptstyle\rm E}H_{\scriptscriptstyle\rm E}$ selection.

\subsection{Impact on shear measurements}\label{sec:shear_measurements}
\begin{figure*}[t]
\centering
\includegraphics[width = \hsize-2.7cm, height = 8.5cm] {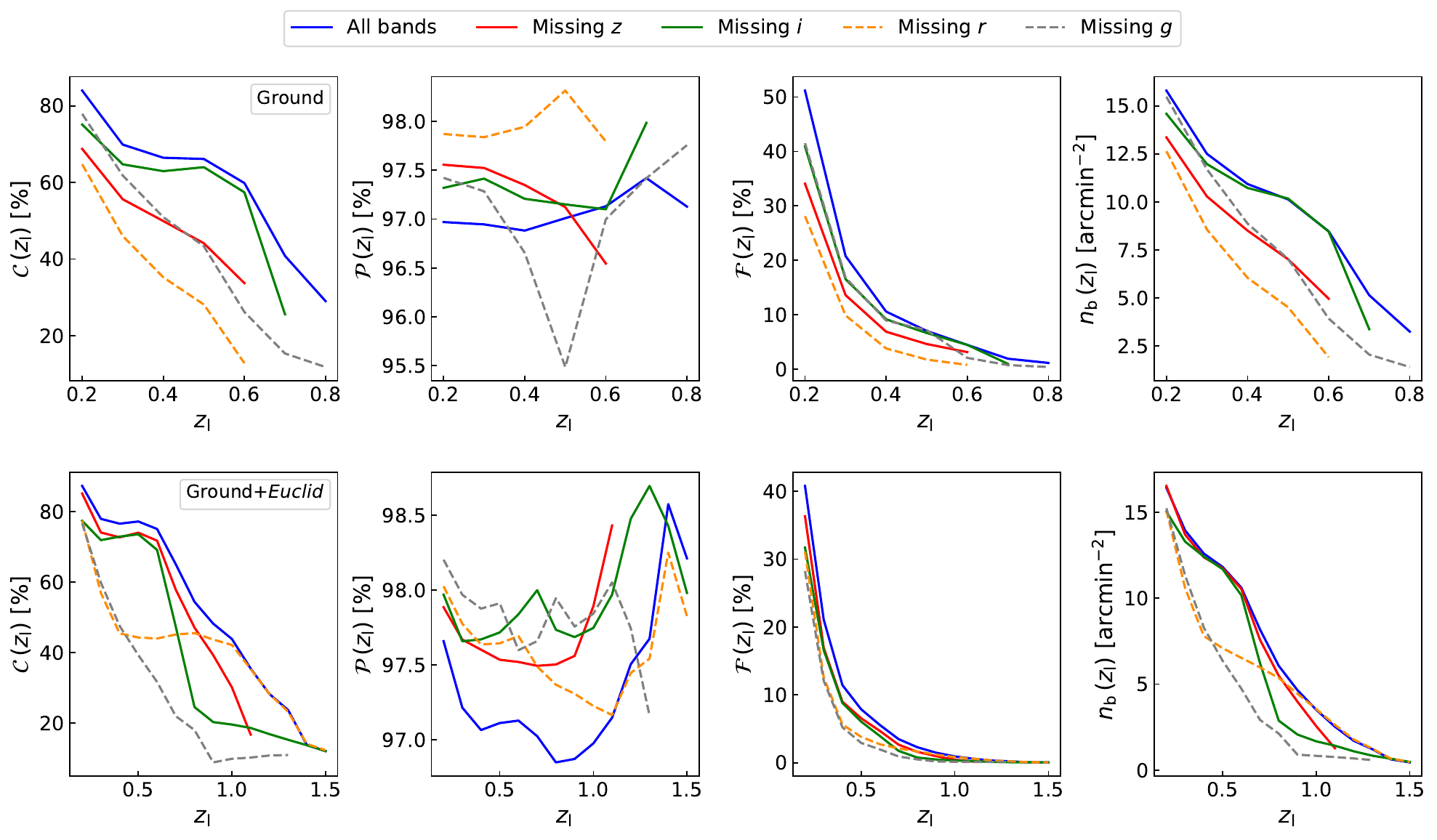}
\includegraphics[width = \hsize-6.5cm, height = 5cm] {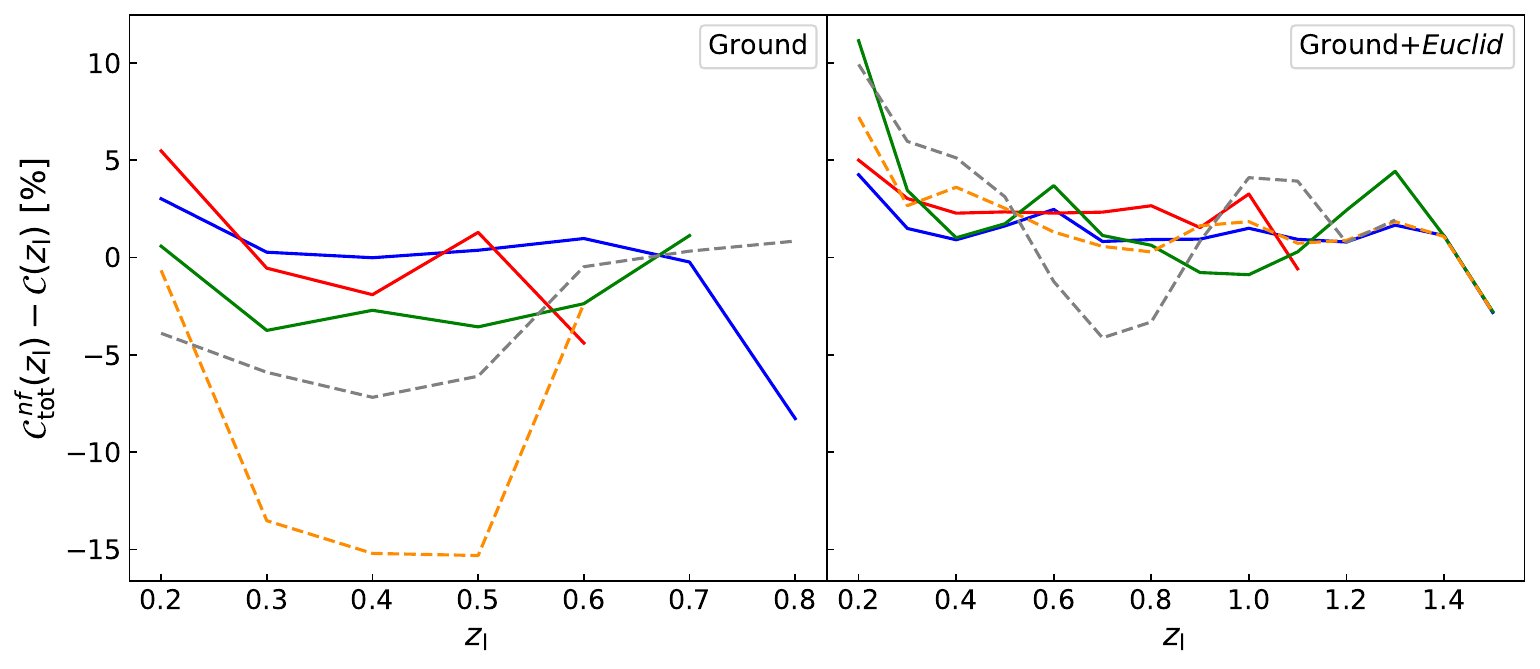}
\caption{Colour selection results, obtained from the \citetalias{Bisigello20} catalogue, in case of missing $z$ (solid red), $i$ (solid green), $r$ (dashed orange), and $g$ (dashed grey) bands. The blue curves represent the results from the $griz$ and $grizY_{\scriptscriptstyle\rm E}J_{\scriptscriptstyle\rm E}H_{\scriptscriptstyle\rm E}$ selections reported in Tables \ref{tab:griz} and \ref{tab:grizYJH}. \textit{Top panels}, from left to right: completeness, purity, foreground failure rate, and background density are shown, in the case of ground-only photometry. \textit{Middle panels}: colour selections from the combination of ground-based and \textit{Euclid} photometry. The plot structure is analogous to that of top panels. \textit{Bottom panels}: difference between $\mathcal{C}^{nf}_{\rm tot}$ and $\mathcal{C}$, for ground-only observations (left panel) and for the combination of ground-based and \textit{Euclid} photometry (right panel).}
\label{fig:missing_bands}
\end{figure*}
In cluster weak-lensing analyses, the inclusion of foreground sources in the shear measurements may significantly dilute the signal \citep{Broadhurst05, Medezinski07,Sifon15,McClintock19}. As discussed in the previous sections, the calibrated colour selections provide $\mathcal{P}(z_{\rm l})<1$. To assess the impact of impurities on shear measurements, we can express the cluster reduced tangential shear unaffected by contamination as follows \citep{Dietrich19}:
\begin{equation}\label{eq:true_g}
g_{\rm t, true}(z_{\rm l}) = \frac{g_{\rm t}(z_{\rm l})}{\mathcal{P}(z_{\rm l})}\,,
\end{equation}
where $g_{\rm t}(z_{\rm l})$ is the measured cluster reduced tangential shear at redshift $z_{\rm l}$. As the calibrated colour selections yield $\mathcal{P}(z_{\rm l})>0.97$, we expect at most a 3\% bias on the reduced tangential shear. In addition, as discussed in Sect.\ \ref{sec:real_data}, $\mathcal{P}(z_{\rm l})$ derived from different observed datasets with only ground-based photometry shows a scatter below 1 percent point. This scatter in $\mathcal{P}(z_{\rm l})$ is lower than the systematic uncertainty on galaxy shape measurements for stage III surveys, as we shall discuss in the following. We remark that $\mathcal{P}(z_{\rm l})$ is derived from reference fields, while galaxy clusters are overdense compared to the cosmic mean. Thus, contamination from cluster galaxies must be properly accounted for in Eq.\ \eqref{eq:true_g} \citep[see, e.g.][]{Gruen14,Dietrich19}. Nevertheless, such contamination is consistent with zero in the typical cluster-centric radial range adopted for mass calibration, namely at radii larger than 300 $h^{-1}$kpc \citep[see, e.g.][]{hsc_med+al18b,Bellagamba19}. \\
\indent Furthermore, galaxy shear calibration is usually statistically derived, based on observed and simulated galaxy samples. Nevertheless, through galaxy cluster background selections, some galaxy populations may be systematically excluded. This may invalidate the statistical estimate of the shape multiplicative bias, namely $m$, depending on the shear measurement technique and on the actual properties of the data \citep{hey+al12,mil+al13,hil+al16}. \\
\indent The typical uncertainty on $m$ found for stage III surveys ranges in the interval $\delta m \in [1\times 10^{-2},3 \times 10^{-2}]$ \citep[see, e.g.][]{jar+al16,Melchior17,Giblin21}. To assess the impact of colour selections on $m$, we considered the shape catalogues of \citet{hey+al12}, based on CFHTLS, and of \citet{hsc_man+al18}, based on the HSC Subaru Strategic Program \citep[HSC-SSP;][]{hsc_miy+al18,hsc_aih+al18}. Throughout this section, we adopted a lens redshift of $z_{\rm l}=0.5$. By applying the $griz$ selection calibrated in this work, we derived a shift in the mean shear multiplicative bias of $\Delta m=  7\times10^{-3}$ in CFHTLS and of $\Delta m=-2 \times10^{-3}$ in HSC-SSP. In addition, the \citet{Oguri12} and \citet{hsc_med+al18b} colour selections provide $\Delta m=  -3\times10^{-3}$ and $\Delta m=  -1\times10^{-2}$ from CFHTLS, respectively, while from HSC-SSP we obtained $\Delta m=  -5\times10^{-3}$ and $\Delta m=  -7\times10^{-3}$, respectively. Thus, galaxy population differences due to colour cuts provide systematic effects that are within the typical $m$ uncertainty in stage III surveys. By combining colour and photo-$z$ selections, we expect $\Delta m$ to become closer to zero. In \textit{Euclid}-like surveys, shear has to be calibrated within an accuracy of $2\times 10^{-3}$ \citep{cro+al13}. As we discussed in Sect.\ \ref{sec:photoz_sel}, the combination of $grizY_{\scriptscriptstyle\rm E}J_{\scriptscriptstyle\rm E}H_{\scriptscriptstyle\rm E}$ and photo-$z$ selections leads to 90\% background completeness in the Euclid Wide Survey, on average; thus, we may expect the bias on $m$ to be subdominant with respect to the required shear accuracy. Indeed, let us assume that 90\% of galaxies, selected through the combination of $grizY_{\scriptscriptstyle\rm E}J_{\scriptscriptstyle\rm E}H_{\scriptscriptstyle\rm E}$ and photo-$z$ selections, have an average $m$ similar to that derived from stage III surveys, namely $\langle m\rangle=0.01$. We assume that the remaining 10\% of galaxies have a very biased $m$, namely $\langle m\rangle=0.02$, compared to the selected population. This would imply a systematic error of $\Delta m = 10^{-3}$ in the average $m$ of the selected population. We will delve deeper into these variations in $m$ by examining the first data releases of the \textit{Euclid} surveys.

\subsection{Selection efficiency with missing bands}\label{sec:missing_bands}
\begin{figure*}[t]
\centering
\includegraphics[width = \hsize-2.3cm, height = 4.8cm] {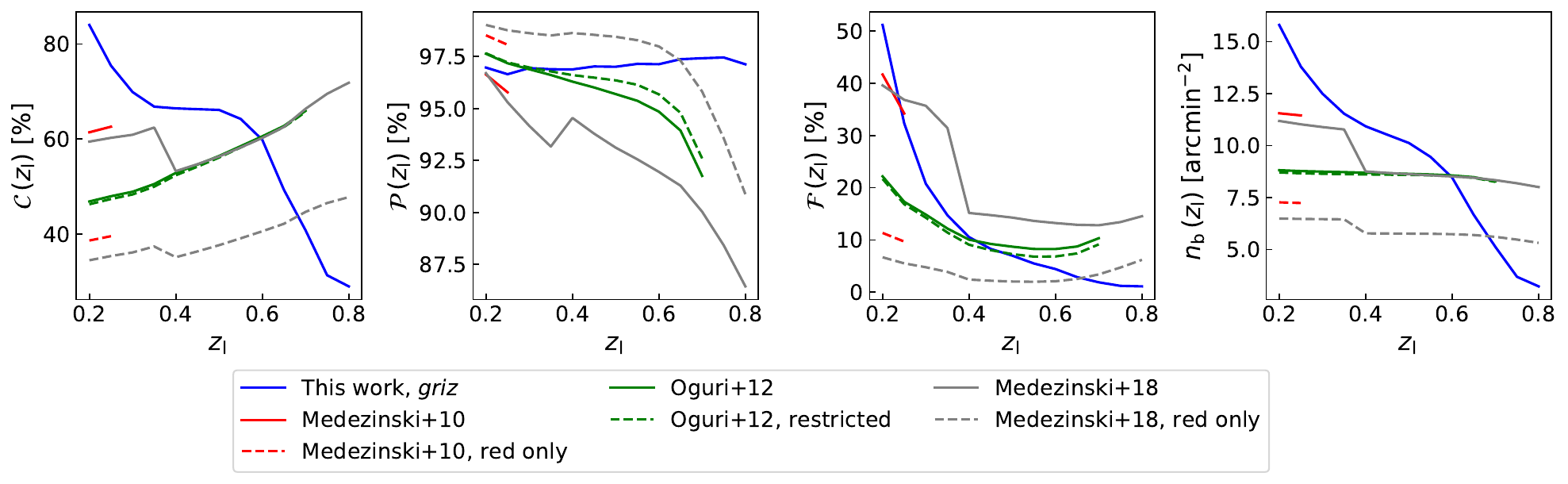}
\caption{Comparison of the $griz$ selection described in Sect.\ \ref{sec:results:calibration} with literature colour selections. From left to right: completeness, purity, foreground failure rate, and background density, derived from the \citetalias{Bisigello20} catalogue. The blue solid lines represent the $griz$ selection derived in this work. The red curves refer to the \citet{Medezinski10} selection, where the solid lines are given by Eqs. \eqref{eq:med10_red} -- \eqref{eq:med10_blue1}, while the dashed lines are given by Eq.\ \eqref{eq:med10_red}. The green curves represent the results from the \citet{Oguri12} selection, where the solid lines are given by Eqs. \eqref{eq:Oguri_1} -- \eqref{eq:Oguri_4}, while the dashed lines are given by Eqs. \eqref{eq:Oguri_1} -- \eqref{eq:Oguri_3}. The grey curves refer to the \citet{hsc_med+al18b} selection, where the solid lines are given by Eqs. \eqref{eq:Medezinski18_red} -- \eqref{eq:Medezinski18_blue}, while the dashed lines are given by Eq.\ \eqref{eq:Medezinski18_red}.}
\label{fig:literature}
\end{figure*}
In this work, we derived colour selections based on $griz$ and $grizY_{\scriptscriptstyle\rm E}J_{\scriptscriptstyle\rm E}H_{\scriptscriptstyle\rm E}$ photometry. In some cases, however, the full ground-based $griz$ photometry may be not available. For example, the DES Year 3 galaxy shape catalogue was not based on $g$ band \citep{Gatti21}, due to issues in the point spread function estimation \citep{Jarvis21}. Thus, we investigated the efficiency of $griz$ and $grizY_{\scriptscriptstyle\rm E}J_{\scriptscriptstyle\rm E}H_{\scriptscriptstyle\rm E}$ selections in the case of a missing band, based on the \citetalias{Bisigello20} calibration sample described in Sect.\ \ref{sec:data}. In performing this test, we excluded the colour conditions in Tables \ref{tab:griz} and \ref{tab:grizYJH} containing the chosen missing bands. In Fig.\ \ref{fig:missing_bands} we show that, in the case of ground-only observations, the absence of the $r$ band implies the largest completeness decrease, providing $\mathcal{C}(z_{\rm l})<60\%$. In addition, the $z_{\rm l}$ range is substantially reduced, corresponding to $z_{\rm l}\in[0.2,0.6]$. Also the absence of $i$ and $z$ bands implies a reduction of the maximum $z_{\rm l}$ for the ground-based selection, corresponding to $z_{\rm l}=0.7$ and $z_{\rm l}=0.6$, respectively, and a completeness decrease of up to 10 and 20 percent points, respectively. On average, a 20 percent point drop in completeness is found in absence of $g$ band photometry. Nevertheless, in the latter case the $z_{\rm l}$ range is not reduced. We remark that the considered samples differ from case to case, as they contain only galaxies with photometry available in the required bands. \\
\indent In Fig.\ \ref{fig:missing_bands} we show the effect of missing photometric bands on the combination of ground-based and \textit{Euclid} observations. In this case, the lack of $r$ band does not imply changes in $\mathcal{C}(z_{\rm l})$ for $z_{\rm l}>1$. In the absence of $i$ band, $\mathcal{C}(z_{\rm l})$ significantly decreases for $z_{\rm l}\gtrsim0.7$, being below $30\%$, while the $z_{\rm l}$ range is not reduced. A $z_{\rm l}$ range reduction is obtained in the case of missing $z$ or $g$ bands, as we derived $z_{\rm l}\in[0.2,1.1]$ and $z_{\rm l}\in[0.2,1.3]$, respectively. On average, in the case of the combination of ground-based and \textit{Euclid} observations, the largest completeness decrease is caused by the lack of the $g$ band. \\
\indent In this section, we defined colour selections with missing $g$, $r$, $i$, or $z$ band, as subsets of the colour conditions defining the $griz$ and $grizY_{\scriptscriptstyle\rm E}J_{\scriptscriptstyle\rm E}H_{\scriptscriptstyle\rm E}$ selections. In order to assess the difference between the selections defined by such subsets and those that would be derived from the colour selection calibration described in Sect.\ \ref{sec:method}, we compute $\mathcal{C}^{nf}_{\rm tot}$ for each case. In Fig.\ \ref{fig:missing_bands}, we show the difference between $\mathcal{C}^{nf}_{\rm tot}$ and $\mathcal{C}$, the latter being derived by subsets of the colour conditions defining $griz$ and $grizY_{\scriptscriptstyle\rm E}J_{\scriptscriptstyle\rm E}H_{\scriptscriptstyle\rm E}$ selections. In the ground-only case, the lack of $r$ band provides the largest $\mathcal{C}$ underestimation, as $\mathcal{C}^{nf}_{\rm tot}-\mathcal{C}\sim15$ percent points for $z_{\rm l}\in[0.3,0.5]$. Nevertheless, in case of other missing bands, the average $\mathcal{C}^{nf}_{\rm tot}-\mathcal{C}$ is close to 0. The same holds for the combination of ground-based and \textit{Euclid} photometry. We conclude that $griz$ and $grizY_{\scriptscriptstyle\rm E}J_{\scriptscriptstyle\rm E}H_{\scriptscriptstyle\rm E}$ selections provide robust results in the case of a missing band, except for ground-only observations without the $r$ band, for which a dedicated calibration might be needed.

\section{Comparison with literature ground-based selections}\label{sec:literature}
\begin{figure}[t]
\centering
\includegraphics[width = 0.8 \hsize, height = 7cm] {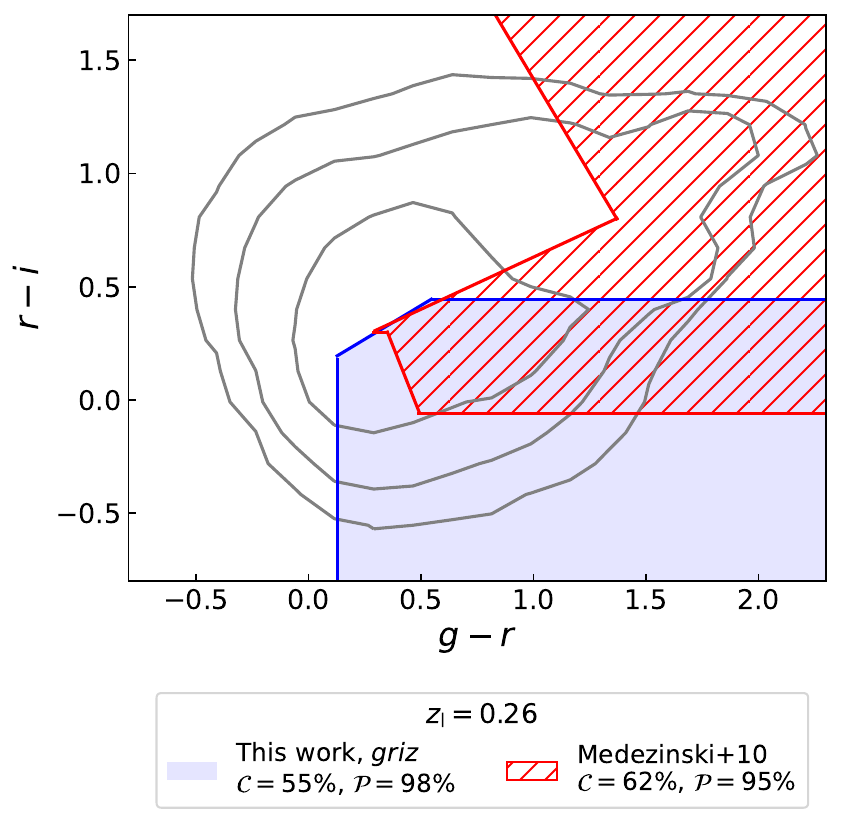}

\vspace{0.6cm}

\includegraphics[width = 0.8 \hsize, height = 7cm] {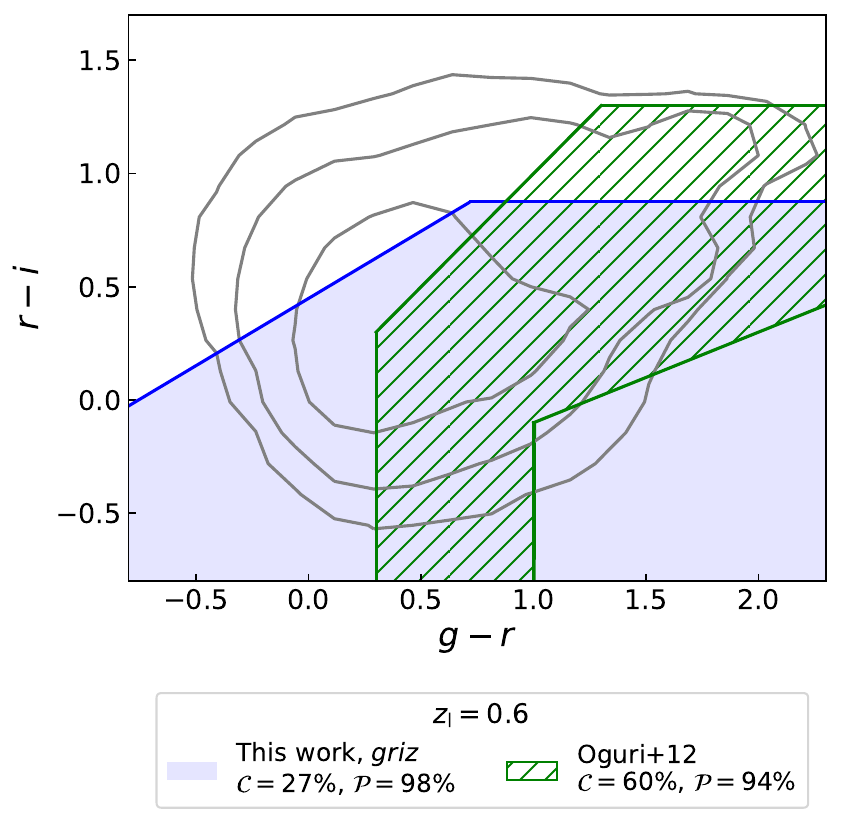}
\caption{Comparison of colour selections in the $(r-i)$ - $(g-r)$ colour-colour space. The solid grey contours indicate the 68\%, 95\%, and 99\% galaxy number density in the calibration sample, described in Sect.\ \ref{sec:data}. The blue shaded areas represent the regions excluded by applying the $griz$ selection calibrated in Sect.\ \ref{sec:results:calibration}. Completeness and purity of the selections are reported in the legends. For the $griz$ selection, $\mathcal{C}$ and $\mathcal{P}$ are computed by considering the colour conditions in Tab.\ \ref{tab:griz} which are defined in the $(r-i)$ - $(g-r)$ space. \textit{Top panel}: the red hatched area shows the region excluded through the \citet{Medezinski10} selection, and $z_{\rm l}=0.26$ is assumed. \textit{Bottom panel}: the green hatched area shows the region excluded through the \citet{Oguri12} selection, and $z_{\rm l}=0.6$ is assumed.}
\label{fig:literature:CCspaces}
\end{figure}
\begin{figure*}[t]
\centering
\includegraphics[width = \hsize-3cm, height = 5cm] {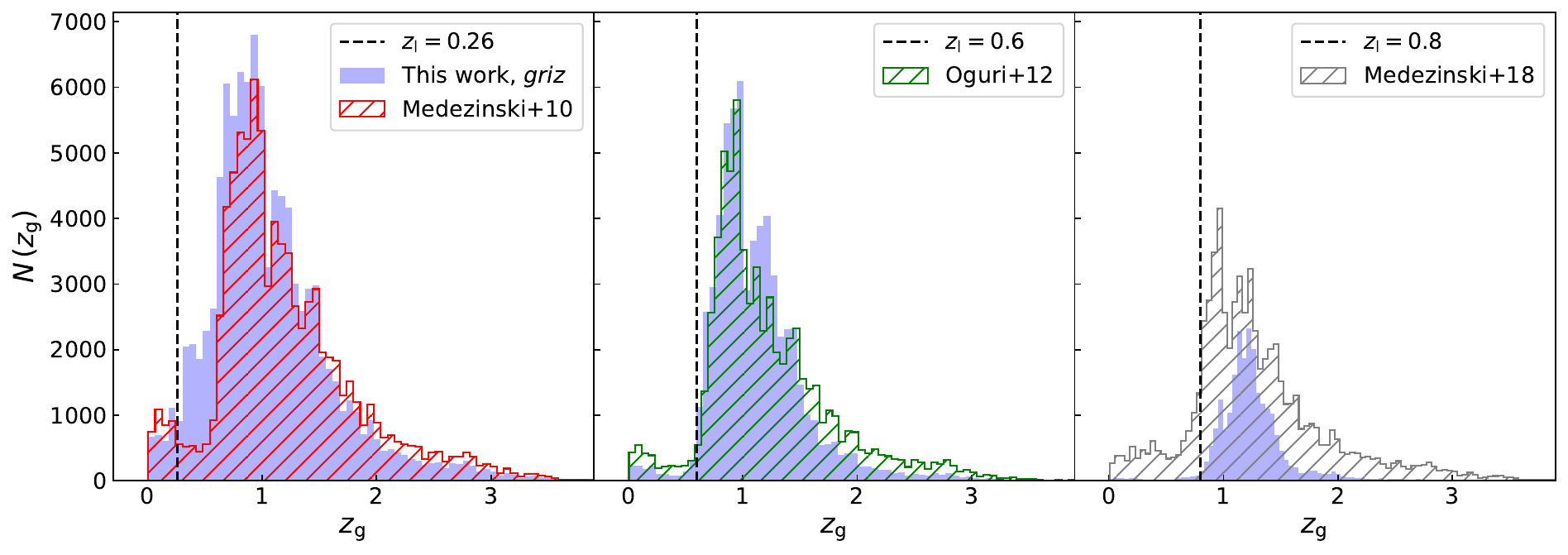}
\caption{From left to right: number of selected galaxies as a function of $z_{\rm g}$ assuming $z_{\rm l}=0.26$, $z_{\rm l}=0.6$, and $z_{\rm l}=0.8$. The $z_{\rm l}$ values are represented by vertical black dashed lines. The blue histograms represent the $griz$ selection calibrated in this work. The galaxy redshift distributions derived with the \citet{Medezinski10}, \citet{Oguri12}, and \citet{hsc_med+al18b} selections are represented by red, green, and grey hatched histograms, respectively.}
\label{fig:literature:Nz}
\end{figure*}
Based on the \citetalias{Bisigello20} sample considered in Sect.\ \ref{sec:results:calibration}, we compared our $griz$ colour selection to those derived by \citet{Medezinski10}, \citet{Oguri12}, and \citet{hsc_med+al18b}, which are also implemented in \texttt{COMB-CL}. As detailed below, for each of these selections, we considered two versions. One includes all the colour conditions provided by the corresponding authors, while the other comprises only a subsample of such conditions, providing lower foreground contamination. \texttt{COMB-CL} includes both versions of each colour selection. \\ 
\indent \citet{Medezinski10} derived colour selections for three massive clusters, identified through deep Subaru imaging, by maximising their weak-lensing signal. \texttt{COMB-CL} provides the selection calibrated for the A1703 cluster at redshift $z_{\rm l}\sim0.26$, as this is the one based on $gri$ photometry. This selection is expressed as follows,
\begin{align}\label{eq:med10_red}
\big[\, &(g-r) < 2.17\,(r-i) - 0.37\,\, \land\nonumber\\
& (g-r) < -0.6\,(r-i) + 1.85\,\,\land\,\, (r-i) > 0.3 \,\big] \,\, \lor
\end{align}
\begin{align}\label{eq:med10_blue1}
\big[\,&(g-r) < -0.4\,(r-i) + 0.47 \,\,\land\,\, (r-i) < 0.3 \,\big]\,\,\lor\,\,\nonumber\\
&(r-i) < -0.06\,,
\end{align}
where $\land$ is the logical `and' operator. Notably, Eq.\ \eqref{eq:med10_red} defines the red background sample, namely the galaxies redder than cluster galaxies, while Eq.\  \eqref{eq:med10_blue1} defines the blue background sample. In Fig.\ \ref{fig:literature}, we can see that this selection provides $\mathcal{C}(z_{\rm l})=60\%$, which is $20$ percent points lower than that provided by the $griz$ selection calibrated in this work. In addition, from the \citet{Medezinski10} selection we derived $\mathcal{P}(z_{\rm l})>96\%$, which is slightly lower than that obtained from the $griz$ selection discussed in Sect.\ \ref{sec:results:calibration}. Foreground contamination can be attenuated by considering the red sample selection only, namely Eq.\ \eqref{eq:med10_red}, as shown in Fig.\ \ref{fig:literature}. In this case, however, the completeness is lowered by 20 percent points. In Fig.\ \ref{fig:literature:CCspaces} (upper panel), we show a comparison between the selection by \citet{Medezinski10}, namely Eqs.\ \eqref{eq:med10_red} and \eqref{eq:med10_blue1}, and our $griz$ selection in the $(r-i)$ - $(g-r)$ colour-colour space, by assuming $z_{\rm l}=0.26$. Within this colour-colour space, we obtained $\mathcal{C}(z_{\rm l})=55\%$ and $\mathcal{P}(z_{\rm l})=98\%$ from the $griz$ selection, while through the \citet{Medezinski10} selection we found $\mathcal{C}(z_{\rm l})=62\%$ and $\mathcal{P}(z_{\rm l})=95\%$. We remark that the full set of colour conditions defining the $griz$ selection yields 80\% completeness for $z_{\rm l}=0.26$, and that a calibration based on $gri$ bands only would yield larger completeness values in the $(r-i)$ - $(g-r)$ space (see Sect.\ \ref{sec:missing_bands}). In addition, Fig.\ \ref{fig:literature:CCspaces} shows that the $griz$ selection extends the selected region defined by Eq.\ \eqref{eq:med10_red}, thus enhancing the red background sample compared to \citet{Medezinski10}. On the other hand, the $griz$ selection shows a more conservative definition of the blue background sample, compared to Eq.\ \eqref{eq:med10_blue1}. \\
\indent \citet{Oguri12} calibrated a selection based on $gri$ photometry from the COSMOS catalogue by \citet{Ilbert09}, providing reliable results for lenses at redshift $z_{\rm l}\lesssim 0.7$. This selection is expressed as
\begin{align}
(g-r)<0.3  \,\,\lor\label{eq:Oguri_1}
\end{align}
\begin{align}
(r-i)>1.3 \,\,\lor\label{eq:Oguri_2}
\end{align}
\begin{align}
(r-i)>(g-r) \,\,\lor\label{eq:Oguri_3}
\end{align}
\begin{align}
(g-r)>1 \,\,\land\,\, (r-i) < 0.4\,(g-r) - 0.5.\label{eq:Oguri_4}
\end{align}
The inclusion of Eq.\ \eqref{eq:Oguri_4} does not provide significant improvement in the completeness, causing a lower selection purity \citep{cov+al14}. In fact, Fig.\ \ref{fig:literature} shows that the selection including Eqs.\ \eqref{eq:Oguri_1} -- \eqref{eq:Oguri_4} provides sub-percent improvements in $\mathcal{C}(z_{\rm l})$, compared to the selection including Eqs.\ \eqref{eq:Oguri_1} -- \eqref{eq:Oguri_3} only, while $\mathcal{P}(z_{\rm l})$ and $\mathcal{F}(z_{\rm l})$ are up to 1 percent point lower and higher, respectively. With respect to the $griz$ selection calibrated in this work, the \citet{Oguri12} selection provides a purity which is up to $5$ percent points lower. This explains the higher completeness values for $z_{\rm l}>0.6$. For lower $z_{\rm l}$, the \citet{Oguri12} selection provides a completeness up to 35 percent points lower, which is expected as the selection by \citet{Oguri12} was calibrated for clusters at $z_{\rm l}\sim0.7$. Similarly to the comparison with the \citet{Medezinski10} selection discussed above, in Fig.\ \ref{fig:literature:CCspaces} (lower panel) we compare the $griz$ and \citet{Oguri12} selections in the $(r-i)$ - $(g-r)$ colour-colour space, assuming $z_{\rm l}=0.6$. We obtained $\mathcal{C}(z_{\rm l})=27\%$ and $\mathcal{P}(z_{\rm l})=98\%$ from the $griz$ selection, while with the \citet{Oguri12} selection we found $\mathcal{C}(z_{\rm l})=60\%$ and $\mathcal{P}(z_{\rm l})=94\%$. With respect to what we found by comparing $griz$ and \citet{Medezinski10} selections at $z_{\rm l}=0.26$, the decrease in completeness due to a purity enhancement is much larger at $z_{\rm l}=0.6$. This depends on the overlap of the foreground and background galaxy distributions in the $(r-i)$ - $(g-r)$ space. In addition, we remark that the calibration process presented in Sect.\ \ref{sec:method} excludes redundant colour conditions. This may partially explain the 27\% completeness found in the case of the $griz$ selection. \\
\indent \citet{hsc_med+al18b} calibrated a colour selection based on HSC observations, including $griz$ bands, following an approach similar to \citet{Medezinski10}. This colour selection is expressed as
\begin{align}\label{eq:Medezinski18_red}
\Bigg[\,&(g-i) < 2.276 \, (r-z) - 0.152 +  a(z_{\rm l}) \,\,\land \nonumber\\
& (g-i) < \frac{1}{2.276} \, (r-z) + b(z_{\rm l}) \left(1+\frac{1}{2.276^2}\right) - \frac{0.152}{2.276^2} \,\,\land \nonumber\\
& (r-z) > 0.5 \,\,\land \nonumber\\
& z > 21\,\Bigg]\,\,\lor
\end{align}
\begin{align}\label{eq:Medezinski18_blue}
\Bigg\{\,&(r-z) < -0.0248\,z + 1.604+c(z_{\rm l}) \,\,\lor \nonumber\\ &\Bigg[\,(g-i)<\frac{1}{2.276}\,(r-z)+d(z_{\rm l}) \left(1+\frac{1}{2.276^2}\right) - \frac{0.152}{2.276^2} \,\,\land \nonumber\\
&(g-i)<4\,\Bigg]\,\,\lor \nonumber\\
&(r-z) < 0.5\,\,\lor \nonumber\\
& z> 22\,\Bigg\}\,,
\end{align}
where 
\begin{equation}
a(z_{\rm l})=
\begin{cases}
-0.7\,\,\,\text{if $z_{\rm l}<0.4$}\\
-0.8\,\,\,\text{if $z_{\rm l}\geq0.4$}
\end{cases},
\end{equation}
\begin{equation}
b(z_{\rm l})=
\begin{cases}
4.0\,\,\,\text{if $z_{\rm l}<0.4$}\\
1.7\,\,\,\text{if $z_{\rm l}\geq0.4$}
\end{cases},
\end{equation}
\begin{equation}
c(z_{\rm l})=
\begin{cases}
-0.8\,\,\,\text{if $z_{\rm l}<0.4$}\\
-0.9\,\,\,\text{if $z_{\rm l}\geq0.4$}
\end{cases},
\end{equation}
\begin{equation}
d(z_{\rm l})=
\begin{cases}
0.5\,\,\,\text{if $z_{\rm l}<0.4$}\\
0.3\,\,\,\text{if $z_{\rm l}\geq0.4$}
\end{cases}.
\end{equation}
Similarly to \citet{Medezinski10}, Eqs. \eqref{eq:Medezinski18_red} and \eqref{eq:Medezinski18_blue} define the red and blue background samples, respectively. This selection provides a much larger amount of contamination compared to that derived in this work, reaching $\mathcal{P}(z_{\rm l})<90\%$ for $z_{\rm l}>0.7$, as shown in Fig.\ \ref{fig:literature}. However, by considering the red sample selection only, the purity improves by up to $6$ percent points. In the latter case, compared to the $griz$ selection detailed in Table \ref{tab:griz}, $\mathcal{C}(z_{\rm l})$ is up to $50$ percent points lower for $z_{\rm l}<0.7$. For higher $z_{\rm l}$, the lower purity from the \citet{hsc_med+al18b} selection allows for higher completeness values.\\
\indent In Fig.\ \ref{fig:literature:Nz} we show the redshift distributions of the galaxies selected through the $griz$, \citet{Medezinski10}, \citet{Oguri12}, and \citet{hsc_med+al18b} selections, assuming different $z_{\rm l}$ values. At $z_{\rm l}=0.26$, the $griz$ selection shows a larger number of galaxies, of the order of 10$^3$, which are correctly identified as background objects with $z_{\rm g}<0.6$, compared to \citet{Medezinski10}. This results in the larger completeness of the $griz$ selection shown in Fig.\ \ref{fig:literature}. At $z_{\rm l}=0.6$, where the $griz$ and \citet{Oguri12} selections have the same completeness (see Fig.\ \ref{fig:literature}), the $griz$ selection is less complete at $z_{\rm g}>1.5$ and more complete at lower redshifts, compared to \citet{Oguri12}. At $z_{\rm l}=0.8$, where the $griz$ selection is remarkably purer than that by \citet{hsc_med+al18b}, a notable incompleteness of the $griz$ selection is evident at any $z_{\rm g}$, compared to \citet{hsc_med+al18b}. In fact, for the case of the $griz$ selection, Fig.\ \ref{fig:literature:Nz} shows that the number of rejected galaxies at high redshift increases with the number of excluded foreground galaxies. This reflects the overlapping of foreground and background galaxy distributions in the considered colour-colour spaces.

\section{Discussion and conclusions}\label{sec:summary}
We developed a method to derive optimal galaxy colour selections for cluster weak-lensing analyses, given any set of photometric bands. To this aim, we considered all the available colour-colour combinations. Based on the galaxy catalogue by \citetalias{Bisigello20}, we calibrated selections based on ground-based $griz$ and \textit{Euclid} $Y_{\scriptscriptstyle\rm E}J_{\scriptscriptstyle\rm E}H_{\scriptscriptstyle\rm E}$ bands, with purity higher than $97\%$. Specifically, we showed that the $griz$ selection provides a completeness between 30\% and 84\%, in the lens limiting redshift range $z_{\rm l}\in[0.2,0.8]$. The inclusion of \textit{Euclid} $Y_{\scriptscriptstyle\rm E}J_{\scriptscriptstyle\rm E}H_{\scriptscriptstyle\rm E}$ bands, leading to a $grizY_{\scriptscriptstyle\rm E}J_{\scriptscriptstyle\rm E}H_{\scriptscriptstyle\rm E}$ selection, improves the completeness by up to $25$ percent points in this $z_{\rm l}$ range, allowing for a galaxy selection up to $z_{\rm l}=1.5$. In addition, for the first time in the literature, we expressed such selections as a continuous function of $z_{\rm l}$. In the following, we summarise the main results obtained from the tests presented in Sects.\ \ref{sec:results} and \ref{sec:literature}.
\begin{itemize}
    \item The calibrated colour selections, described in Sect.\ \ref{sec:results:calibration}, are stable with respect to changes in the sample limiting magnitudes and redshift.
    \item By applying the $griz$ selection to the VMLS catalogue by \citet{Moutard16} and to the COSMOS20 catalogue by \citet{Weaver22}, we derived completeness and purity estimates that are consistent with those obtained from the calibration sample by \citetalias{Bisigello20}. Consequently, the calibrated selections provide stable results by assuming alternative photometric aperture definitions, obtained from different ground-based telescopes.
    \item The application of $griz$ and $grizY_{\scriptscriptstyle\rm E}J_{\scriptscriptstyle\rm E}H_{\scriptscriptstyle\rm E}$ selections to the simulated \textit{Euclid} Flagship galaxy catalogue v2.1.10 (Euclid Collaboration in prep.) provided a purity of around $99\%$, on average, which is higher than that obtained from the \citetalias{Bisigello20} catalogue. The completeness from the Flagship and \citetalias{Bisigello20} samples is compatible in the $griz$ selection case, while the $grizY_{\scriptscriptstyle\rm E}J_{\scriptscriptstyle\rm E}H_{\scriptscriptstyle\rm E}$ selection provides up to $50$ percent points higher completeness from Flagship. We verified that this discrepancy does not depend on magnitude limits. In addition, we found no significant differences in the star forming galaxy fraction from the Flagship and \citetalias{Bisigello20} samples. A calibration of the $grizY_{\scriptscriptstyle\rm E}J_{\scriptscriptstyle\rm E}H_{\scriptscriptstyle\rm E}$ selection based on the Euclid Deep Survey will allow for a more thorough investigation of these completeness differences.
    \item Based on the Flagship sample, we combined the calibrated colour selections with photo-$z$ selections based on the $p(z_{\rm g})$ shape. We showed that in this case the completeness is up to $95\%$.
    \item We found no significant systematic effects on the multiplicative shear bias due to colour selections for stage III surveys. The first \textit{Euclid} data releases will provide further insights into the influence of background selections on this bias.
    \item The calibrated colour selections provide robust results in the case of a missing band from ground-based observations, except for those without the $r$ band, for which a dedicated calibration might be needed.
    \item Compared to the ground-based colour selections provided by \citet{Medezinski10}, \citet{Oguri12}, and \citet{hsc_med+al18b}, the $griz$ selection derived in this work is purer at high redshift and more complete at low redshift.
\end{itemize}
One potential enhancement to the calibration presented in this work could entail the inclusion of a magnitude dependence in the colour cuts. This might mitigate the impact of large photometric scatter at faint magnitudes \citep[see, e.g.][]{Schrabback21}. In addition, enhancing the set of photometric bands in the calibration sample, for example by including the LSST $y$ band, could remarkably improve the effectiveness of the colour selections. The performance of colour selections could further improve through dedicated calibration samples. Ongoing spectroscopic programmes are specifically designed to calibrate the relationship between galaxy colours and redshifts to match the depth of the Euclid survey \citep{Saglia22}. \\
\indent Furthermore, the colour selections presented in this work could improve the shear calibration in cluster weak-lensing analyses. The lensing signal from galaxy clusters differs from that of large scale structure in ways that affect both shear and photometric measurements. The dense cluster environment causes increased blending among light sources, due to both galaxy blends \citep{Simet15,Everett20} and the presence of diffuse intra-cluster light \citep{Gruen19,Kluge20,Sampaio-Santos21}. In addition, cluster lines of sight exhibit characteristically stronger shear, especially at small scales \citep{McClintock19,Ingoglia22}. These effects can lead to distinct biases in shear measurements compared to those obtained from calibrations primarily designed for cosmic shear analyses. Through the combination of colour and photo-$z$ selections, cluster shear calibration and mass bias can be assessed based on dedicated, multi-band cluster image simulations \citep[see, e.g.][]{Hernandez20}.

%\section*{Acknowledgements}
\begin{acknowledgements}
The \textit{Euclid} Consortium acknowledges the European Space Agency and a number of agencies and institutes that have supported the development of \textit{Euclid}, in particular the Academy of Finland, the Agenzia Spaziale Italiana, the Belgian Science Policy, the Canadian \textit{Euclid} Consortium, the French Centre National d’Etudes Spatiales, the Deutsches Zentrum für Luft- und Raumfahrt, the Danish Space Research Institute, the Fundação para a Ciência e a Tecnologia, the Ministerio de Ciencia e Innovación, the National Aeronautics and Space Administration, the National Astronomical Observatory of Japan, the Netherlandse Onderzoekschool Voor Astronomie, the Norwegian Space Agency, the Romanian Space Agency, the State Secretariat for Education, Research and Innovation (SERI) at the Swiss Space Office (SSO), and the United Kingdom Space Agency. A complete and detailed list is available on the \textit{Euclid} web site (\url{http://www.euclid-ec.org})\\
\indent We thank W.\ Hartley for his valuable advice, which remarkably enhanced the quality of this work. GC and LM acknowledge the support from the grant ASI n.2018-23-HH.0. LM and FM acknowledge the financial contribution from the grant PRIN-MUR 2022 20227RNLY3 “The concordance cosmological model: stress-tests with galaxy clusters” supported by Next Generation EU. LB acknowledges financial support from PRIN-MIUR $2017 - 20173$ML3WW\_001. MS acknowledges financial contributions from contract ASI-INAF n.2017-14-H.0, contract INAF mainstream project 1.05.01.86.10, and INAF Theory Grant 2023: Gravitational lensing detection of matter distribution at galaxy cluster boundaries and beyond (1.05.23.06.17). GC thanks the support from INAF Theory Grant 2022: Illuminating Dark Matter using Weak Lensing by Cluster Satellites, PI: Carlo Giocoli.\\
%We thank Micol Bolzonella for providing the codes for the calculation of photometric noise for various data releases of \textit{Euclid} and LSST, applied to Flagship v2.1.10 data. 
\indent This work has made use of CosmoHub \citep{cosmohub2,cosmohub1}. CosmoHub has been developed by the Port d'Informació Científica (PIC), maintained through a collaboration of the Institut de Física d'Altes Energies (IFAE) and the Centro de Investigaciones Energéticas, Medioambientales y Tecnológicas (CIEMAT) and the Institute of Space Sciences (CSIC \& IEEC), and was partially funded by the "Plan Estatal de Investigación Científica y Técnica y de Innovación" program of the Spanish government.
\end{acknowledgements}

\bibliography{aanda}

%\onecolumn
%\appendix
\onecolumn
\begin{appendix}

\section{Colour selection parameterisation}
In Tables \ref{tab:griz} and \ref{tab:grizYJH} we report the parameterisation of $griz$ and $grizY_{\scriptscriptstyle\rm E}J_{\scriptscriptstyle\rm E}H_{\scriptscriptstyle\rm E}$ selections, respectively, described in Sect.\ \ref{sec:results:calibration}.

\begin{table*}[ht]
\tiny
\caption{\label{tab:griz}Calibrated colour selection based on $griz$ photometry. The listed colour conditions are combined through the $\lor$ logical operator.}
  \centering
  %\resizebox{\linewidth}{!}{
    \begin{tabular}{ | c | c | c | } 
      \hline
      \rule{0pt}{2.7ex}
      Colour condition & Parameters & $z_{\rm l}$ range \\[2.4pt]
      \hline
      \rule{0pt}{2.7ex}
      $(g-r) < s\,(r-z) + c$ & $s= 9.39 \,z_{\rm l}^3 -7.93 \,z_{\rm l}^2 + 0.84 \,z_{\rm l} + 1.69$;\,\, $c = -31.8 \,z_{\rm l}^3 + 35.71 \,z_{\rm l}^2 -12.78 \,z_{\rm l} + 1.04$ & $[0.2,0.7]$\\[2.4pt]

      \hline
      \rule{0pt}{2.7ex}
      $(g-r) < s\,(i-z) + c$ & $s= 7.64 \,z_{\rm l}^2 -7.49 \,z_{\rm l} + 2.75$;\,\, $c = -2.53 \,z_{\rm l}^2 + 1.75 \,z_{\rm l} -0.32$ & $[0.2,0.8]$\\[2.4pt]

      \hline
      \rule{0pt}{2.7ex}
      $(r-i) > c$ & $c = 1.28 \,z_{\rm l} + 0.11$ & $[0.2,0.6]$\\[2.4pt]

      \hline
      \rule{0pt}{2.7ex}
      $(g-r) < s\,(r-i) + c$ & $s= 2.07 \,z_{\rm l}^2 -1.75 \,z_{\rm l} + 1.99$;\,\, $c = -25.25 \,z_{\rm l}^3 + 28.86 \,z_{\rm l}^2 -11.72 \,z_{\rm l} + 1.34$ & $[0.2,0.6]$\\[2.4pt]

      \hline
      \rule{0pt}{2.7ex}
      $(r-i) < s\,(r-z) + c$ & $s= 8.05 \,z_{\rm l}^3 -14.37 \,z_{\rm l}^2 + 6.87 \,z_{\rm l} + 0.73$;\,\, $c = -8.42 \,z_{\rm l}^3 + 16.11 \,z_{\rm l}^2  -9.54 \,z_{\rm l} + 0.68$ & $[0.2,0.8]$\\[2.4pt]

      \hline
      \rule{0pt}{2.7ex}
      $(g-i) > s\,(i-z) + c$ & $s= -9.17 \,z_{\rm l} + 3.24$;\,\, $c = 7.07 \,z_{\rm l} -0.45$ & $[0.2,0.3]$\\[2.4pt]

      \hline
      \rule{0pt}{2.7ex}
      $(g-i) < s\,(r-z) + c$ & $s= -47.54 \,z_{\rm l}^4 + 84.36 \,z_{\rm l}^3 -53.03 \,z_{\rm l}^2 + 13.64 \,z_{\rm l} + 0.48$;\,\, $c = 56.05 \,z_{\rm l}^4 -107.76 \,z_{\rm l}^3 + 72.88 \,z_{\rm l}^2 -21.09 \,z_{\rm l} + 1.93$ & $[0.2,0.8]$\\[2.4pt]

      \hline
      \rule{0pt}{2.7ex}
      $(g-z) < s\,(r-z) + c$ & $s= 1.70$;\,\, $c = -21.04 \,z_{\rm l}^4 + 16.37 \,z_{\rm l}^3 + 1.47 \,z_{\rm l}^2  -3.46 \,z_{\rm l} + 0.59$ & $[0.2,0.7]$\\[2.4pt]

      \hline
      \rule{0pt}{2.7ex}
      $(g-r) < s\,(g-z) + c$ & $s= -8.94 \,z_{\rm l}^3 + 13.28 \,z_{\rm l}^2 -6.55 \,z_{\rm l} + 1.73$;\,\, $c = -2.53 \,z_{\rm l}^2 + 2.24 \,z_{\rm l} -0.80$ & $[0.2,0.7]$\\[2.4pt]

      \hline
      \rule{0pt}{2.7ex}
      $(g-r) < c$ & $c = 1.46 \,z_{\rm l}^2 -1.43 \,z_{\rm l} + 0.40$ & $[0.2,0.5]$\\[2.4pt]

      \hline
      \rule{0pt}{2.7ex}
      $(r-i) > s\,(i-z) + c$ & $s= 8.45 \,z_{\rm l}^2 -6.93 \,z_{\rm l} + 1.67$;\,\, $c = -2.53 \,z_{\rm l}^2 + 3.48 \,z_{\rm l} -0.35$ & $[0.2,0.5]$\\[2.4pt]

      \hline
      \rule{0pt}{2.7ex}
      $(i-z) > c$ & $c = -1.53 \,z_{\rm l}^2 + 2.15 \,z_{\rm l} -0.01$ & $[0.2,0.6]$\\[2.4pt]

      \hline
      \rule{0pt}{2.7ex}
      $(g-z) > s\,(r-z) + c$ & $s= -0.58 \,z_{\rm l} - 1.42$;\,\, $c = -10.10 \,z_{\rm l}^2 + 15.76 \,z_{\rm l} -0.52$ & $[0.2,0.5]$\\[2.4pt]

      \hline
      \rule{0pt}{2.7ex}
      $(g-i) < s\,(r-i) + c$ & $s= 0.24 \,z_{\rm l} + 1.53$;\,\, $c = -0.91 \,z_{\rm l} + 0.33$ & $[0.2,0.6]$\\[2.4pt]
      
      \hline
    \end{tabular}
    %}
  \tablefoot{In the first column, the colour conditions are listed. The parameters of such conditions are shown in the second column, while in the last column the $z_{\rm l}$ ranges are listed. From top to bottom, the $i$th row corresponds to the $i$th iteration of the iterative process described in Sect.\ \ref{sec:method}.}
\end{table*}

\begin{table*}[ht!]
\tiny
\caption{\label{tab:grizYJH}Calibrated colour selection based on $grizY_{\scriptscriptstyle\rm E}J_{\scriptscriptstyle\rm E}H_{\scriptscriptstyle\rm E}$ photometry. The listed colour conditions are combined through the $\lor$ logical operator.}
  \centering
  %\resizebox{\linewidth}{!}{
    \begin{tabular}{ | c | c | c | } 
      \hline
      \rule{0pt}{2.7ex}
      Colour condition & Parameters & $z_{\rm l}$ range \\[2.4pt]
      \hline
      \rule{0pt}{3.ex}
      $(g-i) < s\,(r-H_{\scriptscriptstyle\rm E}) + c$ & $s= 30.81 \,z_{\rm l}^4 -79.74 \,z_{\rm l}^3 + 73.28 \,z_{\rm l}^2 -28.85 \,z_{\rm l} + 5.07$;\,\, $c = -31.2 \,z_{\rm l}^4 + 78.62 \,z_{\rm l}^3 -71.16 \,z_{\rm l}^2 + 27.57 \,z_{\rm l} -4.41$ & $[0.2,1.0]$\\[2.4pt]

      \hline
      \rule{0pt}{3.ex}
      $(g-i) < s\,(z-H_{\scriptscriptstyle\rm E}) + c$ & $s= 1.41 \,z_{\rm l}^2 -2.42 \,z_{\rm l} + 2.1$;\,\, $c = -2.42 \,z_{\rm l}^3 + 3.67 \,z_{\rm l}^2 -1.57 \,z_{\rm l} -0.28$ & $[0.2,1.3]$\\[2.4pt]

      \hline
      \rule{0pt}{3.ex}
      $(g-r) < s\,(r-z) + c$ & $s= 0.39 \,z_{\rm l} + 1.33$;\,\, $c = -5.23 \,z_{\rm l}^2 + 2.6 \,z_{\rm l} -0.83$ & $[0.2,0.7]$\\[2.4pt]

      \hline
      \rule{0pt}{3.ex}
      $(g-Y_{\scriptscriptstyle\rm E}) < s\,(z-H_{\scriptscriptstyle\rm E}) + c$ & $s= -0.01 \,z_{\rm l} + 1.65$;\,\, $c = 0.37 \,z_{\rm l}^2 -1.45 \,z_{\rm l} + 0.08$ & $[0.2,1.5]$\\[2.4pt]

      \hline
      \rule{0pt}{3.ex}
      $(r-i) > s\,(J_{\scriptscriptstyle\rm E}-H_{\scriptscriptstyle\rm E}) + c$ & $s= 1.64 \,z_{\rm l} - 1.33$;\,\, $c = -0.72 \,z_{\rm l}^2 + 1.49 \,z_{\rm l} + 0.41$ & $[0.2,0.6]$\\[2.4pt]

      \hline
      \rule{0pt}{3.ex}
      $(g-i) < s\,(z-J_{\scriptscriptstyle\rm E}) + c$ & $s= -0.05 \,z_{\rm l}^4 -2.44 \,z_{\rm l}^3 + 6.19 \,z_{\rm l}^2 -4.35 \,z_{\rm l} + 2.31$;\,\, $c = -0.29 \,z_{\rm l}^2 -0.35 \,z_{\rm l} -0.21$ & $[0.2,1.3]$\\[2.4pt]

      \hline
      \rule{0pt}{3.ex}
      $(g-i) < s\,(i-J_{\scriptscriptstyle\rm E}) + c$ & $s= -2.17 \,z_{\rm l}^3 + 5.01 \,z_{\rm l}^2 -3.25 \,z_{\rm l} + 2.11$;\,\, $c = -2.79 \,z_{\rm l}^3 + 4.06 \,z_{\rm l}^2 -2.7 \,z_{\rm l} -0.18$ & $[0.2,1.1]$\\[2.4pt]

      \hline
      \rule{0pt}{3.ex}
      $(g-r) < s\,(r-H_{\scriptscriptstyle\rm E}) + c$ & $s= -25.93 \,z_{\rm l}^3 + 42.07 \,z_{\rm l}^2 -22.02 \,z_{\rm l} + 4.2$;\,\, $c = 14.96 \,z_{\rm l}^3 -30.3 \,z_{\rm l}^2 + 18.03 \,z_{\rm l} -3.52$ & $[0.2,0.7]$\\[2.4pt]

      \hline
      \rule{0pt}{3.ex}
      $(g-i) > s\,(i-Y_{\scriptscriptstyle\rm E}) + c$ & $s= -5.31 \,z_{\rm l} + 1.99$;\,\, $c = 7.07 \,z_{\rm l} -0.45$ & $[0.2,0.3]$\\[2.4pt]

      \hline
      \rule{0pt}{3.ex}
      $(g-r) < s\,(i-Y_{\scriptscriptstyle\rm E}) + c$ & $s= -26.87 \,z_{\rm l}^4 + 63.55 \,z_{\rm l}^3 -51.34 \,z_{\rm l}^2 + 17.08 \,z_{\rm l} -0.8$;\,\, $c = 18.25 \,z_{\rm l}^4 -46.35 \,z_{\rm l}^3 + 38.77 \,z_{\rm l}^2 -13.53 \,z_{\rm l} + 1.31$ & $[0.2,1.0]$\\[2.4pt]

      \hline
      \rule{0pt}{3.ex}
      $(g-r) < s\,(Y_{\scriptscriptstyle\rm E}-H_{\scriptscriptstyle\rm E}) + c$ & $s= 2.07 \,z_{\rm l}^2 -2.39 \,z_{\rm l} + 1.76$;\,\, $c = -10.29 \,z_{\rm l}^3 + 11.18 \,z_{\rm l}^2 -4.19 \,z_{\rm l} + 0.33$ & $[0.2,0.7]$\\[2.4pt]

      \hline
      \rule{0pt}{3.ex}
      $(g-r) < s\,(i-H_{\scriptscriptstyle\rm E}) + c$ & $s= -4.02 \,z_{\rm l}^3 + 8.91 \,z_{\rm l}^2 -5.87 \,z_{\rm l} + 1.75$;\,\, $c = -2.55 \,z_{\rm l}^3 + 0.24 \,z_{\rm l}^2 + 1.44 \,z_{\rm l} -0.61$ & $[0.2,0.9]$\\[2.4pt]

      \hline
      \rule{0pt}{3.ex}
      $(g-z) < s\,(z-H_{\scriptscriptstyle\rm E}) + c$ & $s= 1.43 \,z_{\rm l}^4 -5.51 \,z_{\rm l}^3 + 7.29 \,z_{\rm l}^2 -3.79 \,z_{\rm l} + 2.23$;\,\, $c = 0.32 \,z_{\rm l}^2 -1.55 \,z_{\rm l} -0.17$ & $[0.2,1.4]$\\[2.4pt]

      \hline
      \rule{0pt}{3.ex}
      $(g-Y_{\scriptscriptstyle\rm E}) < s\,(i-H_{\scriptscriptstyle\rm E}) + c$ & $s= -2.25 \,z_{\rm l}^4 + 6.28 \,z_{\rm l}^3 -6.63 \,z_{\rm l}^2 + 3.09 \,z_{\rm l} + 1.16$;\,\, $c = -1.86 \,z_{\rm l}^3 + 3.75 \,z_{\rm l}^2 -3.43 \,z_{\rm l} + 0.26$ & $[0.2,1.1]$\\[2.4pt]

      \hline
      \rule{0pt}{3.ex}
      $(g-z) < s\,(i-J_{\scriptscriptstyle\rm E}) + c$ & $s= -6.33 \,z_{\rm l}^4 + 15.05 \,z_{\rm l}^3 -11.67 \,z_{\rm l}^2 + 3.25 \,z_{\rm l} + 1.41$;\,\, $c = 8.54 \,z_{\rm l}^4 -23.53 \,z_{\rm l}^3 + 22.33 \,z_{\rm l}^2 -9.11 \,z_{\rm l} + 0.74$ & $[0.2,1.2]$\\[2.4pt]

      \hline
      \rule{0pt}{3.ex}
      $(r-z) > s\,(i-z) + c$ & $s= -5.60 \,z_{\rm l}^2 +2.31 \,z_{\rm l} - 0.61$;\,\, $c = 4.69 \,z_{\rm l}^2 -0.67 \,z_{\rm l} + 0.80$ & $[0.2,0.7]$\\[2.4pt]

      \hline
      \rule{0pt}{3.ex}
      $(g-J_{\scriptscriptstyle\rm E}) < s\,(r-H_{\scriptscriptstyle\rm E}) + c$ & $s= 2.76 \,z_{\rm l}^2 -2.91 \,z_{\rm l} + 2.14$;\,\, $c = -21.51 \,z_{\rm l}^3 + 23.09 \,z_{\rm l}^2 -7.14 \,z_{\rm l} + 0.12$ & $[0.2,0.7]$\\[2.4pt]

      \hline
      \rule{0pt}{3.ex}
      $(r-i) > s\,(z-J_{\scriptscriptstyle\rm E}) + c$ & $s= -1.21 \,z_{\rm l}^2 + 0.8 \,z_{\rm l} + 0.2$;\,\, $c = 1.01 \,z_{\rm l} + 0.15$ & $[0.2,0.5]$\\[2.4pt]

      \hline
      \rule{0pt}{3.ex}
      $(g-z) < s\,(i-H_{\scriptscriptstyle\rm E}) + c$ & $s= -3.66 \,z_{\rm l}^4 + 12.87 \,z_{\rm l}^3 -16.26 \,z_{\rm l}^2 + 7.83 \,z_{\rm l} + 0.36$;\,\, $c = 6.29 \,z_{\rm l}^4 -21.87 \,z_{\rm l}^3 + 26.63 \,z_{\rm l}^2 -13.19 \,z_{\rm l} + 1.43$ & $[0.2,1.3]$\\[2.4pt]

      \hline
      \rule{0pt}{3.ex}
      $(g-i) < s\,(r-J_{\scriptscriptstyle\rm E}) + c$ & $s= -2.48 \,z_{\rm l}^3 + 6.11 \,z_{\rm l}^2 -5.46 \,z_{\rm l} + 2.59$;\,\, $c = 11.18 \,z_{\rm l}^4 -27.18 \,z_{\rm l}^3 + 20.47 \,z_{\rm l}^2 -4.48 \,z_{\rm l} -0.72$ & $[0.2,1.0]$\\[2.4pt]

      \hline
      \rule{0pt}{3.ex}
      $(g-z) < s\,(g-J_{\scriptscriptstyle\rm E}) + c$ & $s= -2.16 \,z_{\rm l}^3 + 5.88 \,z_{\rm l}^2 -5.06 \,z_{\rm l} + 2.12$;\,\, $c = 2.33 \,z_{\rm l}^3 -6.82 \,z_{\rm l}^2 + 6.18 \,z_{\rm l} -2.32$ & $[0.2,1.3]$\\[2.4pt]

      \hline
      \rule{0pt}{3.ex}
      $(i-Y_{\scriptscriptstyle\rm E}) > s\,(z-Y_{\scriptscriptstyle\rm E}) + c$ & $s= -0.02 \,z_{\rm l} + 0.42$;\,\, $c = 2.81 \,z_{\rm l}^3 -5.05 \,z_{\rm l}^2 + 3.36 \,z_{\rm l} + 0.08$ & $[0.2,0.8]$\\[2.4pt]

      \hline
      \rule{0pt}{3.ex}
      $(g-r) < s\,(z-J_{\scriptscriptstyle\rm E}) + c$ & $s= 13.22 \,z_{\rm l}^4 + 31.92 \,z_{\rm l}^3 -25.03 \,z_{\rm l}^2 + 7.6 \,z_{\rm l} + 0.42$;\,\, $c = 14.13 \,z_{\rm l}^4 -34.81 \,z_{\rm l}^3 + 27.79 \,z_{\rm l}^2 -9.36 \,z_{\rm l} + 0.83$ & $[0.2,1.1]$\\[2.4pt]

      \hline
      \rule{0pt}{3.ex}
      $(g-i) < s\,(i-H_{\scriptscriptstyle\rm E}) + c$ & $s= -11.75 \,z_{\rm l}^4 + 33.51 \,z_{\rm l}^3 -33.07 \,z_{\rm l}^2 + 12.73 \,z_{\rm l} -0.33$;\,\, $c = 10.74 \,z_{\rm l}^4 -35.01 \,z_{\rm l}^3 + 37.81 \,z_{\rm l}^2 -16.31 \,z_{\rm l} + 1.62$ & $[0.2,1.1]$\\[2.4pt]

      \hline
      \rule{0pt}{3.ex}
      $(g-J_{\scriptscriptstyle\rm E}) < s\,(i-H_{\scriptscriptstyle\rm E}) + c$ & $s= -0.03 \,z_{\rm l} + 1.7$;\,\, $c = 21.04 \,z_{\rm l}^4 -48.34 \,z_{\rm l}^3 + 38.51 \,z_{\rm l}^2 -13.25 \,z_{\rm l} + 1.21$ & $[0.2,0.9]$\\[2.4pt]

      \hline
      \rule{0pt}{3.ex}
      $(r-i) < s\,(z-J_{\scriptscriptstyle\rm E}) + c$ & $s= -0.66 \,z_{\rm l}^3 + 1.27 \,z_{\rm l}^2 -0.3 \,z_{\rm l} + 1.37$;\,\, $c = 0.58 \,z_{\rm l}^2 -1.84 \,z_{\rm l} -0.38$ & $[0.2,1.3]$\\[2.4pt]

      \hline
      \rule{0pt}{3.ex}
      $(i-Y_{\scriptscriptstyle\rm E}) < s\,(i-H_{\scriptscriptstyle\rm E}) + c$ & $s= 2.90 \,z_{\rm l} + 0.49$;\,\, $c = -6.06 \,z_{\rm l} + 0.66$ & $[0.2,0.3]$\\[2.4pt]

      \hline
      \rule{0pt}{3.ex}
      $(r-z) > s\,(Y_{\scriptscriptstyle\rm E}-J_{\scriptscriptstyle\rm E}) + c$ & $s= 0.58 \,z_{\rm l} + 0.29$;\,\, $c = 1.01 \,z_{\rm l} + 0.45$ & $[0.2,0.6]$\\[2.4pt]

      \hline
      \rule{0pt}{3.ex}
      $(g-r) < s\,(z-H_{\scriptscriptstyle\rm E}) + c$ & $s= 1.1 \,z_{\rm l}^3 + 0.69 \,z_{\rm l}^2 -1.14 \,z_{\rm l} + 1.09$;\,\, $c = -8.67 \,z_{\rm l}^3 + 9.14 \,z_{\rm l}^2 -3.35 \,z_{\rm l} + 0.24$ & $[0.2,0.9]$\\[2.4pt]

      \hline
      \rule{0pt}{3.ex}
      $(g-H_{\scriptscriptstyle\rm E}) < s\,(r-H_{\scriptscriptstyle\rm E}) + c$ & $s= 0.78 \,z_{\rm l}^2 -1.03 \,z_{\rm l} + 1.89$;\,\, $c = -13.09 \,z_{\rm l}^3 + 15.15 \,z_{\rm l}^2 -5.42 \,z_{\rm l} + 0.22$ & $[0.2,0.7]$\\[2.4pt]

      \hline
      \rule{0pt}{3.ex}
      $(g-Y_{\scriptscriptstyle\rm E}) < s\,(r-J_{\scriptscriptstyle\rm E}) + c$ & $s= 12.07 \,z_{\rm l}^3 -15.52 \,z_{\rm l}^2 + 5.71 \,z_{\rm l} + 1.08$;\,\, $c = -33.67 \,z_{\rm l}^3 + 42.93 \,z_{\rm l}^2 -17.52 \,z_{\rm l} + 1.61$ & $[0.2,0.7]$\\[2.4pt]
      
      \hline
    \end{tabular}
    %}
  \tablefoot{The table structure is analogous to that of Table \ref{tab:griz}.}
\end{table*}

\end{appendix}

\end{document}

%% file: authors.tex
%%%% please do not edit the author list -- contact ECEB Bureau for changes
\author{Euclid Collaboration:
G.~F.~Lesci\orcid{0000-0002-4607-2830}\inst{\ref{aff1},\ref{aff2}}\thanks{\email{giorgio.lesci2@unibo.it}}
\and M.~Sereno\orcid{0000-0003-0302-0325}\inst{\ref{aff2},\ref{aff3}}
\and M.~Radovich\orcid{0000-0002-3585-866X}\inst{\ref{aff4}}
\and G.~Castignani\orcid{0000-0001-6831-0687}\inst{\ref{aff1},\ref{aff2}}
\and L.~Bisigello\orcid{0000-0003-0492-4924}\inst{\ref{aff5},\ref{aff4}}
\and F.~Marulli\orcid{0000-0002-8850-0303}\inst{\ref{aff1},\ref{aff2},\ref{aff3}}
\and L.~Moscardini\orcid{0000-0002-3473-6716}\inst{\ref{aff1},\ref{aff2},\ref{aff3}}
\and L.~Baumont\orcid{0000-0002-1518-0150}\inst{\ref{aff6}}
\and G.~Covone\orcid{0000-0002-2553-096X}\inst{\ref{aff7},\ref{aff8},\ref{aff9}}
\and S.~Farrens\orcid{0000-0002-9594-9387}\inst{\ref{aff6}}
\and C.~Giocoli\orcid{0000-0002-9590-7961}\inst{\ref{aff2},\ref{aff10}}
\and L.~Ingoglia\orcid{0000-0002-7587-0997}\inst{\ref{aff1}}
\and S.~Miranda~La~Hera\inst{\ref{aff6}}
\and M.~Vannier\inst{\ref{aff11}}
\and A.~Biviano\orcid{0000-0002-0857-0732}\inst{\ref{aff12},\ref{aff13}}
\and S.~Maurogordato\inst{\ref{aff11}}
\and N.~Aghanim\inst{\ref{aff14}}
\and A.~Amara\inst{\ref{aff15}}
\and S.~Andreon\orcid{0000-0002-2041-8784}\inst{\ref{aff16}}
\and N.~Auricchio\orcid{0000-0003-4444-8651}\inst{\ref{aff2}}
\and M.~Baldi\orcid{0000-0003-4145-1943}\inst{\ref{aff17},\ref{aff2},\ref{aff3}}
\and S.~Bardelli\inst{\ref{aff2}}
\and R.~Bender\orcid{0000-0001-7179-0626}\inst{\ref{aff18},\ref{aff19}}
\and C.~Bodendorf\inst{\ref{aff18}}
\and D.~Bonino\inst{\ref{aff20}}
\and E.~Branchini\orcid{0000-0002-0808-6908}\inst{\ref{aff21},\ref{aff22}}
\and M.~Brescia\orcid{0000-0001-9506-5680}\inst{\ref{aff7},\ref{aff8},\ref{aff9}}
\and J.~Brinchmann\orcid{0000-0003-4359-8797}\inst{\ref{aff23}}
\and S.~Camera\orcid{0000-0003-3399-3574}\inst{\ref{aff24},\ref{aff25},\ref{aff20}}
\and V.~Capobianco\orcid{0000-0002-3309-7692}\inst{\ref{aff20}}
\and C.~Carbone\orcid{0000-0003-0125-3563}\inst{\ref{aff26}}
\and J.~Carretero\orcid{0000-0002-3130-0204}\inst{\ref{aff27},\ref{aff28}}
\and S.~Casas\orcid{0000-0002-4751-5138}\inst{\ref{aff29}}
\and F.~J.~Castander\orcid{0000-0001-7316-4573}\inst{\ref{aff30},\ref{aff31}}
\and M.~Castellano\orcid{0000-0001-9875-8263}\inst{\ref{aff32}}
\and S.~Cavuoti\orcid{0000-0002-3787-4196}\inst{\ref{aff8},\ref{aff9}}
\and A.~Cimatti\inst{\ref{aff33}}
\and G.~Congedo\orcid{0000-0003-2508-0046}\inst{\ref{aff34}}
\and C.~J.~Conselice\inst{\ref{aff35}}
\and L.~Conversi\orcid{0000-0002-6710-8476}\inst{\ref{aff36},\ref{aff37}}
\and Y.~Copin\orcid{0000-0002-5317-7518}\inst{\ref{aff38}}
\and L.~Corcione\orcid{0000-0002-6497-5881}\inst{\ref{aff20}}
\and F.~Courbin\orcid{0000-0003-0758-6510}\inst{\ref{aff39}}
\and H.~M.~Courtois\orcid{0000-0003-0509-1776}\inst{\ref{aff40}}
\and A.~Da~Silva\orcid{0000-0002-6385-1609}\inst{\ref{aff41},\ref{aff42}}
\and H.~Degaudenzi\orcid{0000-0002-5887-6799}\inst{\ref{aff43}}
\and A.~M.~Di~Giorgio\inst{\ref{aff44}}
\and J.~Dinis\inst{\ref{aff42},\ref{aff41}}
\and F.~Dubath\orcid{0000-0002-6533-2810}\inst{\ref{aff43}}
\and C.~A.~J.~Duncan\inst{\ref{aff35},\ref{aff45}}
\and X.~Dupac\inst{\ref{aff37}}
\and S.~Dusini\orcid{0000-0002-1128-0664}\inst{\ref{aff46}}
\and M.~Farina\inst{\ref{aff44}}
\and S.~Ferriol\inst{\ref{aff38}}
\and P.~Fosalba\orcid{0000-0002-1510-5214}\inst{\ref{aff31},\ref{aff47}}
\and S.~Fotopoulou\inst{\ref{aff48}}
\and M.~Frailis\orcid{0000-0002-7400-2135}\inst{\ref{aff12}}
\and E.~Franceschi\orcid{0000-0002-0585-6591}\inst{\ref{aff2}}
\and P.~Franzetti\inst{\ref{aff26}}
\and M.~Fumana\orcid{0000-0001-6787-5950}\inst{\ref{aff26}}
\and S.~Galeotta\orcid{0000-0002-3748-5115}\inst{\ref{aff12}}
\and B.~Garilli\orcid{0000-0001-7455-8750}\inst{\ref{aff26}}
\and B.~Gillis\orcid{0000-0002-4478-1270}\inst{\ref{aff34}}
\and A.~Grazian\orcid{0000-0002-5688-0663}\inst{\ref{aff4}}
\and F.~Grupp\inst{\ref{aff18},\ref{aff49}}
\and S.~V.~H.~Haugan\orcid{0000-0001-9648-7260}\inst{\ref{aff50}}
\and I.~Hook\orcid{0000-0002-2960-978X}\inst{\ref{aff51}}
\and F.~Hormuth\inst{\ref{aff52}}
\and A.~Hornstrup\orcid{0000-0002-3363-0936}\inst{\ref{aff53},\ref{aff54}}
\and P.~Hudelot\inst{\ref{aff55}}
\and K.~Jahnke\orcid{0000-0003-3804-2137}\inst{\ref{aff56}}
\and M.~K\"ummel\orcid{0000-0003-2791-2117}\inst{\ref{aff19}}
\and S.~Kermiche\orcid{0000-0002-0302-5735}\inst{\ref{aff57}}
\and A.~Kiessling\orcid{0000-0002-2590-1273}\inst{\ref{aff58}}
\and M.~Kilbinger\orcid{0000-0001-9513-7138}\inst{\ref{aff59}}
\and B.~Kubik\inst{\ref{aff38}}
\and M.~Kunz\orcid{0000-0002-3052-7394}\inst{\ref{aff60}}
\and H.~Kurki-Suonio\orcid{0000-0002-4618-3063}\inst{\ref{aff61},\ref{aff62}}
\and S.~Ligori\orcid{0000-0003-4172-4606}\inst{\ref{aff20}}
\and P.~B.~Lilje\orcid{0000-0003-4324-7794}\inst{\ref{aff50}}
\and V.~Lindholm\orcid{0000-0003-2317-5471}\inst{\ref{aff61},\ref{aff62}}
\and I.~Lloro\inst{\ref{aff63}}
\and E.~Maiorano\orcid{0000-0003-2593-4355}\inst{\ref{aff2}}
\and O.~Mansutti\orcid{0000-0001-5758-4658}\inst{\ref{aff12}}
\and O.~Marggraf\orcid{0000-0001-7242-3852}\inst{\ref{aff64}}
\and K.~Markovic\orcid{0000-0001-6764-073X}\inst{\ref{aff58}}
\and N.~Martinet\orcid{0000-0003-2786-7790}\inst{\ref{aff65}}
\and R.~Massey\orcid{0000-0002-6085-3780}\inst{\ref{aff66}}
\and E.~Medinaceli\orcid{0000-0002-4040-7783}\inst{\ref{aff2}}
\and M.~Melchior\inst{\ref{aff67}}
\and Y.~Mellier\inst{\ref{aff68},\ref{aff55}}
\and M.~Meneghetti\orcid{0000-0003-1225-7084}\inst{\ref{aff2},\ref{aff3}}
\and E.~Merlin\orcid{0000-0001-6870-8900}\inst{\ref{aff32}}
\and G.~Meylan\inst{\ref{aff39}}
\and M.~Moresco\orcid{0000-0002-7616-7136}\inst{\ref{aff1},\ref{aff2}}
\and E.~Munari\orcid{0000-0002-1751-5946}\inst{\ref{aff12}}
\and R.~Nakajima\inst{\ref{aff64}}
\and S.-M.~Niemi\inst{\ref{aff69}}
\and C.~Padilla\orcid{0000-0001-7951-0166}\inst{\ref{aff27}}
\and S.~Paltani\inst{\ref{aff43}}
\and F.~Pasian\inst{\ref{aff12}}
\and K.~Pedersen\inst{\ref{aff70}}
\and V.~Pettorino\inst{\ref{aff71}}
\and S.~Pires\inst{\ref{aff6}}
\and G.~Polenta\orcid{0000-0003-4067-9196}\inst{\ref{aff72}}
\and M.~Poncet\inst{\ref{aff73}}
\and L.~A.~Popa\inst{\ref{aff74}}
\and L.~Pozzetti\orcid{0000-0001-7085-0412}\inst{\ref{aff2}}
\and F.~Raison\orcid{0000-0002-7819-6918}\inst{\ref{aff18}}
\and R.~Rebolo\inst{\ref{aff75},\ref{aff76}}
\and A.~Renzi\orcid{0000-0001-9856-1970}\inst{\ref{aff5},\ref{aff46}}
\and J.~Rhodes\inst{\ref{aff58}}
\and G.~Riccio\inst{\ref{aff8}}
\and E.~Romelli\orcid{0000-0003-3069-9222}\inst{\ref{aff12}}
\and M.~Roncarelli\orcid{0000-0001-9587-7822}\inst{\ref{aff2}}
\and E.~Rossetti\inst{\ref{aff17}}
\and R.~Saglia\orcid{0000-0003-0378-7032}\inst{\ref{aff19},\ref{aff18}}
\and D.~Sapone\orcid{0000-0001-7089-4503}\inst{\ref{aff77}}
\and B.~Sartoris\inst{\ref{aff19},\ref{aff12}}
\and M.~Schirmer\orcid{0000-0003-2568-9994}\inst{\ref{aff56}}
\and P.~Schneider\orcid{0000-0001-8561-2679}\inst{\ref{aff64}}
\and A.~Secroun\orcid{0000-0003-0505-3710}\inst{\ref{aff57}}
\and G.~Seidel\orcid{0000-0003-2907-353X}\inst{\ref{aff56}}
\and S.~Serrano\orcid{0000-0002-0211-2861}\inst{\ref{aff31},\ref{aff30},\ref{aff78}}
\and C.~Sirignano\orcid{0000-0002-0995-7146}\inst{\ref{aff5},\ref{aff46}}
\and G.~Sirri\orcid{0000-0003-2626-2853}\inst{\ref{aff3}}
\and J.~Skottfelt\orcid{0000-0003-1310-8283}\inst{\ref{aff79}}
\and L.~Stanco\orcid{0000-0002-9706-5104}\inst{\ref{aff46}}
\and J.-L.~Starck\orcid{0000-0003-2177-7794}\inst{\ref{aff59}}
\and P.~Tallada-Cresp\'{i}\orcid{0000-0002-1336-8328}\inst{\ref{aff80},\ref{aff28}}
\and A.~N.~Taylor\inst{\ref{aff34}}
\and H.~I.~Teplitz\orcid{0000-0002-7064-5424}\inst{\ref{aff81}}
\and I.~Tereno\inst{\ref{aff41},\ref{aff82}}
\and R.~Toledo-Moreo\orcid{0000-0002-2997-4859}\inst{\ref{aff83}}
\and F.~Torradeflot\orcid{0000-0003-1160-1517}\inst{\ref{aff28},\ref{aff80}}
\and I.~Tutusaus\orcid{0000-0002-3199-0399}\inst{\ref{aff84}}
\and E.~A.~Valentijn\inst{\ref{aff85}}
\and L.~Valenziano\orcid{0000-0002-1170-0104}\inst{\ref{aff2},\ref{aff86}}
\and T.~Vassallo\orcid{0000-0001-6512-6358}\inst{\ref{aff19},\ref{aff12}}
\and A.~Veropalumbo\orcid{0000-0003-2387-1194}\inst{\ref{aff16},\ref{aff22}}
\and Y.~Wang\orcid{0000-0002-4749-2984}\inst{\ref{aff81}}
\and J.~Weller\inst{\ref{aff19},\ref{aff18}}
\and A.~Zacchei\orcid{0000-0003-0396-1192}\inst{\ref{aff12},\ref{aff13}}
\and G.~Zamorani\orcid{0000-0002-2318-301X}\inst{\ref{aff2}}
\and J.~Zoubian\inst{\ref{aff57}}
\and E.~Zucca\orcid{0000-0002-5845-8132}\inst{\ref{aff2}}
\and M.~Bolzonella\orcid{0000-0003-3278-4607}\inst{\ref{aff2}}
\and E.~Bozzo\inst{\ref{aff43}}
\and C.~Colodro-Conde\inst{\ref{aff75}}
\and D.~Di~Ferdinando\inst{\ref{aff3}}
\and J.~Graci\'{a}-Carpio\inst{\ref{aff18}}
\and S.~Marcin\inst{\ref{aff67}}
\and N.~Mauri\orcid{0000-0001-8196-1548}\inst{\ref{aff33},\ref{aff3}}
\and C.~Neissner\orcid{0000-0001-8524-4968}\inst{\ref{aff27},\ref{aff28}}
\and A.~A.~Nucita\inst{\ref{aff87},\ref{aff88},\ref{aff89}}
\and Z.~Sakr\orcid{0000-0002-4823-3757}\inst{\ref{aff90},\ref{aff84},\ref{aff91}}
\and V.~Scottez\inst{\ref{aff68},\ref{aff92}}
\and M.~Tenti\orcid{0000-0002-4254-5901}\inst{\ref{aff3}}
\and M.~Viel\orcid{0000-0002-2642-5707}\inst{\ref{aff13},\ref{aff12},\ref{aff93},\ref{aff94},\ref{aff95}}
\and M.~Wiesmann\inst{\ref{aff50}}
\and Y.~Akrami\orcid{0000-0002-2407-7956}\inst{\ref{aff96},\ref{aff97}}
\and S.~Anselmi\orcid{0000-0002-3579-9583}\inst{\ref{aff5},\ref{aff46},\ref{aff98}}
\and C.~Baccigalupi\orcid{0000-0002-8211-1630}\inst{\ref{aff93},\ref{aff12},\ref{aff94},\ref{aff13}}
\and M.~Ballardini\orcid{0000-0003-4481-3559}\inst{\ref{aff99},\ref{aff100},\ref{aff2}}
\and S.~Borgani\orcid{0000-0001-6151-6439}\inst{\ref{aff101},\ref{aff13},\ref{aff12},\ref{aff94}}
\and A.~S.~Borlaff\orcid{0000-0003-3249-4431}\inst{\ref{aff102},\ref{aff103},\ref{aff104}}
\and S.~Bruton\inst{\ref{aff105}}
\and C.~Burigana\orcid{0000-0002-3005-5796}\inst{\ref{aff106},\ref{aff86}}
\and R.~Cabanac\orcid{0000-0001-6679-2600}\inst{\ref{aff84}}
\and A.~Calabro\orcid{0000-0003-2536-1614}\inst{\ref{aff32}}
\and A.~Cappi\inst{\ref{aff2},\ref{aff11}}
\and C.~S.~Carvalho\inst{\ref{aff82}}
\and T.~Castro\orcid{0000-0002-6292-3228}\inst{\ref{aff12},\ref{aff94},\ref{aff13},\ref{aff95}}
\and G.~Ca\~{n}as-Herrera\orcid{0000-0003-2796-2149}\inst{\ref{aff69},\ref{aff107}}
\and K.~C.~Chambers\orcid{0000-0001-6965-7789}\inst{\ref{aff108}}
\and A.~R.~Cooray\orcid{0000-0002-3892-0190}\inst{\ref{aff109}}
\and J.~Coupon\inst{\ref{aff43}}
\and O.~Cucciati\orcid{0000-0002-9336-7551}\inst{\ref{aff2}}
\and S.~Davini\inst{\ref{aff22}}
\and S.~de~la~Torre\inst{\ref{aff65}}
\and G.~De~Lucia\orcid{0000-0002-6220-9104}\inst{\ref{aff12}}
\and G.~Desprez\inst{\ref{aff110}}
\and S.~Di~Domizio\orcid{0000-0003-2863-5895}\inst{\ref{aff21},\ref{aff22}}
\and H.~Dole\orcid{0000-0002-9767-3839}\inst{\ref{aff14}}
\and A.~D\'{i}az-S\'{a}nchez\orcid{0000-0003-0748-4768}\inst{\ref{aff111}}
\and J.~A.~Escartin~Vigo\inst{\ref{aff18}}
\and S.~Escoffier\orcid{0000-0002-2847-7498}\inst{\ref{aff57}}
\and I.~Ferrero\orcid{0000-0002-1295-1132}\inst{\ref{aff50}}
\and F.~Finelli\orcid{0000-0002-6694-3269}\inst{\ref{aff2},\ref{aff86}}
\and L.~Gabarra\inst{\ref{aff5},\ref{aff46}}
\and K.~Ganga\orcid{0000-0001-8159-8208}\inst{\ref{aff112}}
\and J.~Garc\'ia-Bellido\orcid{0000-0002-9370-8360}\inst{\ref{aff96}}
\and F.~Giacomini\orcid{0000-0002-3129-2814}\inst{\ref{aff3}}
\and G.~Gozaliasl\orcid{0000-0002-0236-919X}\inst{\ref{aff113},\ref{aff61}}
\and S.~Gwyn\orcid{0000-0001-8221-8406}\inst{\ref{aff114}}
\and H.~Hildebrandt\orcid{0000-0002-9814-3338}\inst{\ref{aff115}}
\and M.~Huertas-Company\orcid{0000-0002-1416-8483}\inst{\ref{aff75},\ref{aff116},\ref{aff117},\ref{aff118}}
\and A.~Jimenez~Mu\~{n}oz\orcid{0009-0004-5252-185X}\inst{\ref{aff119}}
\and J.~J.~E.~Kajava\orcid{0000-0002-3010-8333}\inst{\ref{aff120},\ref{aff121}}
\and V.~Kansal\inst{\ref{aff122},\ref{aff123}}
\and C.~C.~Kirkpatrick\inst{\ref{aff124}}
\and L.~Legrand\orcid{0000-0003-0610-5252}\inst{\ref{aff60}}
\and A.~Loureiro\orcid{0000-0002-4371-0876}\inst{\ref{aff125},\ref{aff126}}
\and J.~Macias-Perez\orcid{0000-0002-5385-2763}\inst{\ref{aff119}}
\and M.~Magliocchetti\orcid{0000-0001-9158-4838}\inst{\ref{aff44}}
\and G.~Mainetti\inst{\ref{aff127}}
\and R.~Maoli\orcid{0000-0002-6065-3025}\inst{\ref{aff128},\ref{aff32}}
\and M.~Martinelli\orcid{0000-0002-6943-7732}\inst{\ref{aff32},\ref{aff129}}
\and C.~J.~A.~P.~Martins\orcid{0000-0002-4886-9261}\inst{\ref{aff130},\ref{aff23}}
\and S.~Matthew\inst{\ref{aff34}}
\and M.~Maturi\orcid{0000-0002-3517-2422}\inst{\ref{aff90},\ref{aff131}}
\and L.~Maurin\orcid{0000-0002-8406-0857}\inst{\ref{aff14}}
\and R.~B.~Metcalf\orcid{0000-0003-3167-2574}\inst{\ref{aff1},\ref{aff2}}
\and M.~Migliaccio\inst{\ref{aff132},\ref{aff133}}
\and P.~Monaco\orcid{0000-0003-2083-7564}\inst{\ref{aff101},\ref{aff12},\ref{aff94},\ref{aff13}}
\and G.~Morgante\inst{\ref{aff2}}
\and S.~Nadathur\orcid{0000-0001-9070-3102}\inst{\ref{aff15}}
\and L.~Patrizii\inst{\ref{aff3}}
\and A.~Pezzotta\inst{\ref{aff18}}
\and C.~Porciani\inst{\ref{aff64}}
\and D.~Potter\orcid{0000-0002-0757-5195}\inst{\ref{aff134}}
\and M.~P\"{o}ntinen\orcid{0000-0001-5442-2530}\inst{\ref{aff61}}
\and P.~Reimberg\orcid{0000-0003-3410-0280}\inst{\ref{aff68}}
\and P.-F.~Rocci\inst{\ref{aff14}}
\and A.~G.~S\'anchez\orcid{0000-0003-1198-831X}\inst{\ref{aff18}}
\and A.~Schneider\orcid{0000-0001-7055-8104}\inst{\ref{aff134}}
\and M.~Schultheis\inst{\ref{aff11}}
\and E.~Sefusatti\orcid{0000-0003-0473-1567}\inst{\ref{aff12},\ref{aff13},\ref{aff94}}
\and P.~Simon\inst{\ref{aff64}}
\and A.~Spurio~Mancini\orcid{0000-0001-5698-0990}\inst{\ref{aff135}}
\and S.~A.~Stanford\orcid{0000-0003-0122-0841}\inst{\ref{aff136}}
\and J.~Steinwagner\inst{\ref{aff18}}
\and G.~Testera\inst{\ref{aff22}}
\and R.~Teyssier\inst{\ref{aff137}}
\and S.~Toft\inst{\ref{aff54},\ref{aff138},\ref{aff139}}
\and S.~Tosi\orcid{0000-0002-7275-9193}\inst{\ref{aff21},\ref{aff22},\ref{aff16}}
\and A.~Troja\orcid{0000-0003-0239-4595}\inst{\ref{aff5},\ref{aff46}}
\and M.~Tucci\inst{\ref{aff43}}
\and J.~Valiviita\orcid{0000-0001-6225-3693}\inst{\ref{aff61},\ref{aff62}}
\and D.~Vergani\orcid{0000-0003-0898-2216}\inst{\ref{aff2}}}

%%%% please do not edit the affiliation list -- contact ECEB Bureau for changes
\institute{Dipartimento di Fisica e Astronomia "Augusto Righi" - Alma Mater Studiorum Universit\`a di Bologna, via Piero Gobetti 93/2, 40129 Bologna, Italy\label{aff1}
\and
INAF-Osservatorio di Astrofisica e Scienza dello Spazio di Bologna, Via Piero Gobetti 93/3, 40129 Bologna, Italy\label{aff2}
\and
INFN-Sezione di Bologna, Viale Berti Pichat 6/2, 40127 Bologna, Italy\label{aff3}
\and
INAF-Osservatorio Astronomico di Padova, Via dell'Osservatorio 5, 35122 Padova, Italy\label{aff4}
\and
Dipartimento di Fisica e Astronomia "G. Galilei", Universit\`a di Padova, Via Marzolo 8, 35131 Padova, Italy\label{aff5}
\and
Universit\'e Paris-Saclay, Universit\'e Paris Cit\'e, CEA, CNRS, AIM, 91191, Gif-sur-Yvette, France\label{aff6}
\and
Department of Physics "E. Pancini", University Federico II, Via Cinthia 6, 80126, Napoli, Italy\label{aff7}
\and
INAF-Osservatorio Astronomico di Capodimonte, Via Moiariello 16, 80131 Napoli, Italy\label{aff8}
\and
INFN section of Naples, Via Cinthia 6, 80126, Napoli, Italy\label{aff9}
\and
Istituto Nazionale di Fisica Nucleare, Sezione di Bologna, Via Irnerio 46, 40126 Bologna, Italy\label{aff10}
\and
Universit\'e C\^{o}te d'Azur, Observatoire de la C\^{o}te d'Azur, CNRS, Laboratoire Lagrange, Bd de l'Observatoire, CS 34229, 06304 Nice cedex 4, France\label{aff11}
\and
INAF-Osservatorio Astronomico di Trieste, Via G. B. Tiepolo 11, 34143 Trieste, Italy\label{aff12}
\and
IFPU, Institute for Fundamental Physics of the Universe, via Beirut 2, 34151 Trieste, Italy\label{aff13}
\and
Universit\'e Paris-Saclay, CNRS, Institut d'astrophysique spatiale, 91405, Orsay, France\label{aff14}
\and
Institute of Cosmology and Gravitation, University of Portsmouth, Portsmouth PO1 3FX, UK\label{aff15}
\and
INAF-Osservatorio Astronomico di Brera, Via Brera 28, 20122 Milano, Italy\label{aff16}
\and
Dipartimento di Fisica e Astronomia, Universit\`a di Bologna, Via Gobetti 93/2, 40129 Bologna, Italy\label{aff17}
\and
Max Planck Institute for Extraterrestrial Physics, Giessenbachstr. 1, 85748 Garching, Germany\label{aff18}
\and
Universit\"ats-Sternwarte M\"unchen, Fakult\"at f\"ur Physik, Ludwig-Maximilians-Universit\"at M\"unchen, Scheinerstrasse 1, 81679 M\"unchen, Germany\label{aff19}
\and
INAF-Osservatorio Astrofisico di Torino, Via Osservatorio 20, 10025 Pino Torinese (TO), Italy\label{aff20}
\and
Dipartimento di Fisica, Universit\`a di Genova, Via Dodecaneso 33, 16146, Genova, Italy\label{aff21}
\and
INFN-Sezione di Genova, Via Dodecaneso 33, 16146, Genova, Italy\label{aff22}
\and
Instituto de Astrof\'isica e Ci\^encias do Espa\c{c}o, Universidade do Porto, CAUP, Rua das Estrelas, PT4150-762 Porto, Portugal\label{aff23}
\and
Dipartimento di Fisica, Universit\`a degli Studi di Torino, Via P. Giuria 1, 10125 Torino, Italy\label{aff24}
\and
INFN-Sezione di Torino, Via P. Giuria 1, 10125 Torino, Italy\label{aff25}
\and
INAF-IASF Milano, Via Alfonso Corti 12, 20133 Milano, Italy\label{aff26}
\and
Institut de F\'{i}sica d'Altes Energies (IFAE), The Barcelona Institute of Science and Technology, Campus UAB, 08193 Bellaterra (Barcelona), Spain\label{aff27}
\and
Port d'Informaci\'{o} Cient\'{i}fica, Campus UAB, C. Albareda s/n, 08193 Bellaterra (Barcelona), Spain\label{aff28}
\and
Institute for Theoretical Particle Physics and Cosmology (TTK), RWTH Aachen University, 52056 Aachen, Germany\label{aff29}
\and
Institute of Space Sciences (ICE, CSIC), Campus UAB, Carrer de Can Magrans, s/n, 08193 Barcelona, Spain\label{aff30}
\and
Institut d'Estudis Espacials de Catalunya (IEEC), Carrer Gran Capit\'a 2-4, 08034 Barcelona, Spain\label{aff31}
\and
INAF-Osservatorio Astronomico di Roma, Via Frascati 33, 00078 Monteporzio Catone, Italy\label{aff32}
\and
Dipartimento di Fisica e Astronomia "Augusto Righi" - Alma Mater Studiorum Universit\`a di Bologna, Viale Berti Pichat 6/2, 40127 Bologna, Italy\label{aff33}
\and
Institute for Astronomy, University of Edinburgh, Royal Observatory, Blackford Hill, Edinburgh EH9 3HJ, UK\label{aff34}
\and
Jodrell Bank Centre for Astrophysics, Department of Physics and Astronomy, University of Manchester, Oxford Road, Manchester M13 9PL, UK\label{aff35}
\and
European Space Agency/ESRIN, Largo Galileo Galilei 1, 00044 Frascati, Roma, Italy\label{aff36}
\and
ESAC/ESA, Camino Bajo del Castillo, s/n., Urb. Villafranca del Castillo, 28692 Villanueva de la Ca\~nada, Madrid, Spain\label{aff37}
\and
University of Lyon, Univ Claude Bernard Lyon 1, CNRS/IN2P3, IP2I Lyon, UMR 5822, 69622 Villeurbanne, France\label{aff38}
\and
Institute of Physics, Laboratory of Astrophysics, Ecole Polytechnique F\'ed\'erale de Lausanne (EPFL), Observatoire de Sauverny, 1290 Versoix, Switzerland\label{aff39}
\and
UCB Lyon 1, CNRS/IN2P3, IUF, IP2I Lyon, 4 rue Enrico Fermi, 69622 Villeurbanne, France\label{aff40}
\and
Departamento de F\'isica, Faculdade de Ci\^encias, Universidade de Lisboa, Edif\'icio C8, Campo Grande, PT1749-016 Lisboa, Portugal\label{aff41}
\and
Instituto de Astrof\'isica e Ci\^encias do Espa\c{c}o, Faculdade de Ci\^encias, Universidade de Lisboa, Campo Grande, 1749-016 Lisboa, Portugal\label{aff42}
\and
Department of Astronomy, University of Geneva, ch. d'Ecogia 16, 1290 Versoix, Switzerland\label{aff43}
\and
INAF-Istituto di Astrofisica e Planetologia Spaziali, via del Fosso del Cavaliere, 100, 00100 Roma, Italy\label{aff44}
\and
Department of Physics, Oxford University, Keble Road, Oxford OX1 3RH, UK\label{aff45}
\and
INFN-Padova, Via Marzolo 8, 35131 Padova, Italy\label{aff46}
\and
Institut de Ciencies de l'Espai (IEEC-CSIC), Campus UAB, Carrer de Can Magrans, s/n Cerdanyola del Vall\'es, 08193 Barcelona, Spain\label{aff47}
\and
School of Physics, HH Wills Physics Laboratory, University of Bristol, Tyndall Avenue, Bristol, BS8 1TL, UK\label{aff48}
\and
University Observatory, Faculty of Physics, Ludwig-Maximilians-Universit{\"a}t, Scheinerstr. 1, 81679 Munich, Germany\label{aff49}
\and
Institute of Theoretical Astrophysics, University of Oslo, P.O. Box 1029 Blindern, 0315 Oslo, Norway\label{aff50}
\and
Department of Physics, Lancaster University, Lancaster, LA1 4YB, UK\label{aff51}
\and
von Hoerner \& Sulger GmbH, Schlo{\ss}Platz 8, 68723 Schwetzingen, Germany\label{aff52}
\and
Technical University of Denmark, Elektrovej 327, 2800 Kgs. Lyngby, Denmark\label{aff53}
\and
Cosmic Dawn Center (DAWN), Denmark\label{aff54}
\and
Institut d'Astrophysique de Paris, UMR 7095, CNRS, and Sorbonne Universit\'e, 98 bis boulevard Arago, 75014 Paris, France\label{aff55}
\and
Max-Planck-Institut f\"ur Astronomie, K\"onigstuhl 17, 69117 Heidelberg, Germany\label{aff56}
\and
Aix-Marseille Universit\'e, CNRS/IN2P3, CPPM, Marseille, France\label{aff57}
\and
Jet Propulsion Laboratory, California Institute of Technology, 4800 Oak Grove Drive, Pasadena, CA, 91109, USA\label{aff58}
\and
AIM, CEA, CNRS, Universit\'{e} Paris-Saclay, Universit\'{e} de Paris, 91191 Gif-sur-Yvette, France\label{aff59}
\and
Universit\'e de Gen\`eve, D\'epartement de Physique Th\'eorique and Centre for Astroparticle Physics, 24 quai Ernest-Ansermet, CH-1211 Gen\`eve 4, Switzerland\label{aff60}
\and
Department of Physics, P.O. Box 64, 00014 University of Helsinki, Finland\label{aff61}
\and
Helsinki Institute of Physics, Gustaf H{\"a}llstr{\"o}min katu 2, University of Helsinki, Helsinki, Finland\label{aff62}
\and
NOVA optical infrared instrumentation group at ASTRON, Oude Hoogeveensedijk 4, 7991PD, Dwingeloo, The Netherlands\label{aff63}
\and
Universit\"at Bonn, Argelander-Institut f\"ur Astronomie, Auf dem H\"ugel 71, 53121 Bonn, Germany\label{aff64}
\and
Aix-Marseille Universit\'e, CNRS, CNES, LAM, Marseille, France\label{aff65}
\and
Department of Physics, Institute for Computational Cosmology, Durham University, South Road, DH1 3LE, UK\label{aff66}
\and
University of Applied Sciences and Arts of Northwestern Switzerland, School of Engineering, 5210 Windisch, Switzerland\label{aff67}
\and
Institut d'Astrophysique de Paris, 98bis Boulevard Arago, 75014, Paris, France\label{aff68}
\and
European Space Agency/ESTEC, Keplerlaan 1, 2201 AZ Noordwijk, The Netherlands\label{aff69}
\and
Department of Physics and Astronomy, University of Aarhus, Ny Munkegade 120, DK-8000 Aarhus C, Denmark\label{aff70}
\and
Universit\'e Paris-Saclay, Universit\'e Paris Cit\'e, CEA, CNRS, Astrophysique, Instrumentation et Mod\'elisation Paris-Saclay, 91191 Gif-sur-Yvette, France\label{aff71}
\and
Space Science Data Center, Italian Space Agency, via del Politecnico snc, 00133 Roma, Italy\label{aff72}
\and
Centre National d'Etudes Spatiales -- Centre spatial de Toulouse, 18 avenue Edouard Belin, 31401 Toulouse Cedex 9, France\label{aff73}
\and
Institute of Space Science, Str. Atomistilor, nr. 409 M\u{a}gurele, Ilfov, 077125, Romania\label{aff74}
\and
Instituto de Astrof\'isica de Canarias, Calle V\'ia L\'actea s/n, 38204, San Crist\'obal de La Laguna, Tenerife, Spain\label{aff75}
\and
Departamento de Astrof\'isica, Universidad de La Laguna, 38206, La Laguna, Tenerife, Spain\label{aff76}
\and
Departamento de F\'isica, FCFM, Universidad de Chile, Blanco Encalada 2008, Santiago, Chile\label{aff77}
\and
Satlantis, University Science Park, Sede Bld 48940, Leioa-Bilbao, Spain\label{aff78}
\and
Centre for Electronic Imaging, Open University, Walton Hall, Milton Keynes, MK7~6AA, UK\label{aff79}
\and
Centro de Investigaciones Energ\'eticas, Medioambientales y Tecnol\'ogicas (CIEMAT), Avenida Complutense 40, 28040 Madrid, Spain\label{aff80}
\and
Infrared Processing and Analysis Center, California Institute of Technology, Pasadena, CA 91125, USA\label{aff81}
\and
Instituto de Astrof\'isica e Ci\^encias do Espa\c{c}o, Faculdade de Ci\^encias, Universidade de Lisboa, Tapada da Ajuda, 1349-018 Lisboa, Portugal\label{aff82}
\and
Universidad Polit\'ecnica de Cartagena, Departamento de Electr\'onica y Tecnolog\'ia de Computadoras,  Plaza del Hospital 1, 30202 Cartagena, Spain\label{aff83}
\and
Institut de Recherche en Astrophysique et Plan\'etologie (IRAP), Universit\'e de Toulouse, CNRS, UPS, CNES, 14 Av. Edouard Belin, 31400 Toulouse, France\label{aff84}
\and
Kapteyn Astronomical Institute, University of Groningen, PO Box 800, 9700 AV Groningen, The Netherlands\label{aff85}
\and
INFN-Bologna, Via Irnerio 46, 40126 Bologna, Italy\label{aff86}
\and
Department of Mathematics and Physics E. De Giorgi, University of Salento, Via per Arnesano, CP-I93, 73100, Lecce, Italy\label{aff87}
\and
INAF-Sezione di Lecce, c/o Dipartimento Matematica e Fisica, Via per Arnesano, 73100, Lecce, Italy\label{aff88}
\and
INFN, Sezione di Lecce, Via per Arnesano, CP-193, 73100, Lecce, Italy\label{aff89}
\and
Institut f\"ur Theoretische Physik, University of Heidelberg, Philosophenweg 16, 69120 Heidelberg, Germany\label{aff90}
\and
Universit\'e St Joseph; Faculty of Sciences, Beirut, Lebanon\label{aff91}
\and
Junia, EPA department, 41 Bd Vauban, 59800 Lille, France\label{aff92}
\and
SISSA, International School for Advanced Studies, Via Bonomea 265, 34136 Trieste TS, Italy\label{aff93}
\and
INFN, Sezione di Trieste, Via Valerio 2, 34127 Trieste TS, Italy\label{aff94}
\and
ICSC - Centro Nazionale di Ricerca in High Performance Computing, Big Data e Quantum Computing, Via Magnanelli 2, Bologna, Italy\label{aff95}
\and
Instituto de F\'isica Te\'orica UAM-CSIC, Campus de Cantoblanco, 28049 Madrid, Spain\label{aff96}
\and
CERCA/ISO, Department of Physics, Case Western Reserve University, 10900 Euclid Avenue, Cleveland, OH 44106, USA\label{aff97}
\and
Laboratoire Univers et Th\'eorie, Observatoire de Paris, Universit\'e PSL, Universit\'e Paris Cit\'e, CNRS, 92190 Meudon, France\label{aff98}
\and
Dipartimento di Fisica e Scienze della Terra, Universit\`a degli Studi di Ferrara, Via Giuseppe Saragat 1, 44122 Ferrara, Italy\label{aff99}
\and
Istituto Nazionale di Fisica Nucleare, Sezione di Ferrara, Via Giuseppe Saragat 1, 44122 Ferrara, Italy\label{aff100}
\and
Dipartimento di Fisica - Sezione di Astronomia, Universit\`a di Trieste, Via Tiepolo 11, 34131 Trieste, Italy\label{aff101}
\and
NASA Ames Research Center, Moffett Field, CA 94035, USA\label{aff102}
\and
Kavli Institute for Particle Astrophysics \& Cosmology (KIPAC), Stanford University, Stanford, CA 94305, USA\label{aff103}
\and
Bay Area Environmental Research Institute, Moffett Field, California 94035, USA\label{aff104}
\and
Minnesota Institute for Astrophysics, University of Minnesota, 116 Church St SE, Minneapolis, MN 55455, USA\label{aff105}
\and
INAF, Istituto di Radioastronomia, Via Piero Gobetti 101, 40129 Bologna, Italy\label{aff106}
\and
Institute Lorentz, Leiden University, PO Box 9506, Leiden 2300 RA, The Netherlands\label{aff107}
\and
Institute for Astronomy, University of Hawaii, 2680 Woodlawn Drive, Honolulu, HI 96822, USA\label{aff108}
\and
Department of Physics \& Astronomy, University of California Irvine, Irvine CA 92697, USA\label{aff109}
\and
Department of Astronomy \& Physics and Institute for Computational Astrophysics, Saint Mary's University, 923 Robie Street, Halifax, Nova Scotia, B3H 3C3, Canada\label{aff110}
\and
Departamento F\'isica Aplicada, Universidad Polit\'ecnica de Cartagena, Campus Muralla del Mar, 30202 Cartagena, Murcia, Spain\label{aff111}
\and
Universit\'e Paris Cit\'e, CNRS, Astroparticule et Cosmologie, 75013 Paris, France\label{aff112}
\and
Department of Computer Science, Aalto University, PO Box 15400, Espoo, FI-00 076, Finland\label{aff113}
\and
NRC Herzberg, 5071 West Saanich Rd, Victoria, BC V9E 2E7, Canada\label{aff114}
\and
Ruhr University Bochum, Faculty of Physics and Astronomy, Astronomical Institute (AIRUB), German Centre for Cosmological Lensing (GCCL), 44780 Bochum, Germany\label{aff115}
\and
Instituto de Astrof\'isica de Canarias (IAC); Departamento de Astrof\'isica, Universidad de La Laguna (ULL), 38200, La Laguna, Tenerife, Spain\label{aff116}
\and
Universit\'e PSL, Observatoire de Paris, Sorbonne Universit\'e, CNRS, LERMA, 75014, Paris, France\label{aff117}
\and
Universit\'e Paris-Cit\'e, 5 Rue Thomas Mann, 75013, Paris, France\label{aff118}
\and
Univ. Grenoble Alpes, CNRS, Grenoble INP, LPSC-IN2P3, 53, Avenue des Martyrs, 38000, Grenoble, France\label{aff119}
\and
Department of Physics and Astronomy, Vesilinnantie 5, 20014 University of Turku, Finland\label{aff120}
\and
Serco for European Space Agency (ESA), Camino bajo del Castillo, s/n, Urbanizacion Villafranca del Castillo, Villanueva de la Ca\~nada, 28692 Madrid, Spain\label{aff121}
\and
Centre for Astrophysics \& Supercomputing, Swinburne University of Technology, Victoria 3122, Australia\label{aff122}
\and
ARC Centre of Excellence for Dark Matter Particle Physics, Melbourne, Australia\label{aff123}
\and
Department of Physics and Helsinki Institute of Physics, Gustaf H\"allstr\"omin katu 2, 00014 University of Helsinki, Finland\label{aff124}
\and
Oskar Klein Centre for Cosmoparticle Physics, Department of Physics, Stockholm University, Stockholm, SE-106 91, Sweden\label{aff125}
\and
Astrophysics Group, Blackett Laboratory, Imperial College London, London SW7 2AZ, UK\label{aff126}
\and
Centre de Calcul de l'IN2P3/CNRS, 21 avenue Pierre de Coubertin 69627 Villeurbanne Cedex, France\label{aff127}
\and
Dipartimento di Fisica, Sapienza Universit\`a di Roma, Piazzale Aldo Moro 2, 00185 Roma, Italy\label{aff128}
\and
INFN-Sezione di Roma, Piazzale Aldo Moro, 2 - c/o Dipartimento di Fisica, Edificio G. Marconi, 00185 Roma, Italy\label{aff129}
\and
Centro de Astrof\'{\i}sica da Universidade do Porto, Rua das Estrelas, 4150-762 Porto, Portugal\label{aff130}
\and
Zentrum f\"ur Astronomie, Universit\"at Heidelberg, Philosophenweg 12, 69120 Heidelberg, Germany\label{aff131}
\and
Dipartimento di Fisica, Universit\`a di Roma Tor Vergata, Via della Ricerca Scientifica 1, Roma, Italy\label{aff132}
\and
INFN, Sezione di Roma 2, Via della Ricerca Scientifica 1, Roma, Italy\label{aff133}
\and
Institute for Computational Science, University of Zurich, Winterthurerstrasse 190, 8057 Zurich, Switzerland\label{aff134}
\and
Mullard Space Science Laboratory, University College London, Holmbury St Mary, Dorking, Surrey RH5 6NT, UK\label{aff135}
\and
Department of Physics and Astronomy, University of California, Davis, CA 95616, USA\label{aff136}
\and
Department of Astrophysical Sciences, Peyton Hall, Princeton University, Princeton, NJ 08544, USA\label{aff137}
\and
Niels Bohr Institute, University of Copenhagen, Jagtvej 128, 2200 Copenhagen, Denmark\label{aff138}
\and
Cosmic Dawn Center (DAWN)\label{aff139}}